\documentclass[11pt,letterpaper]{article}
\usepackage[T1]{fontenc}
\usepackage[margin=1in]{geometry}
\usepackage{amssymb}

\usepackage{float}
\usepackage{tablefootnote}
\usepackage{graphicx}
\usepackage{newtxtext}
\usepackage{newtxmath}
\usepackage[numbers,sort&compress]{natbib}
\usepackage{hyperref}
\usepackage{xcolor}
\usepackage{comment}
\usepackage{amsmath}
\usepackage{bm}
\usepackage{upgreek}
\usepackage{adjustbox}
\usepackage{tabularx}
\usepackage{booktabs}
\usepackage{verbatim} 
\usepackage{makecell}

\hypersetup{
    colorlinks = true,
    urlcolor   = blue,
    citecolor  = black,
}

\numberwithin{equation}{section}
\newcommand{\backsection}[2][]{\section*{#1}#2\par}

\newif\ifCRreview
\CRreviewtrue

\usepackage{float}
\usepackage{subcaption}
\usepackage[percent]{overpic}

\usepackage{lscape} 
\usepackage{tikz}

\newcommand{\ppx}{\frac{\partial}{\partial x_1}}
\newcommand{\ppy}{\frac{\partial}{\partial x_2}}
\newcommand{\ppz}{\frac{\partial}{\partial x_3}}

\newcommand{\mbl}{\mathbf{L}}

\newcommand{\mblj}{\mathbf{L}_j}
\newcommand{\mblk}{\mathbf{L}_k}

\newcommand{\mblx}{\mathbf{L}_1}
\newcommand{\mbly}{\mathbf{L}_2}
\newcommand{\mblz}{\mathbf{L}_3}

\newcommand{\mbd}{\mathbf{D}}

\usepackage[normalem]{ulem}
\newcommand{\redsout}[1]{\bgroup\markoverwith{\textcolor{red}{\rule[0.5ex]{2pt}{0.4pt}}}\ULon{\textcolor{red}{#1}}}

\title{Resolvent-based modeling of turbulent flow over multilayer compliant surfaces}

\author{Rutvij Bhagwat and Kourosh Shoele\thanks{Corresponding author: \texttt{kshoele@eng.famu.fsu.edu}}\\[0.5em]
\small Department of Mechanical and Aerospace Engineering, Joint College of Engineering\\
\small Florida State University--Florida A\&M University, Tallahassee, FL 32210, USA}
\date{}
\begin{document}
\maketitle
\begin{abstract}
Compliant coatings offer a passive approach to modifying near-wall turbulence and potentially reducing skin-friction drag, but the coupled fluid--solid parameter space is too large to explore through high-fidelity simulations or experiments. We develop a two-way coupled resolvent--Galerkin framework for turbulent channel flow over finite-thickness, layered viscoelastic coatings. The fluid is modeled using the Navier--Stokes equations linearized about a turbulent mean state, while the solid is represented by a through-thickness Galerkin discretization of the elastodynamic equations. The solid response is condensed into a spectral surface-impedance operator at the fluid--solid interface. At \(Re_\tau=180\), stiff panels approach the rigid-wall response, whereas more compliant panels modify the low-rank response associated with the near-wall cycle. A Reynolds-number study shows that panels with fixed nondimensional material stiffness can produce  different responses depending on whether their thickness scales with inner or outer flow length scales. The influence of viscoelastic damping is examined across low- and high-frequency regimes, where stiffness-proportional damping alters the high-frequency fractional-energy distribution, while the low-frequency distribution remains relatively insensitive to damping. For multilayer panels, the fluid-facing material determines the interfacial wave speed, while the subsurface stiffness distribution modifies the panel's effective thickness. Temporal stability analysis identifies regions of parameter space where a steady harmonic response is relevant and regions where instabilities grow. Finally, a leading-order, covariance-based drag-estimation framework relates the coupled response to changes in the mean flow and to the estimated skin-friction drag. Together, these tools provide a basis for screening coating designs and selecting candidates for high-fidelity simulations and experiments.
\end{abstract}

%%======= INTRODUCTION ====================================
\section{Introduction} \label{sec:intro}
Compliant coatings have long been investigated as a passive flow control technique for delaying laminar-turbulent transition, reducing turbulent skin-friction drag, and mitigating flow-induced noise~\citep{gadelhak_2003}. For wall-bounded turbulent flows, interest has focused primarily on drag reduction given its economic significance. Skin friction is reported to contribute approximately 25--45\% of the total drag for aerial platforms~\citep{fischer_general_1974} and 80--90\% for some marine vehicles~\citep{lackenby1964}. Beyond their operational simplicity, passive techniques offer an alternative to active control by avoiding the need for complex sensing and actuation hardware~\citep{gadelhak_2003}, although their practical cost and maintenance depend on the implementation. Consequently, various passive techniques have been studied to reduce skin-friction drag, such as riblets, shark-skin-inspired denticles, superhydrophobic coatings, and large-eddy breakup devices~\citep{perlin_2016}. Compliant surfaces differ from these alternatives in that they deform dynamically in response to turbulent pressure and shear fluctuations, thereby enabling bidirectional coupling between the flow and the underlying solid.

Initial interest in compliant-wall drag reduction was stimulated by the experiments of \citet{kramer_1960, kramer_1962}, who reported drag reductions of up to 60\% for rubber-coated cylindrical bodies. Subsequent attempts to reproduce these results, however, yielded inconsistent outcomes~\citep{annu_review_riley_gad_el_hak_1988, fish_1998}. Early theoretical assessments suggested that drag reductions of this magnitude were unlikely and that, under the assumptions considered, favorable performance would require highly flexible, weakly damped materials with a solid-to-fluid density ratio, 
\(\rho_s/\rho_f\), close to unity~\citep{annu_review_riley_gad_el_hak_1988}. Although compliant panels can delay transition in laminar boundary layers~\citep{Lee_Fisher_Schwarz_1995, Carpenter_Morris_1990, DAVIES_CARPENTER_1997a}, their interaction with turbulent flows is still not well understood, with prior reviews reporting mixed outcomes~\citep{bushnell_1977, annu_review_riley_gad_el_hak_1988, gad_el_hak_1996, gadelhak_2003}. More recently, as a result of advances in experimental diagnostics~\citep{Zhang_Wang_Blake_Katz_2017, Wang_Koley_Katz_2020, Lu_Xiang_Zaki_Katz_2024, Lu_Xiang_Zaki_Katz_2025, huynh_measurements_2020, Greidanus_Delfos_Picken_Westerweel_2022} and access to large-scale high-fidelity simulation capabilities~\citep{Rosti_Brandt_2017, rosti_brandt_2020, Esteghamatian_Katz_Zaki_2022, Koseki_Aswathy_Rosti_2025}, the focus has broadened beyond determining whether compliant walls reduce drag to understanding the physical mechanisms governing the coupled interaction between compliant coatings and turbulent flows.

 \subsection{Previous Studies}
Experimental studies have reported a wide range of responses to wall compliance. For a turbulent boundary layer (TBL) over a single-layer viscoelastic coating, \citet{Lee_Fisher_Schwarz_1993} observed wider spanwise spacing and greater streamwise coherence for near-wall streaks, together with local reductions in skin-friction coefficient and Reynolds stress. 
In separate experimental campaigns, \citet{Kulik1991} and \citet{choi_turbulent_1997} found drag reductions up to 20\% and 7\%, respectively, for relatively stiff coatings, while avoiding the large deformations and surface instabilities associated with softer materials. These and subsequent studies also emphasized the importance of including the frequency dependence of material properties in modeling the interaction~\citep{kulik_pof_2008,kulik_2012_IJHFF,kulik_2018}. Using simultaneous three-dimensional flow and two-dimensional surface-deformation measurements, \citet{Zhang_Wang_Blake_Katz_2017} identified deformation waves traveling at phase speeds of $0.72U_0$ and $U_0$; however, their r.m.s. surface-deformation amplitudes were much smaller than the viscous length scale, and the authors concluded that they did not appreciably affect the flow under the conditions studied. Here, $U_0$ denotes the channel centerline velocity or the boundary-layer freestream velocity, depending on the configuration. 

Subsequent experiments employed softer materials to obtain deformations comparable to the viscous length scale~\citep{Wang_Koley_Katz_2020,Lu_Xiang_Zaki_Katz_2024,Lu_Xiang_Zaki_Katz_2025}. \citet{Lu_Xiang_Zaki_Katz_2024} reported deformation waves traveling at approximately $0.53U_0$ with wavelength close to three times the coating thickness. They also found that their results, as well as those of other studies, follow an almost linear scaling between the r.m.s. deformation normalized by coating thickness and the r.m.s. pressure fluctuation normalized by material shear modulus. \citet{Lu_Xiang_Zaki_Katz_2025} further showed that even though deformation amplitudes increase with Reynolds number, the correlation between the flow and surface deformation weakens. By varying material stiffness, \citet{Greidanus_Delfos_Picken_Westerweel_2022} identified different interaction regimes based on the ratio of the bulk flow velocity to the coating shear-wave speed and reported increased skin-friction drag in strongly coupled cases. These results suggest that stronger two-way coupling can be accompanied by increased drag, depending on the material and flow conditions~\citep{Greidanus_Delfos_Picken_Westerweel_2022, Lu_Xiang_Zaki_Katz_2024}. Although compliant coatings have historically been studied primarily in marine hydrodynamic flows, recent studies have also investigated their use for drag reduction in high-speed aerodynamic flows~\citep{chakravarty_narayanswamy_2024_aiaa,chakravarty_venkat_2024,chakravarty2026resolventanalysisinformviscoelastic}.

High-fidelity simulations have likewise observed mixed drag trends. Typically, compliant walls have been modeled either as spring-backed membranes and other surface-based models \citep{Endo01012002, xu_turbulence_2003, Kim_Choi_2014}, or, more recently, as finite-thickness volume-based models \citep{Rosti_Brandt_2017, rosti_brandt_2020, Koseki_Aswathy_Rosti_2025, Esteghamatian_Katz_Zaki_2022}, following the terminology introduced by \citet{carpenter_1990}. Depending on the model and the parameters of fluid and solid phases, simulations have predicted drag reduction~\citep{Endo01012002, fukagata_2008_ani_comp}, drag increase~\citep{Rosti_Brandt_2017,Esteghamatian_Katz_Zaki_2022}, or little change~\citep{xu_turbulence_2003}. For example, \citet{fukagata_2008_ani_comp} reported an 8\% drag reduction associated with reduced near-wall Reynolds shear stress. At $Re_\tau=138$, \citet{Kim_Choi_2014} found little change in skin friction or near-wall structure for stiffer walls, whereas a softer wall supported quasi-two-dimensional traveling waves and increased total drag through an additional form-drag contribution. Using a finite-thickness volume-based model, \citet{Rosti_Brandt_2017} found that material elasticity generally increased skin friction, and \citet{Esteghamatian_Katz_Zaki_2022} similarly reported greater skin-friction drag at a compliant wall than at a rigid wall. To separate the effects of fluid--structure-interaction (FSI)-induced wall motion from those due to a deformed surface, \citet{Koseki_Aswathy_Rosti_2025} compared compliant walls with rigid rough walls having similar roughness statistics. Although the deformed rigid walls reproduced several trends observed for compliant walls, differences remained between the rigid-wall and fully coupled cases.

High-fidelity simulations become increasingly expensive at application-relevant Reynolds numbers, while experiments do not always provide sufficient access to the flow-field details needed to identify the mechanisms underlying the coupled response. Broad parametric studies can therefore be demanding with either approach. Reduced-order models remain valuable for identifying dominant mechanisms and screening regions of the design space for further investigation.

One class of theoretical models treats the deformation of a compliant solid as a one-way response to prescribed unsteady forcing intended to represent turbulent fluctuations~\citep{Duncan_1986, chase_1991,kulik_pof_2008,kulik_2012_IJHFF}. Within this class, \citet{Benschop_Greidanus_Delfos_Westerweel_Breugem_2019} used the shear-stress and pressure  spectra from a rigid-wall DNS to predict the deformation of a finite-thickness viscoelastic solid while varying its density, compressibility, stiffness, thickness, and viscoelasticity. \citet{rempfer_2001} instead developed a low-dimensional model by projecting the Navier-Stokes equations onto Karhunen-Loève modes obtained from a rigid-wall DNS, wherein the compliant wall was represented by a spring-mass-damper system. In resolvent-based models, the turbulent flow is instead represented through the amplification of disturbances by the linearized dynamics, while the compliant wall is often approximated by a spring--mass--damper system~\citep{Luhar_Sharma_McKeon_2015,luhar_2016_J_turbulence}. These simplified approaches illustrate a recurring compromise: models with detailed structural physics often simplify or prescribe the flow, whereas models that retain more of the turbulent dynamics generally idealize the solid~\citep{duncan_1987_comparative}. The present work builds on these approaches by coupling a resolvent formulation for the turbulent flow with a Galerkin-based model of a finite-thickness multilayered viscoelastic solid, enabling the effects of frequency-dependent viscoelastic response and subsurface architecture on the coupled dynamics to be examined. 

\subsection{Resolvent-based Model}
Resolvent analysis recasts the Navier--Stokes equations linearized about a mean flow as an input--output system in the frequency domain, with nonlinear interactions treated as forcing to the linear dynamics~\citep{McKEON_SHARMA_2010}. It provides an efficient framework for identifying amplification mechanisms and coherent response structures and, when combined with a model or measurement of the forcing, for estimating flow statistics~\citep{jeun_2015,Schmidt_Towne_Rigas_Colonius_Bres_2018,Yeh_Taira_2019,Ribeiro_Yeh_Taira_2023,sun_2019_cavity_resolvent,Liu_Sun_Yeh_Ukeiley_Cattafesta_Taira_2021,Bonne_Brion_Garnier_Bur_Molton_Sipp_Jacquin_2019}. Resolvent-based methods have also been used for flow estimation~\citep{Towne_Lozano-Duran_Yang_2020,Martini_Cavalieri_Jordan_Towne_Lesshafft_2020,Amaral_Cavalieri_Martini_Jordan_Towne_2021,towne_2024_jet_estimation} and control~\citep{Martini_Jung_Cavalieri_Jordan_Towne_2022,jung_resolvent-based_2024,zhou_2026}. While most resolvent-based studies on wall-bounded turbulence have focused on smooth, rigid walls, recently resolvent-based modeling has also been used to study flows over complex surfaces, including rough, permeable, and frequency-tuned surfaces~\citep{morgan_linear_2019,Chavarin_Gómez-de-Segura_García-Mayoral_Luhar_2021,Jafari_McKeon_Arjomandi_2023}.

Most resolvent formulations for compliant walls have represented the interface through a pointwise spring-mass-damper relation. \citet{Luhar_Sharma_McKeon_2015} found that favorably modifying the energetic near-wall cycle required unphysical negative damping, whereas very-large-scale motions could be modified using positive damping; however, walls optimized for these large-scale motions amplified disturbances at other scales. \citet{luhar_2016_J_turbulence} extended this formulation to include effects due to material stiffness, tension and directional anisotropy. \citet{Huynh2021-ha} subsequently showed that the influence of a compliant wall could be represented as deterministic forcing to the corresponding rigid-wall system. \cite{Song_Wu_Wu_Liu_He_2026} used structured input-output analysis along with a spring-mass-damper model for the compliant wall to determine optimal compliant wall parameters and observed that walls optimized for one scale ---either near-wall streaks or very-large-scale motions (VLSMs)--- tend to amplify motions at other scales.  Recently, \cite{chakravarty2026resolventanalysisinformviscoelastic} employed a resolvent-based framework for turbulent boundary layers over isotropic viscoelastic coatings and found that potential reductions in skin-friction drag are strongly constrained by hydroelastic instabilities, with a narrow stable operating range in supersonic flow. While most of the resolvent-based work on the modeling of compliant wall interaction has been restricted to parallel mean flows that vary only in the wall-normal direction, \citet{Pfister_Fabbiane_Marquet_2022} applied resolvent analysis to a finite-length viscoelastic patch in a spatially developing two-dimensional laminar boundary layer.

These formulations linearize the coupled dynamics about an undeformed interface and are most applicable when surface displacement and slope are sufficiently small for nonlinear geometric effects to be neglected. For example, \citet{Huynh2021-ha} reported deformation amplitudes of \(O(10^{-4}\delta)\) at \(u_s/U_0=3.63\). Extending an efficient linear framework to resolve multilayer viscoelastic response and the influence of subsurface architecture remains an important modeling challenge.

\subsection{Outstanding Modeling Challenges}

Existing models of compliant-wall turbulence face three related challenges: capturing the relevant turbulent-flow dynamics, representing the structural response with sufficient physical fidelity, and identifying the dynamical regime in which the coupled system operates. One-way models with prescribed forcing or wall motion cannot capture the feedback through which wall deformation modifies the flow and, in turn, the forcing on the wall.

Many reduced-order formulations represent the solid using a surface-based spring--mass--damper model with wall-normal displacement only, thereby neglecting tangential deformation and explicit shear-stress forcing~\citep{annu_review_riley_gad_el_hak_1988}. The latter may become particularly important at low wavenumbers~\citep{Zhang_Wang_Blake_Katz_2017}. \citet{duncan_1987_comparative} showed that surface-based models can reproduce some responses, such as deformation waves in the solid driven by pressure fluctuations, but may provide only qualitative predictions for others, including the stability of surface waves at high flow speeds. Moreover, it is difficult to translate the effective parameters of the spring-mass-damper models to experiments and real-world applications~\citep{duncan_1987_comparative}.

These limitations are especially restrictive for biological and engineered coatings that are finite-thickness, multilayered, anisotropic or structurally heterogeneous. Dolphin skin, one of the original inspirations for compliant-wall studies, contains layered tissue and embedded features such as dermal ridges and papillae~\citep{carpenter_dolphin_secret}. Similarly, the ``stubbed'' and ``ribbed'' coatings in Kramer's original experiments incorporated fluid-filled inclusions or streamwise structural supports~\citep{kramer_1960,kramer_1962}; moreover, \citet{kramer_1977} noted that  favorable results were observed only after transitioning from single-layer to three-layer coatings.
Anisotropic panels have also shown advantages over isotropic panels for transition-control applications~\citep{Carpenter_Morris_1990}. These observations motivate volume-based continuum formulations in which finite thickness, layering, anisotropy and embedded structures can be represented directly rather than absorbed into effective interface parameters. In the present work, we focus on coatings whose properties vary through the thickness; the treatment of anisotropy and discrete embedded structures is left for future study.

{Even with an adequate structural representation, the drag response depends strongly on the hydroelastic regime.} Compliant-wall systems support several instabilities whose onset and character are  jointly governed by the flow conditions and material properties. \citet{gad_el_hak_1996} summarizes several classification frameworks: classes A, B and C~\citep{Benjamin_1960}; fluid- and solid-based mechanisms~\citep{Carpenter_Garrad_1985,Carpenter_Garrad_1986}; and convective and absolute instabilities~\citep{huerre_monkewitz_1990}. Static-divergence waves and traveling-wave flutter are two commonly encountered flow-induced surface instabilities~\citep{gad_el_hak_1996}. Because these mechanisms alter the amplitude, phase, and propagation of the wall response relative to the turbulent forcing, studies conducted in different regimes need not exhibit comparable drag trends. Characterizing the stability and dynamical regime is therefore important when interpreting and comparing performance across configurations.

Comparison among previous studies is further complicated by limited overlap in their governing parameters. For example, the drag reductions reported for the relatively stiff coatings of \citet{Kulik1991} and \citet{choi_turbulent_1997} cannot be compared directly with results for substantially softer materials because the configurations may occupy different dynamical regimes. Whereas $Re_\tau$ often serves as a useful primary descriptor for rigid-wall turbulence, characterizing the relation between inner and outer scales, compliant-wall systems require additional measures of structural scales and coupling strength. Three groups of parameters are particularly relevant: those describing the turbulent state, such as $Re_\tau$; ratios comparing material thickness with viscous and outer length scales, such as $t^*/\delta_\nu$ and $t^*/\delta$; and ratios comparing characteristic solid and fluid velocity or stiffness scales, such as $u_s/U_0$ and $E/(\rho_f U_0^2)$. Here, $t$ is the coating thickness, $\delta_\nu$ is the viscous length scale, and $\delta$ denotes the boundary-layer thickness for the cited boundary-layer studies or the channel half-height in the present calculations. The quantities $\rho_f$, $u_s$, $U_0$, and $E$ denote the fluid density, solid shear-wave speed, freestream or centerline velocity, and Young's modulus, respectively. Few studies match all three groups simultaneously. Consequently, differences in reported drag may conflate changes in the turbulence state, fluid-solid coupling strength, and structural-wave dynamics rather than indicate fundamentally contradictory behavior. 

Together, these considerations motivate a framework that retains a physically interpretable representation of turbulent amplification, resolves the finite-thickness material response, and permits a systematic exploration of the relevant flow and structural parameters.

\subsection{Paper Outline}

\begin{figure}
    \centering
    \includegraphics[width=\textwidth]{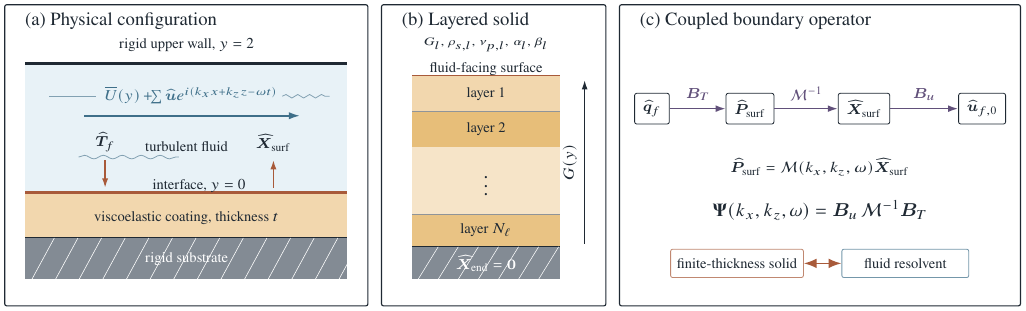}
    \caption{Physical configuration and coupled resolvent--Galerkin formulation.
    (a) Turbulent channel flow over a finite-thickness viscoelastic coating bonded
    to a rigid substrate. Pressure and shear tractions drive all three components
    of the surface displacement. (b) Coating consisting of $N_\ell$ material layers which are further discretized cumulatively into $N_s$ layers using through-thickness Galerkin discretization of
    the coating, allowing the material properties to vary with
    wall-normal position. The lower surface is clamped. (c) Dynamic condensation
    defines the displacement-based surface-impedance operator
    $\mathcal{M}(k_x,k_z,\omega)$, which is introduced into the fluid boundary
    condition through
    $\bm{\Psi}=\bm B_u\mathcal{M}^{-1}\bm B_T$.}
    \label{fig:problemFramework}
\end{figure}

The present work develops a two-way coupled framework that combines a resolvent representation of the turbulent flow with a volume-based Galerkin model of a finite-thickness viscoelastic solid. Consistent kinematic and dynamic interface conditions couple the fluid velocity to the material motion and transmit both pressure and shear-stress fluctuations to the solid. Figure~\ref{fig:problemFramework} summarizes the physical configuration, the through-thickness representation of the viscoelastic coating, and the reduction of the solid-domain equations to a frequency- and wavenumber-dependent interface operator that will be discussed later in this paper.

The presented formulation can be viewed as combining the complementary strengths of the two-way resolvent model of \citet{Luhar_Sharma_McKeon_2015}, which retains the amplification mechanisms of the turbulent flow but idealizes the wall as a spring-mass-damper system, and the one-way model of \citet{Benschop_Greidanus_Delfos_Westerweel_Breugem_2019}, which resolves a finite-thickness viscoelastic layer but prescribes the forcing from rigid-wall flow statistics. The resulting framework retains the efficiency of a linear input-output formulation while accommodating finite thickness, viscoelastic effects, and through-thickness inhomogeneity via layering.

The coupled model is further integrated with a stochastic drag-estimation framework~\citep{Moarref_Jovanovic_2012,Ran_Zare_Jovanovic_2021}. With an assumed forcing model and an eddy-viscosity closure, this extends the analysis beyond individual resolvent modes to estimates of turbulent kinetic energy, mean-flow modification, and changes in skin-friction drag. The framework, therefore, provides a systematic means of relating design variables such as material properties and architecture to the amplification of energetic turbulent structures and to drag-relevant flow statistics.

In \S~\ref{sec:method}, we derive the coupled resolvent formulation and interface conditions. 
Section \S~\ref{sec:validation} verifies the solid and fluid solvers, and \S~\ref{sec:globalStability} examines the temporal stability of the coupled system. Section \S~\ref{sec:results} analyzes the solid response and the coupled resolvent response for single-layer and multilayer coatings. Section \S~\ref{sec:drag-estimation} integrates the resolvent model with a leading-order drag-estimation framework. Finally, \S~\ref{sec:conclusion} summarizes the main findings, design implications, and directions for future work.

\section{Numerical Method}
\label{sec:method}

\subsection{Problem setup and governing equations}
We consider fully developed turbulent channel flow between a rigid upper wall and a lower wall that is either rigid or compliant. The channel height is \(2\delta\), and the mean position of the lower wall is \(y^*=0\). Dimensional quantities are denoted by a superscript \(*\), except for the channel half-height \(\delta\), for which the superscript is omitted to distinguish it from the conventional notation for boundary-layer displacement thickness. The fluid is governed by the incompressible Navier-Stokes equations, 
\begin{align}
        \frac{\partial u_i}{\partial \mathcal{T}} + u_j \frac{\partial u_i}{\partial x_j}   &= -\frac{\partial p}{\partial x_i} + \frac{1}{Re_\tau}\frac{\partial^2 u_i}{\partial x_j\partial x_j}  \,, \label{eq:nse1}
        \\
        \frac{\partial u_i}{\partial x_i} &= 0 \,.
        \label{eq:nse2}
\end{align}
Here, \(u_i\) and \(x_i\) denote the nondimensional velocity component and coordinate in direction \(i\), respectively, and repeated indices imply summation. The variables \(p\) and \(\mathcal{T}\) denote nondimensional pressure and time. The equations are nondimensionalized using the channel half-height \(\delta\), the friction velocity \(u_\tau^*\) of the rigid-wall reference flow, and the fluid density \(\rho_f^*\). The associated pressure and time scales used here are $\rho^*_f {u^*_\tau}^2$ and $\delta/u^*_\tau$, respectively; thus $\mathcal{T}=\mathcal{T}^* u^*_\tau/\delta$ indicates non-dimensional time, whereas $t=t^*/\delta$ denotes coating thickness. The friction Reynolds number $Re_\tau$ is defined as $Re_\tau = u^*_\tau \delta / \nu^*$, where $\nu^*$ is the kinematic viscosity. Quantities expressed in inner units based on the reference friction velocity \(u_\tau^*\) and viscous length \(\nu^*/u_\tau^*\) are denoted by a superscript \(+\). Calculations are performed at $Re_\tau=180$, $550$, $950$, and $2000$. The corresponding mean velocity profiles are taken from the channel-flow DNS database \url{https://turbulence.oden.utexas.edu/data/channels/profiles/}~\citep{delAlamo_Jimenez_2003, DEL_ALAMO_JIMENEZ_ZANDONADE_MOSER_2004, hoyas_and_jimenez,hoyas_jimenez_2008} and are used as rigid-wall reference mean states for the compliant-wall analysis.

\subsection{Linearized Navier--Stokes equations and resolvent model for a compliant wall}
We decompose the flow state as \(\boldsymbol q=\bar{\boldsymbol q}+\boldsymbol q'\), where \(\boldsymbol q=[\boldsymbol u,p]^T\), \(\boldsymbol u=(u,v,w)^T\), and \(p\) are the state vector, velocity vector, and pressure, respectively. Throughout the paper, $\bar{(\cdot)}$ indicates the time-averaged mean flow and  ${(\cdot)'}$ indicates a perturbation from the mean. The linearized Navier--Stokes (LNS) equations with nonlinear forcing can be written as 
\begin{equation}
    \mathbf{B} \frac{\partial \bm{q'}}{\partial \mathcal{T}} = \mathbf{A}\bm{q}' + \bm{f}' \,,
    \label{forced_lde}
\end{equation}
Here, \(\mathbf A\) is the Navier--Stokes operator linearized about \(\bar{\boldsymbol q}\), and \(\boldsymbol f'\) contains the fluctuating nonlinear terms and any externally imposed momentum forcing. The forcing component associated with the continuity equation is zero. The descriptor matrix $\mathbf{B}$ has identity blocks for the velocity equations and a zero block for the algebraic incompressibility constraint. We consider a statistically stationary, fully developed turbulent flow whose mean velocity is independent of the wall-parallel coordinates and time. Fourier decomposition in the wall-parallel directions and time expresses the disturbance field as a superposition of modes: 
\begin{equation}
    \bm{q}'(x,y,z,\mathcal{T}) = \sum_{\bm{k},\omega} \hat{\bm q}_{\bm k}(y) e^{i(k_x x + k_z z - \omega\mathcal{T})}.
    \label{eq:modalAnsatz}
\end{equation}
Here, $\bm{k}$ represents the wavenumber-frequency triplet $(k_x,k_z,\omega)$, where $k_x$, $k_z$, and $\omega$ are the streamwise wavenumber, spanwise wavenumber, and angular frequency, respectively; $\hat{\Box}$ is used to denote variables after the Fourier transform. Explicit forms of the operators $\mathbf{A}$ and $\mathbf{B}$ are given in Appendix~\ref{appA}. 
 For a stable matrix pencil ($\mathbf{A}$, $\mathbf{B}$), the  equation~\eqref{forced_lde} gives the asymptotic frequency-domain response 
\begin{equation}
\bm{\hat{q}_{k}}
 =
   \left(
 -i \omega \mathbf{B}
   -  
  \mathbf{A} \right)^{-1}
 \bm{\hat{f}_{k}} =\mathbf{H}(k_x,k_z,\omega)  \bm{\hat{f}_{k}} \,,
\end{equation}
Here, $\mathbf{H}(k_x,k_z,\omega)$ is the resolvent operator, and $\hat{\bm f}$ is the Fourier transform of $\bm f'$ and drives the perturbation dynamics about the mean flow~\citep{McKEON_SHARMA_2010, McKeon_2017}. 

{
To evaluate amplification using the fluid kinetic-energy norm, we
distinguish the full-state resolvent $\mathbf{H}$ from the
input--output operator that maps momentum forcing to velocity
response. Let $\hat{\bm f}_{u,\bm k}$ denote the three-component
momentum forcing, such that
$\hat{\bm f}_{\bm k}=\mathbf{B}_f\hat{\bm f}_{u,\bm k}$, and let
$\hat{\bm u}_{\bm k}=\mathbf{C}_u\hat{\bm q}_{\bm k}$. In the
primitive-variable ordering $\bm q=[\bm u,p]^T$, the input and
output maps have the block forms
\begin{equation}
    \mathbf{B}_f =
    \begin{bmatrix}
        \mathbf{I}_{3\times3} \\
        \mathbf{0}_{1\times3}
    \end{bmatrix},
    \qquad
        \mathbf{C}_u =
    \begin{bmatrix}
        \mathbf{I}_{3\times3} & \mathbf{0}_{3\times1}
    \end{bmatrix},
\end{equation}
Here, \(\mathbf B_f\) and \(\mathbf C_u\) are the forcing-input and velocity-output maps, respectively, and \(\mathbf I_{3\times3}\) is the identity in the three velocity components.
Thus, the forcing acts on the momentum equations but not on the
incompressibility constraint. If boundary conditions are imposed
by row replacement, the corresponding rows of the discrete input
map $\mathbf{B}_f$ are set to zero. The velocity input--output
resolvent is
\begin{equation}
    \mathbf{H}_u
    =
    \mathbf{C}_u\mathbf{H}\mathbf{B}_f
    =
    \mathbf{C}_u
    \left(-i\omega\mathbf{B}-\mathbf{A}\right)^{-1}
    \mathbf{B}_f,
    \qquad
    \hat{\bm u}_{\bm k}
    =
    \mathbf{H}_u\hat{\bm f}_{u,\bm k}.
    \label{eqn:velocity-resolvent}
\end{equation}
To introduce the input--output interpretation, we first consider the singular value decomposition:
\begin{equation}
    \mathbf{H}_u
    =    \mathbf{U}\mathbf{\Sigma}\mathbf{V}^{\dagger},
    \label{resolvent_unweighted}
\end{equation}
This yields a gain-based decomposition in the unweighted Euclidean
norm. The columns of $\mathbf{V}$ define the optimal
forcing directions and the columns of $\mathbf{U}$ represent
the resulting velocity-response modes. The singular values
in $\mathbf{\Sigma}$ give the corresponding amplification
levels, and $\dagger$ denotes the conjugate transpose.
For the results presented here, amplification is evaluated using the fluid kinetic-energy norm for the velocity response and the corresponding \(L^2\) norm for the momentum forcing~\citep{Schmidt_Towne_Rigas_Colonius_Bres_2018}.
After wall-normal discretization, the same positive-definite
weight matrix $\mathbf{W}$ is used for both spaces:
\begin{equation}
    \langle \bm a,\bm b\rangle_{\mathbf{W}}
    =
    \bm a^{\dagger}\mathbf{W}\bm b,
    \qquad
    \mathbf{W}
    =
    \operatorname{diag}
    \left(\mathbf{W}_y,\mathbf{W}_y,\mathbf{W}_y\right).
\end{equation}
Here, $\bm a$ and $\bm b$ are discretized three-component
velocity or momentum-forcing fields, and $\mathbf{W}_y$
contains the positive Clenshaw-Curtis quadrature weights along the wall-normal direction. The quadrature weights are applied consistently to the velocity and forcing degrees of freedom retained in the discrete input--output operator.
For the velocity response, the squared norm is proportional
to the perturbation kinetic energy. The weighted resolvent
operator is
\begin{equation}
    \tilde{\mathbf{H}}
    =
    \mathbf{W}^{1/2}\mathbf{H}_u\mathbf{W}^{-1/2}.
\end{equation}
Its singular value decomposition is
\begin{equation}
    \tilde{\mathbf{H}}
    =
    \tilde{\mathbf{U}}
    \tilde{\mathbf{\Sigma}}
    \tilde{\mathbf{V}}^{\dagger}.
\end{equation}

The physical momentum-forcing and velocity-response modes
are recovered by premultiplying the columns of
$\tilde{\mathbf{V}}$ and $\tilde{\mathbf{U}}$, respectively,
by $\mathbf{W}^{-1/2}$. These physical modes have unit norm
under the corresponding weighted inner product. The singular
values in $\tilde{\mathbf{\Sigma}}$ quantify the velocity-norm
gain from unit-norm momentum forcing; their squares give the
corresponding ratios of squared norms.

These response modes identify structures that are strongly amplified by the linearized dynamics, with their contribution to the turbulent field dependent on the resolvent gain and the forcing statistics. The gains and modes reported below refer to this weighted decomposition.}

The solid model and its coupling to the fluid resolvent are described next.

\subsubsection {Wave propagation in the solid domain}
\label{solid_fem}
Following \citet{barbosa_2012}, we model the solid as a layered medium, with the wall-normal coordinate \(y\) defining the thickness direction. Within each layer, the three-dimensional elastic response is governed by 
\begin{equation}
    \rho_s\,\ddot{\bm{{d}}} - \mathbf{L}^T\mathbf{D L}\,\bm{ d} = \bm{b} \,,
    \label{waveeqn}
\end{equation}
where $\rho_s = \rho^*_{s} / \rho^*_{f}$ is the non-dimensional density of the solid, i.e. the density ratio, $\bm d$ is the solid displacement, $\mathbf D$ is the layer constitutive matrix, and $\bm b$ is the body-force density, which is set to zero here. Again, $\rho^*_{f}$, $\delta$, $u^*_\tau$, and $\delta/u^*_\tau$ are used as the reference density-, length-, velocity- and time-scales for non-dimensionalization. Here, $\ddot{\Box}$ implies the second time derivative. Equation \ref{waveeqn} describes the elastic response; material damping is introduced after discretization through the Rayleigh damping matrix in equation~\ref{RayleighDampingEqn}. The differential operator $\mathbf L$ is defined as
\begin{equation}
     \mbl =  \mblx \ppx
     +\mbly \ppy
     +\mblz \ppz \,.
     \label{L1L2L3Eqn}
\end{equation}
\noindent The definitions of operators $\mblx$, $\mbly$, $\mblz$, and $\mbd$ are given in Appendix~\ref{appA2}. Expanding the differential operator, equation \ref{waveeqn} becomes,
\begin{equation}
    \rho_s\ddot{\bm d}
-\sum_{j,k=1}^{3}\mblj^T\frac{\partial}{\partial x_j}
 \left(\mathbf D\mblk\frac{\partial\bm d}{\partial x_k}\right)
=\bm0\,.
\end{equation}
Here, ${(\cdot)}^{T}$ denotes the transpose of the constant matrices $\mathbf L_j$, while the spatial derivatives act on the stress term. 
We discretize the wall-normal direction using finite elements, as described in Appendix~\ref{appA2}. The coating consists of $N_\ell$ material layers and is discretized
using $N_s$ finite elements in the wall-normal direction. For a given Fourier component, the displacement field is approximated as
\begin{equation}
    \hat{\boldsymbol d}(y)=\mathbf N(y)\hat{\boldsymbol X},
    \label{part-discrete}
\end{equation}
where $\hat{\boldsymbol X}\in\mathbb C^{3(N_s+1)\times 1}$ collects the three displacement amplitudes at each of the $N_s+1$ nodes, and $\mathbf N(y)\in\mathbb R^{3\times 3(N_s+1)}$ is the shape-function matrix.
We can then write the solid-domain equations \label{eq:virWork} in the frequency domain for a given wavenumber-frequency triplet ($k_x$, $k_z$, $\omega$) as,
\begin{equation}
  \bm{\hat{P}} = [\mathbf{K}-i\omega \mathbf{C} - \omega^2 \mathbf{M} ] \bm{\hat{X}} \,= \,\mathbb{M}\,\bm{\hat{X}} \,,
  \label{solideqnKCM}
\end{equation}
Here, \(\hat{\boldsymbol X}\) contains the nodal displacement amplitudes, and \(\hat{\boldsymbol P}\) contains the corresponding assembled nodal loads. The dynamic-stiffness matrix, \(\mathbb{M} \in \mathbb{C}^{3(N_s+1)\times 3(N_s+1)}\), combines the stiffness, damping, and inertia contributions and is a matrix which represents the solid system entirely and encodes its dynamics. We note here that given the linear nature of our coupled framework, the solid domain equations follow the same modal ansatz as that of the fluid domain equations given in \ref{eq:modalAnsatz}. The global mass matrix $\mathbf{M}$ and the global stiffness matrix, $\mathbf{K}$ are defined in Appendix \ref{appA2}. Material dissipation is represented by the Rayleigh damping matrix,
\begin{equation}
\mathbf{C}=\alpha\,\mathbf{M}+\beta\,\mathbf{K},
\label{RayleighDampingEqn}
\end{equation}
where $\alpha=\alpha^* \delta / u^*_\tau$ and $\beta=\beta^* u^*_\tau / \delta$ are the {non-dimensional} mass- and stiffness-proportional damping coefficients, respectively. Unless stated otherwise, the same damping coefficients are applied throughout the coating.
The former primarily controls damping at low frequencies, whereas the latter becomes increasingly influential at high frequencies. Unless otherwise stated, the damping coefficients are set as $\alpha=0.2$ and $\beta=0.1$ throughout this paper.

In the frequency domain, the solid-domain equation~\eqref{solideqnKCM} is recast in block-matrix form by partitioning the boundary and internal degrees of freedom. Denoting the tractions and corresponding displacements at the top interface as $\bm{\hat{P}}_{\mathrm{surf}}$ and $\bm{\hat{X}}_{\mathrm{surf}}$, at the bottom rigid support as $\bm{\hat{P}}_{\mathrm{end}}$ and $\bm{\hat{X}}_{\mathrm{end}}$, and within the internal layers as $\bm{\hat{P}}_{\mathrm{int}}$ and $\bm{\hat{X}}_{\mathrm{int}}$, the system takes the form:
\begin{equation}
\left(\begin{array}{c}
\bm{\hat{P}}_\mathrm{surf} \\
\bm{\hat{P}}_\mathrm{int}   \\
\bm{\hat{P}}_\mathrm{end}
\end{array} \right) =
\left(
\begin{array}{c c  c}
\bm{\mathbb{M}}_{11} & \mathbf{\mathbb{M}}_{12} & \mathbf{\mathbb{M}}_{13}\\
\mathbf{\mathbb{M}}_{21} & \mathbf{\mathbb{M}}_{22} & \mathbf{\mathbb{M}}_{23}\\
\mathbf{\mathbb{M}}_{31} & \mathbf{\mathbb{M}}_{32} & \mathbf{\mathbb{M}}_{33}
\end{array} \right)
\left(\begin{array}{c}
\bm{\hat{X}}_\mathrm{surf} \\
\bm{\hat{X}}_\mathrm{int}   \\
\bm{\hat{X}}_\mathrm{end}
\end{array} \right) \,,
\label{FullSolidSystem}
\end{equation}
No external loads are applied to the internal nodes, so $\bm{\hat P}_{\mathrm{int}}=\bm 0$, and the rigid substrate imposes $\bm{\hat X}_{\mathrm{end}}=\bm 0$. Eliminating the internal degrees of freedom gives the surface relation
\begin{equation}
\bm{\hat{P}}_\mathrm{surf} = \underbrace{\left( \mathbf{\mathbb{M}}_{11} - \mathbf{\mathbb{M}}_{12} \mathbf{\mathbb{M}}^{-1}_{22} \mathbf{\mathbb{M}}_{21} \right)}_{\mathbf{\mathcal{M}}}
\bm{\hat{X}}_\mathrm{surf} \,,    
\label{surfequation}
\end{equation}
where $\mathbf{\mathcal M}(k_x,k_z,\omega)\in\mathbb{C}^{3\times3}$ is the effective surface dynamic-stiffness operator. This expression applies where the internal dynamic-stiffness block \(\ \mathbf{\mathbb{M}}_{22}\) is invertible. This condition was verified over the range of frequencies and wavenumbers considered. Alternatively, $\mathbf{\mathcal M}^{-1}$, which is used in \ref{compliantBCs} for the interface boundary condition, can be computed by solving the full solid system \ref{FullSolidSystem} with unit tractions in $\bm{\hat{P}}_\mathrm{surf}$.

\subsubsection{Kinematic and dynamic coupling at the fluid-solid interface}
\label{solidFluidBCs}

The fluid and solid disturbances are assumed to have the common modal dependence
$\exp[i(k_xx+k_zz-\omega\mathcal{T})]$, and the interface conditions are linearized
about the undeformed wall at $y=0$. The non-dimensional displacement of the solid surface is
written as
\begin{equation}
\bm{X}_{\mathrm{surf}}(x,z,\mathcal{T})
=
\bm{\hat X}_{\mathrm{surf}}
e^{i(k_xx+k_zz-\omega\mathcal{T})},
\qquad
\bm{\hat X}_{\mathrm{surf}}
=
\begin{pmatrix}
\hat X_{\mathrm{surf},x} \\
\hat X_{\mathrm{surf},y} \\
\hat X_{\mathrm{surf},z}
\end{pmatrix}.
\end{equation}

Kinematic continuity requires the fluid velocity at the moving interface to
equal the solid-surface velocity,
\begin{equation}
\left.\bm{u}_f\right|_{\mathrm{int}}
=
\frac{\partial \bm{X}_{\mathrm{surf}}}{\partial\mathcal{T}}.
\end{equation}
Here, \(\left.\bm{u}_f\right|_{\mathrm{int}}\) is the fluid velocity evaluated at the instantaneous fluid–solid interface. In the frequency-wavenumber domain, the solid-surface velocity is therefore
$-i\omega\bm{\hat X}_{\mathrm{surf}}$. For a parallel mean flow
$\overline{\bm{u}}=\overline{U}(y)\bm{e}_x$, a first-order expansion of the
fluid velocity about the mean interface position gives
\begin{equation}
\bm{\hat u}_{f,0}
=
\underbrace{
\begin{pmatrix}
-i\omega & -\overline{U}_y(0) & 0 \\
0 & -i\omega & 0 \\
0 & 0 & -i\omega
\end{pmatrix}}_{\mathbf{B}_u}
\bm{\hat X}_{\mathrm{surf}},
\label{fluid_bc}
\end{equation}
where
$\bm{\hat u}_{f,0}=(\hat u,\hat v,\hat w)^T_{y=0}$. The term
$-\overline{U}_y(0)\hat X_{\mathrm{surf},y}$ accounts for displacement of the
interface through the mean shear where \(U_y(0)\) is the wall-normal derivative of the streamwise mean velocity evaluated at \(y=0\).

Dynamic continuity requires the surface load applied to the solid to equal the fluid traction at the interface. With the solid outward normal
$\bm n_s=\bm e_y$ and the traction convention used in
Appendix~\ref{app:gstabProcedure},
\begin{equation}
\bm{\hat P}_{\mathrm{surf}}
=
\bm{\hat T}_f
=
\left.\mathbf{B}_T\bm{\hat q}_f\right|_{y=0},
\qquad
\bm{\hat q}_f=(\hat u,\hat v,\hat w,\hat p)^T,
\label{eq:TractionToFlow}
\end{equation}
where $\mathbf{B}_T$ is the fluid-traction operator defined as
\begin{equation}
\mathbf{B}_T =
\begin{pmatrix}
\frac{\nu_e/\nu}{Re_\tau} D_y & i k_x \frac{\nu_e/\nu}{Re_\tau} & 0 & 0\\
0 & 2 \frac{\nu_e/\nu}{Re_\tau} D_y & 0 & -1\\
0 & i k_z \frac{\nu_e/\nu}{Re_\tau} & \frac{\nu_e/\nu}{Re_\tau} D_y & 0
\end{pmatrix}. 
\label{BT_coupling}
\end{equation}
Here, \(\nu_e/\nu=1+\nu_T/\nu\) is the effective viscosity normalized by the molecular viscosity, and \(\nu_T\) is the eddy viscosity defined in equation \ref{eqn:cess}. For calculations using the standard LNS operator, \(\nu_T=0\), so that \(\nu_e/\nu=1\).

Using the surface-impedance relation~\eqref{surfequation},
\begin{equation}
\bm{\hat X}_{\mathrm{surf}}
=
\mathbf{\mathcal{M}}^{-1}\bm{\hat P}_{\mathrm{surf}},
\end{equation}
the kinematic and dynamic conditions combine to give
\begin{equation}
\bm{\hat u}_{f,0}
=
\left.
\mathbf{B}_u\mathbf{\mathcal{M}}^{-1}\mathbf{B}_T
\bm{\hat q}_f
\right|_{y=0}
\equiv
\left.
\mathbf{\Psi}(k_x,k_z,\omega)\bm{\hat q}_f
\right|_{y=0}.
\label{compliantBCs}
\end{equation}

The operator \(\boldsymbol\Psi\) incorporates the finite-thickness solid response into the fluid boundary condition and includes wall-normal derivatives through the traction operator \(\mathbf B_T\). Together with the fluid governing equations and the remaining boundary conditions, Eq.~\eqref{compliantBCs} defines the coupled fluid–solid resolvent problem.

This condition generalizes the scalar admittance relation of
\citet{Luhar_Sharma_McKeon_2015}, which couples wall-normal velocity and
pressure, by coupling all three fluid-velocity components to the complete
surface traction. Moreover, the dependence of $\mathbf{\Psi}$ on
$(k_x,k_z,\omega)$ arises naturally from the finite-element representation of
the solid. The resolvent constructed using this coupled boundary condition
is referred to hereafter as the coupled resolvent operator.

\section{Solver Verification}
\label{sec:validation}
We verify the fluid and solid implementations separately. The fluid resolvent is compared with established rigid-wall results for laminar and turbulent channel flows, and the solid solver is compared with published results for a finite-thickness viscoelastic coating. 

\subsection{Rigid-wall verification}
\label{subsec:rigidwall}
We first verify the fluid resolvent solver against established rigid-wall results for laminar and turbulent channel flows. The laminar case examines the frequency dependence of the leading resolvent gain,
while the turbulent case compares the singular-value spectrum and the structure of the leading response mode.

\subsubsection{Laminar channel flow}
\begin{figure}
    \centering
\includegraphics{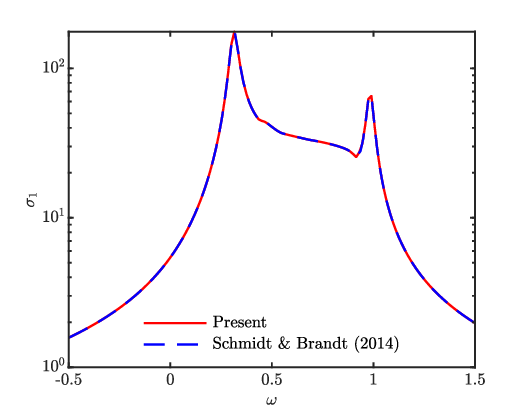}
\caption{Leading resolvent singular value $\sigma_1$ as a function
of frequency $\omega$ for $k_x=1$, $k_z=0$, and $Re=2000$.
The present results are shown by the red solid line, and the
reference implementation of \citet{schmid_brandt_2014} by the
blue dashed line.}
    \label{fig:laminarResolvent}
\end{figure}

For the laminar verification case, we consider incompressible plane
Poiseuille flow in the domain $y\in[-1,1]$. The base velocity profile is
\begin{equation}
    \overline{\bm{u}}(y)
    =
    \begin{pmatrix}
        \overline{U}(y) & 0 & 0
    \end{pmatrix}^{T},
    \qquad
    \overline{U}(y)
    =
    \overline{U}_{\max}(1-y^2).
\end{equation}
Because the base state is homogeneous in the streamwise and spanwise
directions, the disturbance field can be decomposed into Fourier modes of the
form
\begin{equation}
    \bm{q}'(x,y,z,\mathcal{T})
    = \sum_{\bm{k},\omega}
    \hat{\bm q}_{\bm k}(y)
    e^{i(k_xx+k_zz-\omega\mathcal{T})}.
\end{equation}
{The channel half-height and centerline velocity are used as the
reference length and velocity, so that $\delta=1$ and $U_{\max}=1$.
The Reynolds number for this verification case is
$Re=U_{\max}\delta/\nu$.} For $k_x=1$, $k_z=0$, and $Re=2000$, we compute the leading singular
value of the resolvent operator over a range of frequencies. $N_y=128$ points are used to discretize in the wall-normal direction. Figure~\ref{fig:laminarResolvent} shows that the present results
match those obtained using the reference implementation of
\citet{schmid_brandt_2014}. This comparison verifies the construction and
discretization of the fluid-domain resolvent operator for a laminar base flow.

\subsubsection{Turbulent channel flow}
We next verify the solver for a turbulent mean flow using the
channel-flow results of \citet{Moarref_Sharma_Tropp_McKeon_2013}.
The friction Reynolds number is $Re_\tau=2003$, based on the
channel half-height and friction velocity. The mean velocity profile
is taken from the DNS data of \citet{hoyas_and_jimenez}.

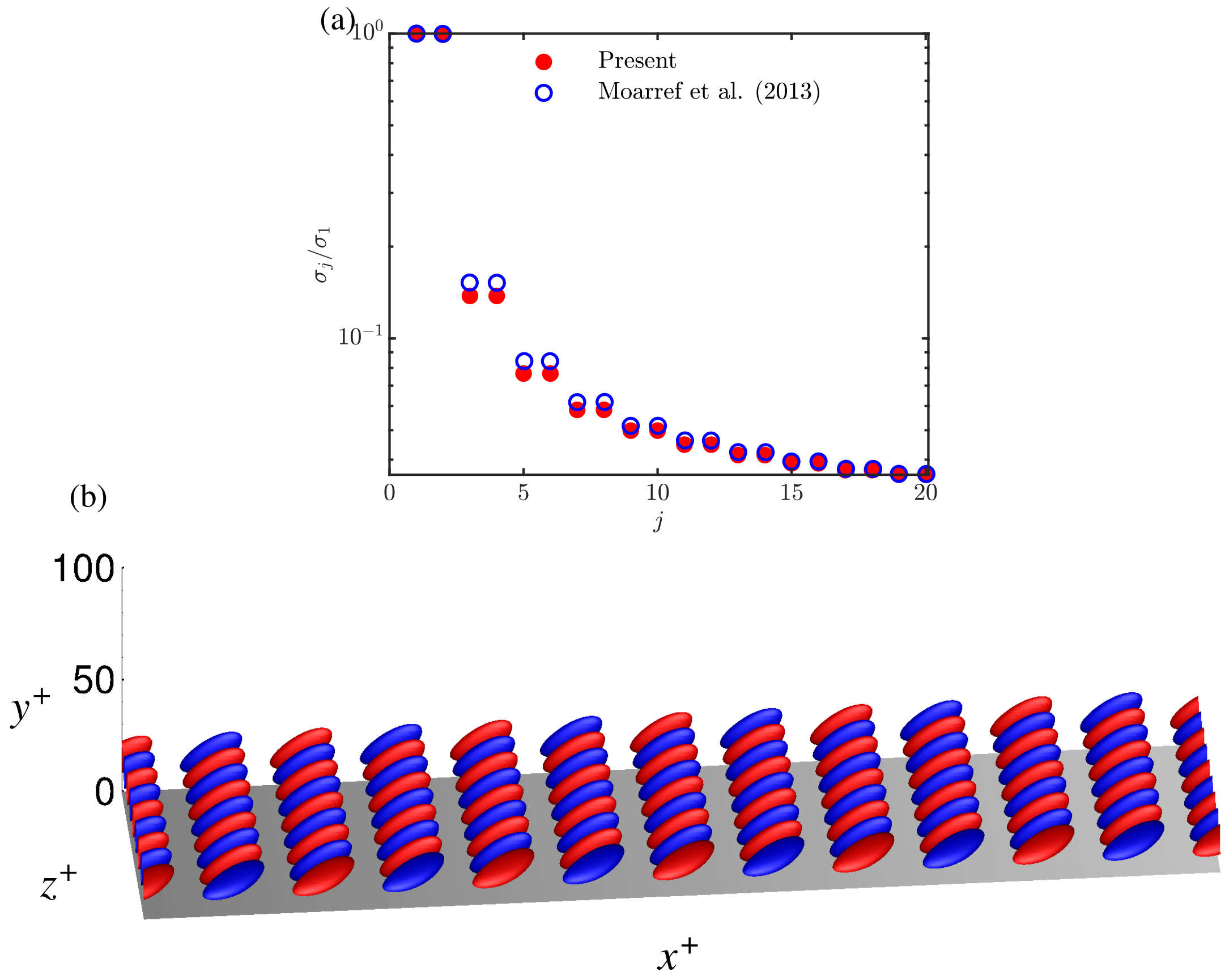
\begin{figure}
    \centering
    \subfloat{
        \begin{overpic}
        {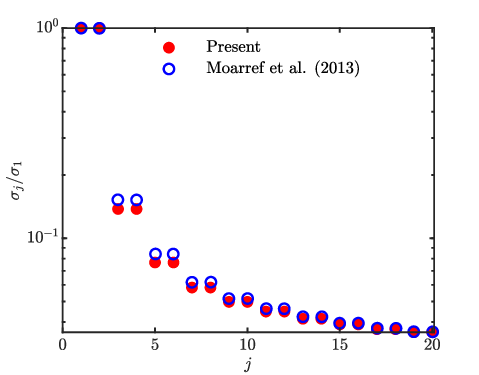}
            \put(3,73){(a)}
        \end{overpic}
    }
    \hspace{0.01\textwidth}
    \subfloat{
        \begin{overpic}[width=0.85\textwidth]
        {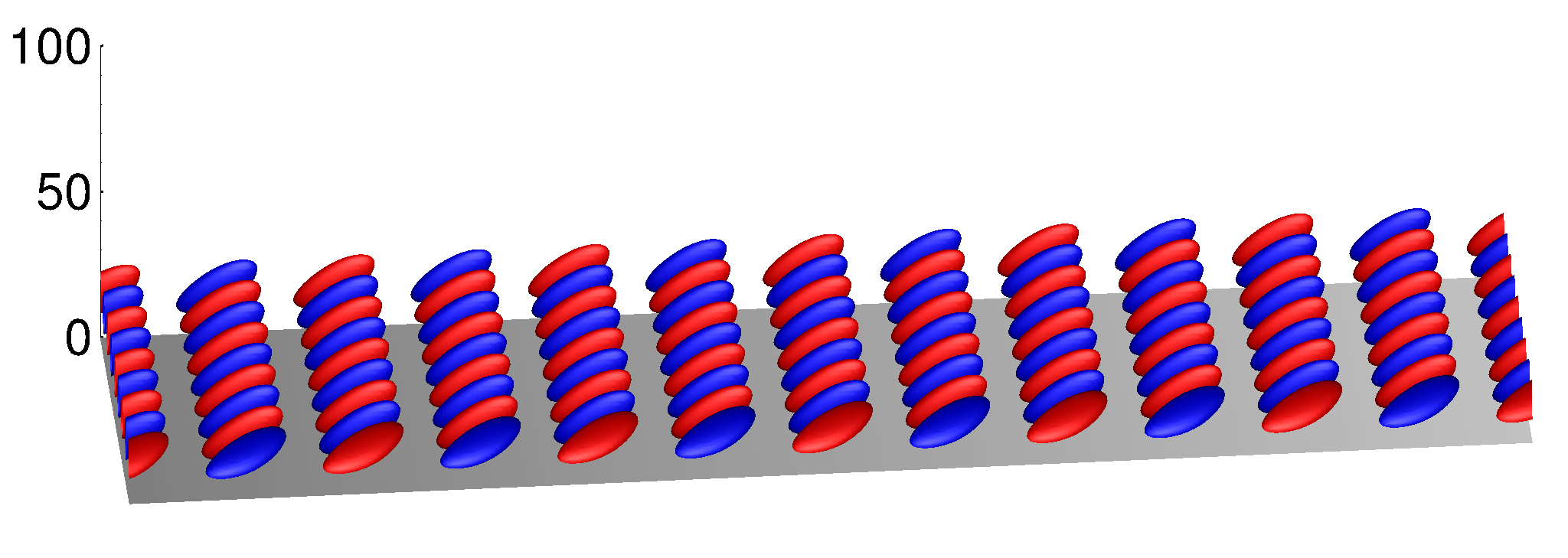}
            \put(2,37.75){(b)}
          \put(-3,20){{\Large$y^+$}}
          \put(51,-0.75){{\Large$x^+$}}
          \put(-0.5,5.3){{\Large$z^+$}}
        \end{overpic}
    }
\caption{Rigid-wall resolvent analysis of turbulent channel flow at
$\lambda_x^+=700$, $\lambda_z^+=100$, $c^+=10$, and
$Re_\tau=2003$: (a) singular-value spectrum $\sigma_j$ as a
function of mode index $j$, compared with the results of
\citet{Moarref_Sharma_Tropp_McKeon_2013}; (b) leading response
mode, shown using isosurfaces of the real part of the streamwise
velocity fluctuation.}
    \label{fig:turbulentResolvent}
\end{figure}

The resolvent operator is evaluated at
$\lambda_x^+=700$, $\lambda_z^+=100$, and $c^+=10$, which are representative
of the energetic near-wall cycle. Figure~\ref{fig:turbulentResolvent}\,(a) compares the resulting
singular-value spectrum with that reported by
\citet{Moarref_Sharma_Tropp_McKeon_2013}. The spectra agree
closely, although small differences remain. One possible
contribution is the mean profile:
\citet{Moarref_Sharma_Tropp_McKeon_2013} used a profile based on
the Cess eddy-viscosity model, whereas the present profile is
obtained directly from DNS data~\citep{hoyas_and_jimenez,
hoyas_jimenez_2008}.

The leading response mode is shown in
figure~\ref{fig:turbulentResolvent}\,(b). Its spatial organization
exhibits the near-wall streak structure expected at these wavelengths
and phase speed~\citep{Kline_Reynolds_Schraub_Runstadler_1967}.
Together, the singular-value comparison and the expected mode
structure support the rigid-wall resolvent implementation for this
turbulent-channel case.

\subsection{Verification of the solid solver}
\label{subsec:solidSolverValidation}
We next assess the solid solver by comparing its predicted surface
response with the finite-thickness viscoelastic-coating results of
\citet{Benschop_Greidanus_Delfos_Westerweel_Breugem_2019}. Their
configuration consists of a viscoelastic layer bonded to a rigid
substrate, matching the basic solid configuration considered here.

The reference model describes material dissipation through the loss factor
$\eta_s=\tan(\phi_s)$, where $\phi_s$ is the loss angle. To match this loss factor at each positive forcing frequency, we set
the mass-proportional damping coefficient to zero and choose the
stiffness-proportional coefficient as
\begin{equation}
    \alpha=0,
    \qquad
    \beta=\frac{\eta_s}{\omega}.
\end{equation}
Thus, $\beta$ is adjusted with frequency for this comparison; it is
not a single constant Rayleigh coefficient over the full frequency
range. The comparison is restricted to two-dimensional disturbances by setting
$k_z=0$, and the loss angle is fixed at $\phi_s=10^\circ$. The relative wave speed is defined as
\begin{equation}
    c_{rs}=\frac{c}{u_{s}},
   \qquad
    u_s=\sqrt{\frac{\breve{G}}{\rho_s}},
\end{equation}
Here, $c$ is the phase speed of the imposed traction wave and $u_s$
is the solid shear-wave speed, and $\breve G=G^*/(\rho_f^*{u_\tau^*}^2)$ is the shear modulus, i.e material stiffness. At $c_{rs}=0.2$, we compare the
following scaled wall-normal surface displacement:
\begin{equation}
    \mathcal{D}_y
    =
    \frac{|\hat X_{\mathrm{surf},y}|/t}
         {\rho_r c_{rs}^2},
    \qquad
    \rho_r=\frac{\rho_f}{\rho_s},
\end{equation}
The response is plotted as a function of $t/\lambda_x$. The inverse
density ratio $\rho_r$ is defined in this form to follow the
convention of
\citet{Benschop_Greidanus_Delfos_Westerweel_Breugem_2019}. It is
related to the density ratio used in \S~\ref{sec:method} by
\begin{equation}
    \rho_s = \frac{\rho_s^*}{\rho_f^*} = \frac{\rho_s}{\rho_f} = \frac{1}{\rho_r}\,.  
\end{equation}
The coating is subjected to in-phase harmonic normal and shear
traction waves, both varying as
$\exp[i(k_xx-\omega\mathcal{T})]$, with amplitudes
\begin{equation}
    {\sigma}_{y,0}=0.0238 \rho_f c^2,
    \qquad
    {\tau}_{xy,0}=0.0041 \rho_f c^2.
\end{equation}

\begin{figure}
    \centering
    \includegraphics[width=1\textwidth]
    {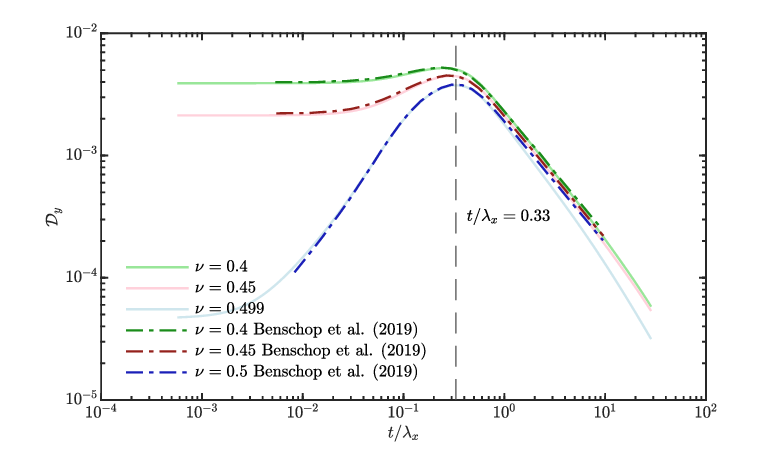}
    \caption{Scaled wall-normal surface displacement
    $\mathcal{D}_y=(|\hat X_{\mathrm{surf},y}|/t)/
    (\rho_r c_{rs}^2)$ at $c_{rs}=0.2$ for several values of Poisson's ratio
    and a fixed loss angle $\phi_s=10^\circ$. The displacement is generated
    by harmonic normal and shear tractions with amplitudes
    ${\sigma}_{y,0}=0.0238\rho_fc^2$ and ${\tau}_{xy,0}=0.0041\rho_fc^2$. Solid curves show
    the present results and dash-dotted curves show the corresponding results of
    \citet{Benschop_Greidanus_Delfos_Westerweel_Breugem_2019}. The vertical
    dashed line corresponds to
    $\lambda_x=3t$.   }
    \label{fig:benschop}
\end{figure}

Figure~\ref{fig:benschop} compares the predicted response with the
reference results for several values of Poisson's ratio at
$\phi_s=10^\circ$. To approximate the incompressible limit used by
\citet{Benschop_Greidanus_Delfos_Westerweel_Breugem_2019}, we use
$\nu_p=0.499$ rather than $\nu_p=0.5$, avoiding the singularity
associated with exact incompressibility in the present formulation.
The predicted displacement agrees closely with the reference results
over the range considered and reproduces the strong dependence on
Poisson's ratio at long wavelengths, particularly for
$\lambda_x\gtrsim 3t$. The vertical dashed line at $t/\lambda_x=1/3$ identifies the
wavelength $\lambda_x=3t$, which has also been discussed in
theoretical and experimental studies of finite-thickness compliant
coatings~\citep{Benschop_Greidanus_Delfos_Westerweel_Breugem_2019,
Lu_Xiang_Zaki_Katz_2024}. The comparison supports the solid
solver's ability to capture the wavelength-dependent surface
response and its sensitivity to material compressibility for the
cases considered.

\section{Coupled fluid--structure stability}
\label{sec:globalStability}

Recent studies~\citep{APSDFD_Dawson_2024, chakravarty2026resolventanalysisinformviscoelastic} have identified
temporally unstable modes in turbulent flows linearized about a mean
state over compliant surfaces. Here, temporal stability analysis
complements the resolvent analysis by identifying conditions under
which perturbations about the prescribed turbulent mean and undeformed
interface grow or decay. The analysis considers small perturbations
with Fourier decomposition in the homogeneous directions.

\subsection{Coupled eigenvalue problem}
Consistent with the Fourier convention in
\eqref{eq:modalAnsatz}, temporal disturbances are expressed as
$\exp[i(k_xx+k_zz)+\lambda_{(e)}\mathcal{T}]$, where
\begin{equation}
 \lambda_{(e)}=-i\omega,\qquad \omega_i=\Re(\lambda_{(e)}),\qquad
 \omega_r=-\Im(\lambda_{(e)}).
 \label{eq:gstabConvention}
\end{equation}
Thus, $\omega_i>0$ indicates temporal growth, whereas
$\omega_i<0$ indicates decay. 
We introduce the solid velocity as an independent variable and write
the structural dynamics in first-order form. In addition, we retain all solid
degrees of freedom in the eigenvalue problem and avoid the
frequency-dependent condensation used in
\eqref{surfequation} to eliminate the potential problem associated with potential singularity of 
$\mathbf{\mathbb{M}}_{22}$ in Eq. \ref{surfequation}. The resulting coupled problem is
\begin{equation}
 \mathbf A_g\bm x=\lambda_{(e)}\mathbf E_g\bm x,\qquad
 \bm x=\begin{bmatrix}\bm q_f^T&\bm X_s^T&\bm V_s^T\end{bmatrix}^{T},
 \label{eq:gstabPencil}
\end{equation}
Here, $\bm q_f$ contains the fluid velocity and pressure, while
$\bm X_s$ and $\bm V_s$ contain the unconstrained solid displacement
and velocity degrees of freedom, including the interface and internal
solid nodes.
The structural operators follow \S~\ref{solid_fem}. The interface
conditions enforce velocity compatibility, including the mean-shear
displacement correction, and transmit pressure and viscous tractions.
Incompressibility and the algebraic boundary conditions make
$\mathbf E_g$ singular. A generalized Schur (QZ) decomposition is
used to separate the finite eigenvalues from the infinite eigenvalues
associated with these constraints~\citep{Moler_Stewart_1973,
manning2008descriptorapproacheliminatingspurious}. The formulation and the checks used
to assess the computed eigenvalues are described in
Appendix~\ref{app:gstabProcedure}.

\subsection{Effect of material stiffness}
\label{sec:gstabRe180Results}
We compute the coupled eigenvalue spectrum for homogeneous
single-layer panels with stiffness values
$G=100,\,10,\,1,\,0.1,$ and $0.01$. The stiffness parameter is
defined as $G=\breve G/\breve G_{\mathrm{ref}}$, where
$\breve G=G^*/(\rho_f^*{u_\tau^*}^2)$ and
$\breve G_{\mathrm{ref}}=121$ is associated with
the baseline compliant-wall case of
\citet{Esteghamatian_Katz_Zaki_2022}, designed to promote
interaction between surface waves and energetic near-wall
pressure fluctuations.
The compliant layers have $t^+=100$,
$\rho_s/\rho_f=1$, $\nu_p=0.45$, and Rayleigh coefficients
$\alpha=0.2$, $\beta=0.1$. The Reynolds number is $Re_\tau=180$. We compute temporal spectra for a range of wavelengths: the domain $10\leq\lambda_x^+,\lambda_z^+\leq10^5$
is sampled uniformly in logarithmic coordinates, with
$\Delta\log_{10}\lambda_x^+=\Delta\log_{10}\lambda_z^+=0.1$.
A rigid-wall case is included as a reference.

We consider two linearized fluid operators: the standard LNS
operator with molecular viscosity $\nu$, and an eddy-viscosity
enhanced operator with prescribed effective viscosity
$\nu_e=\nu+\nu_T$. Both operators use the same turbulent mean
profile. The latter retains the wall-normal gradient of $\nu_e$ but
neglects perturbations of $\nu_T$.

\begin{figure}
\centering
\includegraphics[width=0.8\textwidth]{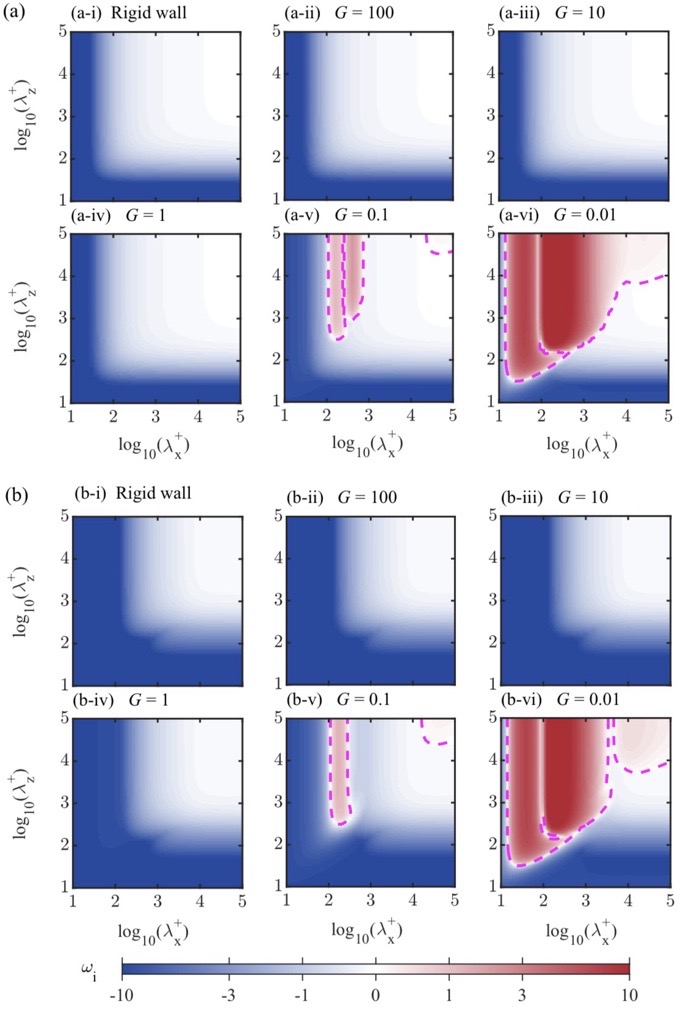}
\caption{Leading temporal growth-rate fields at $Re_\tau=180$ and
$t^+=100$: (a) molecular viscosity $\nu$ and (b) prescribed
effective viscosity $\nu+\nu_T$. Within each group, subpanels
(i)--(vi) correspond to the rigid wall and homogeneous compliant
layers with $G=100,\,10,\,1,\,0.1,$ and $0.01$, respectively.
Each field is evaluated on a $41\times41$ logarithmically spaced
wavelength grid. Blue and red indicate decay and growth,
respectively; black curves indicate zero growth. The color scale is
shared by all panels. Growth rates are normalized by
$u_\tau^*/\delta$, as in table~\ref{tab:gstabProvisional}.}
\label{fig:gstabNuTComposite}
\end{figure}

Figure~\ref{fig:gstabNuTComposite} shows the leading temporal growth
rate,
$\omega_{i,\max}=\max_j\Re(\lambda_{(e),j})$, for both viscosity
models, with the rigid-wall case included for reference. The maximum is taken over the finite eigenvalues retained after the
checks described in Appendix~\ref{app:gstabProcedure}. For the
molecular-viscosity operator, the rigid wall and the stiffest panel
exhibit decay throughout the sampled wavelength domain. As the
panel stiffness decreases, a larger region with large $\lambda_x^+$ and $ \lambda_z^+$ shows positive growth.  
As an example, for $(\lambda_x^+,\lambda_z^+)=(1000,1000)$ and $G=0.01$,
the matched calculations in table~\ref{tab:gstabProvisional} exhibit
positive leading growth under both closures while including eddy viscosity
reduces the spectral maximum by approximately $23\%$.

\begin{table}
\centering
\caption{Leading temporal growth rates for $Re_\tau=180$,
$G=0.01$, $t^+=100$, and
$(\lambda_x^+,\lambda_z^+)=(1000,1000)$. Growth rates are
normalized by $u_\tau^*/\delta$.}
\label{tab:gstabProvisional}
\begin{tabular}{lc}
\toprule
Viscosity & $\omega_{i,\max}$\\
\midrule
$\nu$ & $4.381931$\\
$\nu+\nu_t$ & $3.363345$\\
\bottomrule
\end{tabular}
\end{table}

\label{sec:gstabEvidenceContract}
A stationary stochastic interpretation of the resolvent requires
stable finite dynamics on the compatible state space. In a temporally
unstable regime, the frequency-domain operator may still be evaluated
at a real frequency when the corresponding operator is invertible;
however, its gain does not generally describe a bounded long-time
response to persistent forcing.
The present analysis considers temporal perturbations about a frozen
turbulent mean. It therefore does not establish nonlinear transition
or distinguish convective from absolute instability.
Accordingly, any covariance-based drag estimate must use finite coupled dynamics that are stable over the wavenumbers included in the statistical calculation, with consistent mean flow, viscosity closure, and interface conditions.

\section{Resolvent analysis for single-layer and multi-layer panels}
\label{sec:results}

The coupled resolvent operator describes the response of two interacting dynamical systems: the turbulent flow linearized about its mean state and the finite-thickness, multilayered viscoelastic panel. Its singular values quantify the amplification of the corresponding forcing--response pairs, while the associated response modes identify the flow structures preferentially amplified by the coupled dynamics. 
When the leading singular values are well separated from the
remaining spectrum, the coupled system may admit an effective
low-rank representation, provided that the forcing projects
sufficiently onto the corresponding forcing modes. Still, gain separation
alone does not guarantee a low-rank physical response, particularly
when the forcing preferentially excites suboptimal modes.
 In the present coupled system, this scale selection is further influenced by the compliant boundary condition, which depends explicitly on frequency and wavenumber through the surface-impedance operator. The preferentially amplified scales therefore emerge from the combined dynamics of the turbulent flow and the structural response of the panel, rather than from either subsystem independently.

We use two complementary measures throughout this section. The
singular values $\sigma_j$ quantify the amplification of the
corresponding forcing--response pairs, while
\begin{equation}
    \mathcal{E}
    =
    \frac{\sigma_1^2+\sigma_2^2}
         {\sum_j \sigma_j^2}\,,
    \label{eq:fractionalEnergy}
\end{equation}
measures the fraction of the summed squared singular values contained
in the first two modes. It represents the corresponding fraction of
the modeled response energy when the forcing-mode coefficients are
uncorrelated and have equal variance.
As was pointed out in \citet{Moarref_Sharma_Tropp_McKeon_2013}, for turbulent channel flow conditions, the leading singular values often appear in equal pairs (for wavelengths smaller than the channel half-height) and therefore the fractional energy is computed on the basis of the sum of the squares of the first \emph{two} singular values. 
 Regions of high $\mathcal{E}$ in rigid-wall channel flow have been
associated with scales containing energetic turbulent
motions~\citep{Moarref_Sharma_Tropp_McKeon_2013}. However,
$\mathcal{E}$ measures gain concentration and does not alone
determine the energy of the physical response.  
To examine changes in absolute amplification, we also report the
compliant-to-rigid squared-gain ratios for
the first and second singular values:
\begin{equation}
    \Gamma_1
    =
    \log_{10}
    \left(
    \frac{\sigma_{1,\mathrm{comp}}^2}
         {\sigma_{1,\mathrm{rigid}}^2}
    \right), \qquad \Gamma_2
    =
    \log_{10}
    \left(
    \frac{\sigma_{2,\mathrm{comp}}^2}
         {\sigma_{2,\mathrm{rigid}}^2}
    \right).
    \label{eq:relativeGain}
\end{equation}
We first examine the intrinsic response of a single-layer panel.
We then study how stiffness, thickness, damping, and Reynolds
number affect its coupled response, before considering multilayer
panels with wall-normal variations in stiffness. This sequence
distinguishes features of the isolated solid response from those of
the coupled fluid--solid system.

\subsection{Intrinsic dynamics of a single-layer viscoelastic panel}
\label{soliddynamics}

\begin{figure}
    \centering
    \begin{overpic}{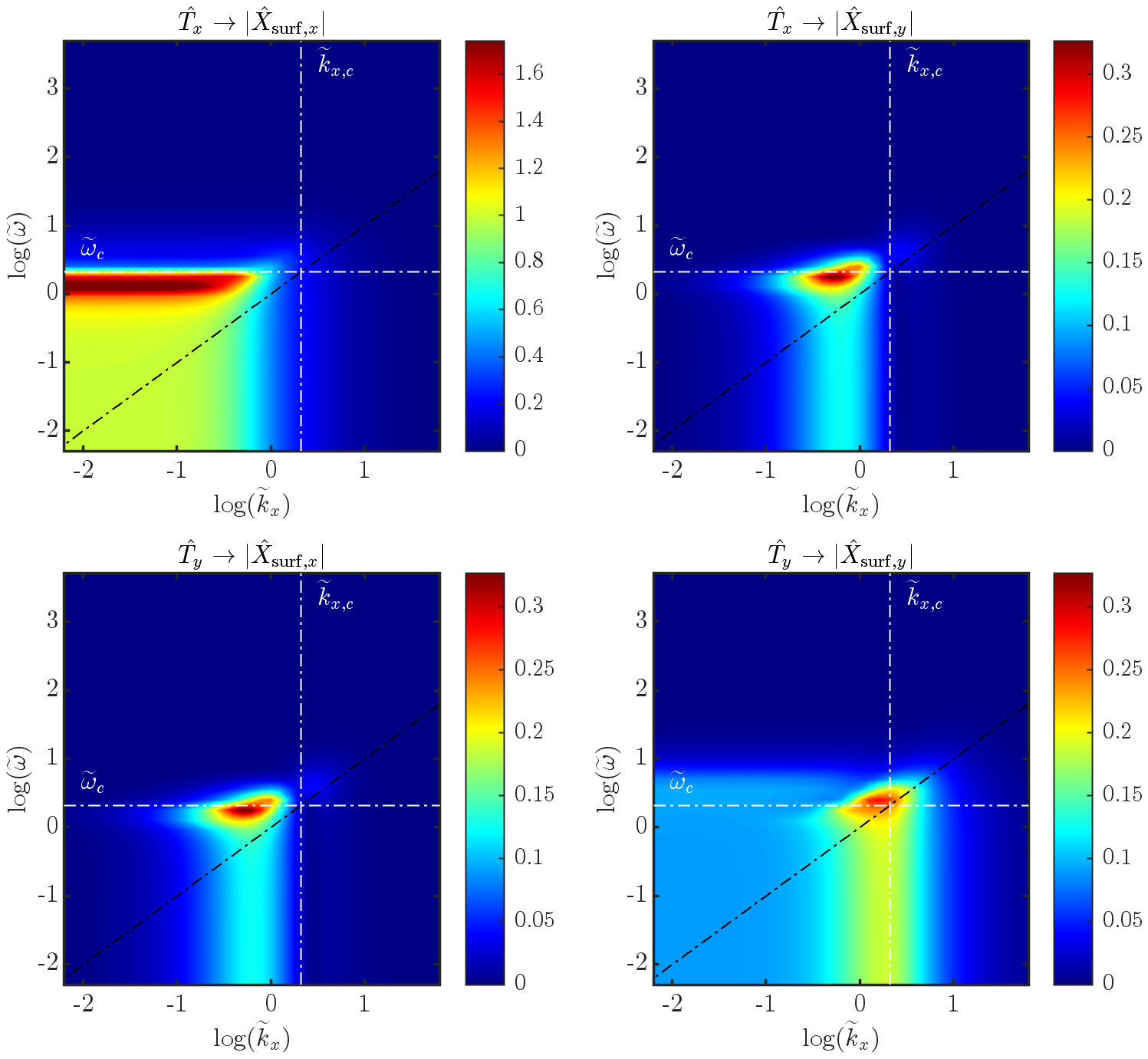}
    \put(2,90){(a)}
    \put(53,90){(b)}
    \put(2,44){(c)}
    \put(53,44){(d)}
    \end{overpic}
    \caption{Amplitudes of selected nonzero components of the surface-compliance operator %
    {
    $|(\mathbf{\mathcal{M}}^{-1})_{ij}|$} for $k_z=0$: (a) $\hat T_x\rightarrow{ |\hat X_{\mathrm{surf},x}|}$; (b) $\hat T_x\rightarrow|\hat X_{\mathrm{surf},y}|$; (c) $\hat T_y\rightarrow{ |\hat X_{\mathrm{surf},x}|}$; and (d) $\hat T_y\rightarrow { | \hat X_{\mathrm{surf},y}|}$. The vertical and horizontal white dash-dotted lines identify $\widetilde{k}_{x,c}=2\pi/(3)$ and the corresponding characteristic shear-wave reference frequency $\widetilde{\omega}_c$, respectively, whereas the diagonal black dash-dotted line denotes the material shear-speed $\widetilde{u}_s$.}
\label{fig:solidCompliance}
\end{figure}
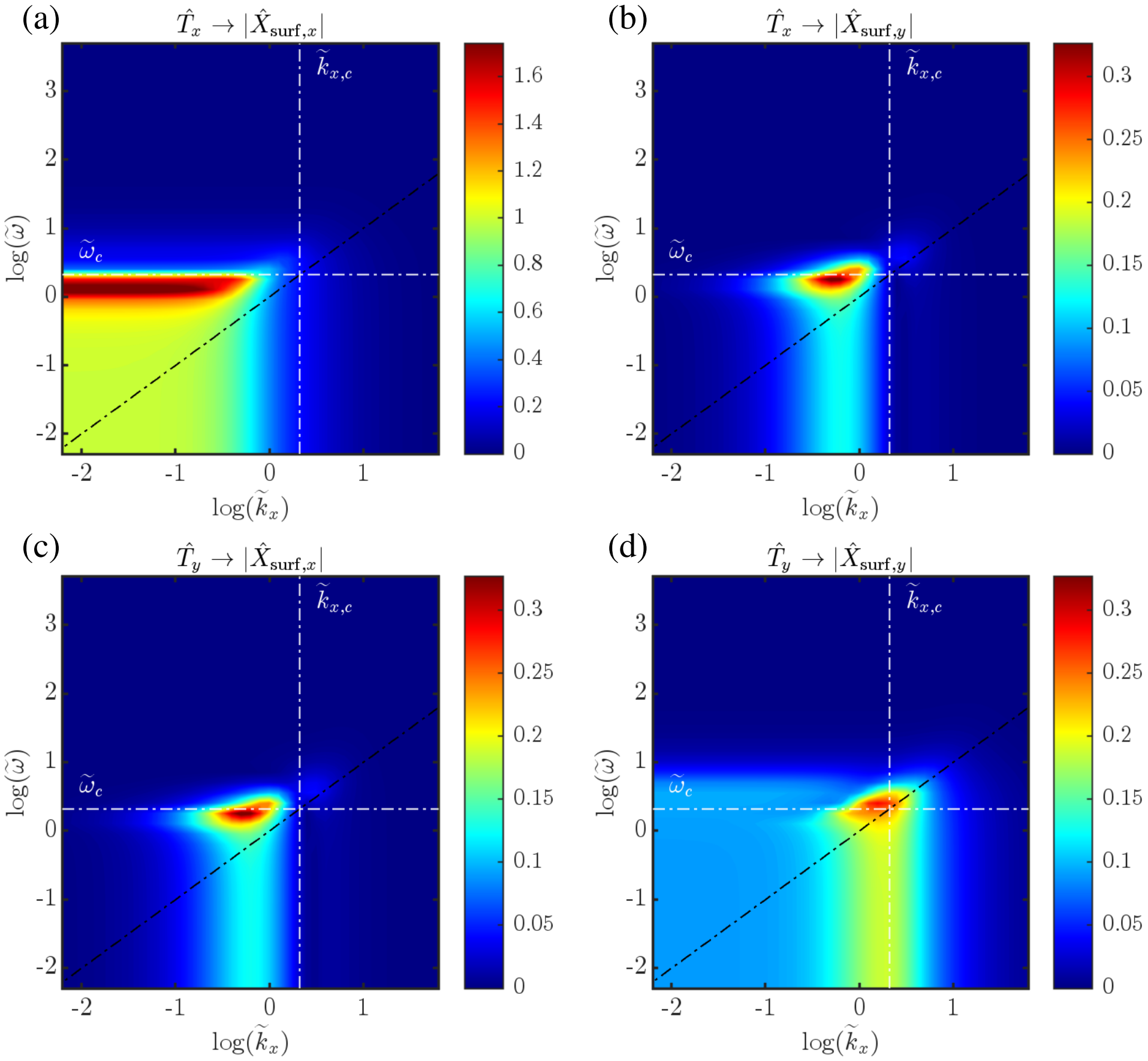

 For harmonic surface loading, the discretized solid equations can be written as
\begin{equation}
    \mathbb{M}(k_x,k_z,\omega)\bm{\hat X}
    =
    \bm{\hat P},
\end{equation}
while the surface response follows from equation~\eqref{surfequation} as
\begin{equation}
    \bm{\hat X}_{\mathrm{surf}}
    =
    \mathbf{\mathcal{M}}^{-1}(k_x,k_z,\omega)
    \bm{\hat P}_{\mathrm{surf}}.
    \label{eq:solidTransferSection4}
\end{equation}
The columns of $\mathcal{M}^{-1}$ give the three-component surface
displacement generated by unit traction in the streamwise,
wall-normal, and spanwise directions, respectively, with the other
traction components set to zero.
Figure~\ref{fig:solidCompliance} shows spectra of representative major components of this transfer function for two-dimensional disturbances. {For the solid-centric analyses, the thickness of the material $t^*$, and the shear-speed of the material $u^*_{s}= \sqrt{G^*/\rho^*_{s}}$ are selected as reference length- and velocity-scales. The non-dimensional wavenumber, frequency, and phase-speed for the solid-only analysis are then defined as,}
\begin{equation}
{
    \widetilde{\omega}
    =
    \left(
\frac{\omega^* t^*}         {u^*_{s}}\right), \qquad 
    \widetilde{k}_x
    =
    k^*_{x} t^*, \qquad \widetilde{c}=\frac{c^*}{u^*_{s}}.
    \label{eq:solidScales1}
    }
\end{equation}

The direct tangential response, $\hat T_x\rightarrow |\hat X_{\mathrm{surf},x}|$, is the strongest of the displayed components. Its largest values occupy a broad low-frequency, low-wavenumber region bounded by a characteristic streamwise wavenumber,
\begin{equation}
    \widetilde{k}_{x,c}=\frac{2\pi}{3},
    \qquad
    \widetilde{\lambda}_{x,c}=3.
    \label{eq:kxc}
\end{equation}
The corresponding shear-wave reference frequency is
$\widetilde{\omega}_c=\widetilde{k}_{x,c}\widetilde{u}_s$, where
$\widetilde{u}_s=1$ under the scaling in equation~\ref{eq:solidScales1}.
The
thickness-associated wavelength $\widetilde{\lambda}_x\approx3$
is consistent with response scales reported for finite-thickness
compliant coatings~\citep{Benschop_Greidanus_Delfos_Westerweel_Breugem_2019,
Esteghamatian_Katz_Zaki_2022,Lu_Xiang_Zaki_Katz_2024}.
In Figure~\ref{fig:solidCompliance}, the horizontal and vertical dash-dotted white lines indicate the characteristic shear-wave frequency $\widetilde{\omega}_c$ and the characteristic streamwise wavenumber $\widetilde{k}_{x,c}$, respectively, while the diagonal dash-dotted black line indicates the material shear speed $u_s$.

The amplitude of the wall-normal response to normal traction, $\hat T_y\rightarrow |\hat X_{\mathrm{surf},y}|$, extends over a wide range of wavenumbers and frequencies, but its strongest response is organized around the panel-thickness-associated characteristic wavenumber and exhibits a peak value near the shear-wave reference frequency (the point where the two dash-dotted white lines and the diagonal dash-dotted black line meet). The cross-coupled components, $\hat T_x\rightarrow |\hat X_{\mathrm{surf},y}|$ and $\hat T_y\rightarrow|\hat X_{\mathrm{surf},x}|$ form narrow high-response bands whose maxima occur near the characteristic frequency, although their preferred wavenumbers are slightly smaller than the characteristic wavenumber $\widetilde{k}_{x,c}$. These cross-coupled terms are naturally retained by the volume-based formulation but would be absent from typical surface-based models, which are restricted to normal displacement driven solely by pressure.

For the isotropic layers considered here, setting $\widetilde{k}_z=0$ decouples the spanwise displacement from the in-plane response. In this case, spanwise traction does not generate streamwise or wall-normal motion, and streamwise or wall-normal traction does not generate spanwise motion. The remaining spanwise direct response also exhibits frequency- and wavenumber-dependent behavior. The important conclusion is that even a single-layer homogeneous panel is not a local, frequency-independent boundary condition and instead acts as a scale-selective filter whose direct and cross-coupled responses depend on the panel thickness, shear-wave speed, and other material properties.

\subsection{Coupled response of a single-layer viscoelastic panel}
\label{isotropicresolvent}
We next examine the interaction between the turbulent flow and the
compliant panel. The coupled response depends on the relationship
between the flow scales, characterized by the Reynolds number, and
the structural scales set by stiffness, thickness, and viscoelastic
damping.
Unless otherwise stated, the baseline calculations in this subsection use $Re_\tau=180$, $t^+=100$, $\nu_p=0.45$, and Rayleigh damping coefficients $\alpha=0.2$ and $\beta=0.1$. The solid/fluid density ratio, $\rho_s/\rho_f$, is maintained at unity, and the viscoelastic solid is discretized using $N_s=42$ through-thickness finite elements.
The fluid is discretized using $N_y=97$, $257$, $385$, and $633$
wall-normal grid points for nominal $Re_\tau=180$, $550$, $950$,
and $2000$, respectively, matching the number of points in the
corresponding DNS mean profiles.
The phase speed is fixed at $c^+=10$ when examining the near-wall cycle.

\subsubsection{Effect of material stiffness}\label{subsec:stiffnessVariation}
Material stiffness not only defines the compliance of the solid, but it also controls the characteristic elastic wave speed $u_s$. 
The ratios $E/(\rho_f U_0^2)$ and $u_s/U_0$ are often used to characterize the relative material and flow scales in turbulent-flow--compliant-panel interactions. As discussed in \S~\ref{sec:gstabRe180Results}, the nondimensional
value $G=1$ corresponds to the reference stiffness
$\breve{G}_{\mathrm{ref}}=121$, based on the compliant material
considered by \citet{Esteghamatian_Katz_Zaki_2022}.

\begin{figure}
    \centering
    \begin{overpic}{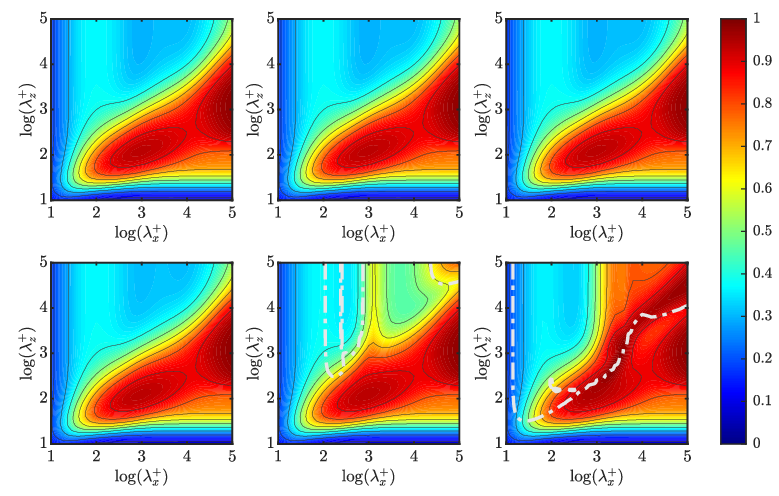}
    \put(2,63){(a)}
    \put(31.5,63){(b)}
    \put(61,63){(c)}    
    \put(2,31.5){(d)}
    \put(31.5,31.5){(e)}
    \put(61,31.5){(f)}    
    \end{overpic}
    \caption{Fractional energy ($\mathcal{E}= (\sigma_1^2 + \sigma_2^2)/\sum^N_{j=1} \sigma_j^2$) contours for $Re_\tau=180$ and material stiffness: (a) rigid; (b) $G=100$; (c)  $G=10$; (d) $G=1$; (e)  $G=0.1$; (f) $G=0.01$. The dash-dotted gray line indicates the boundary between stable and unstable response regimes from linear stability analysis.}
\label{fig:fractionalEnergyGSweep}
\end{figure}

Figure~\ref{fig:fractionalEnergyGSweep} compares the fractional energy $\mathcal{E}$ over a range of wavelengths for a rigid wall, and for compliant panels with varying material stiffness, $G=100$,  $G=10$, $G=1$,  $G=0.1$, and $G=0.01$; the phase-speed is held constant at $c^+=10$, which is representative of the near-wall cycle. The dash-dotted lines indicate the zero-growth contours obtained
from the stability calculations in
\S~\ref{sec:globalStability} and
figure~\ref{fig:gstabNuTComposite}; they separate the sampled
wavelength regions with and without a temporally unstable mode.

For the rigid wall, the dominant low-rank region is identified near
\begin{equation}
    (\lambda_x^+,\lambda_z^+)\approx(1000,125),
\end{equation}
which is consistent with the energetic near-wall cycle~\citep{Kline_Reynolds_Schraub_Runstadler_1967,Moarref_Sharma_Tropp_McKeon_2013}.
The fractional-energy distributions for the stiff compliant panels
remain close to the rigid-wall result. As $G$ decreases, the
distribution of $\mathcal{E}$ changes gradually, with the largest
modifications appearing for the softest panels.
For the wavelength range considered, unstable eigenvalues were observed for the $G=0.1$ and $G=0.01$ panels in the parametric regions above the dash-dotted gray lines as shown in figure~\ref{fig:fractionalEnergyGSweep}\,(e,f). For $G=0.1$, a vertical band of high $\mathcal{E}$-values is observed for streamwise wavelengths slightly larger than $\lambda_x^+\approx1000$. 
Temporal instability is observed in bands with
$\lambda_x^+<1000$ and at very large streamwise and spanwise
wavelengths. The local maximum associated with the near-wall cycle
is slightly distorted but remains qualitatively similar to the
rigid-wall result. 
At $G=0.01$, temporally unstable modes are found over a broad range
of streamwise and spanwise wavelengths, and the fractional-energy
distribution near the near-wall-cycle scales is substantially
modified. 
The reference location of the near-wall peak from the rigid case $(\lambda_x^+,\lambda_z^+)\approx(1000,125)$ remains in the stable region, but a local maximum is not observed in its vicinity. The fractional-energy contours lie close to the stability boundary
in parts of the wavelength plane, indicating that the distribution
of gain among the leading modes changes near this boundary. The
corresponding changes in absolute gain are examined separately
below.

\begin{figure}
    \centering
    \subfloat{
    \begin{overpic}[width=0.8\textwidth]{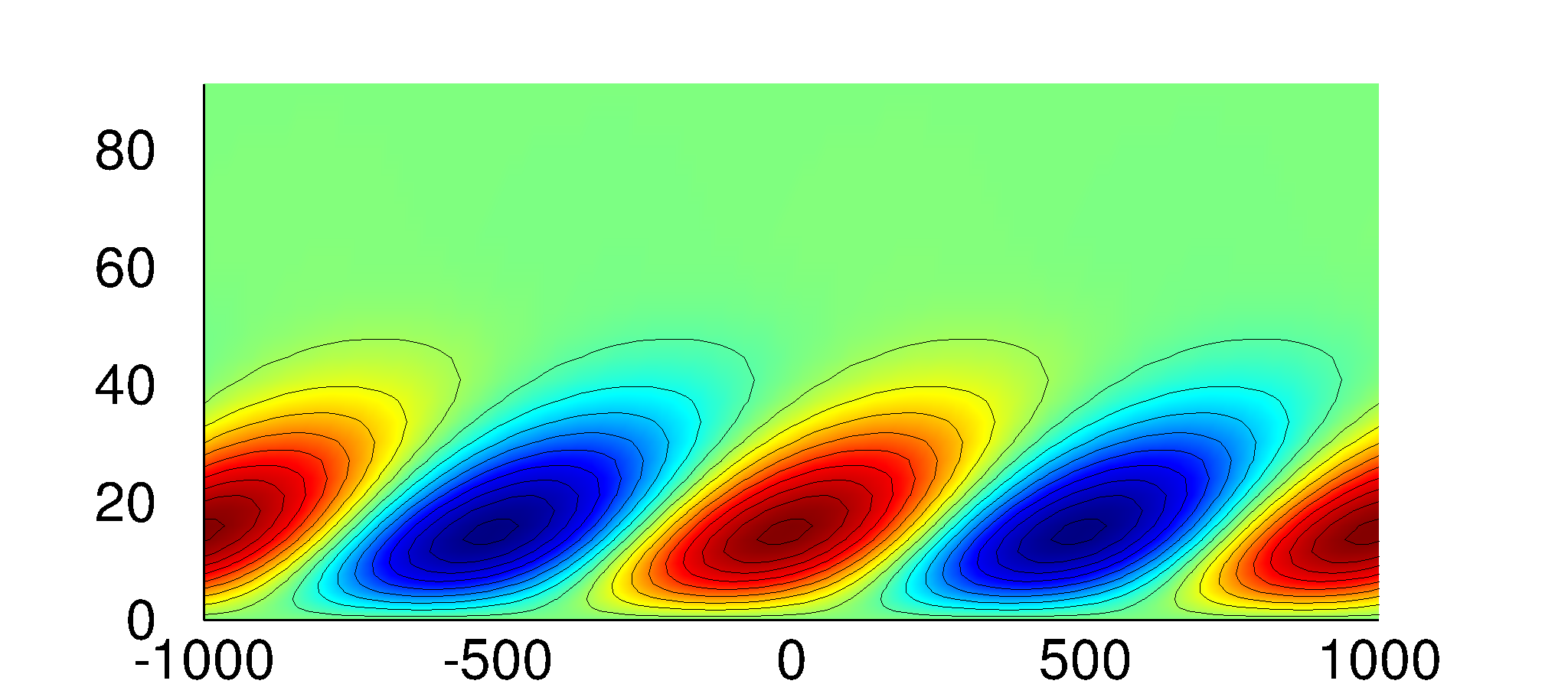}
    \put(0,40){(a)}
    \put(-1,20){{\Large $y^+$}}
    \put(49,-3){{\Large $x^+$}}
    \end{overpic}
    }
    \\
    \vspace{0.4cm}
    \subfloat{
    \begin{overpic}[width=0.8\textwidth]{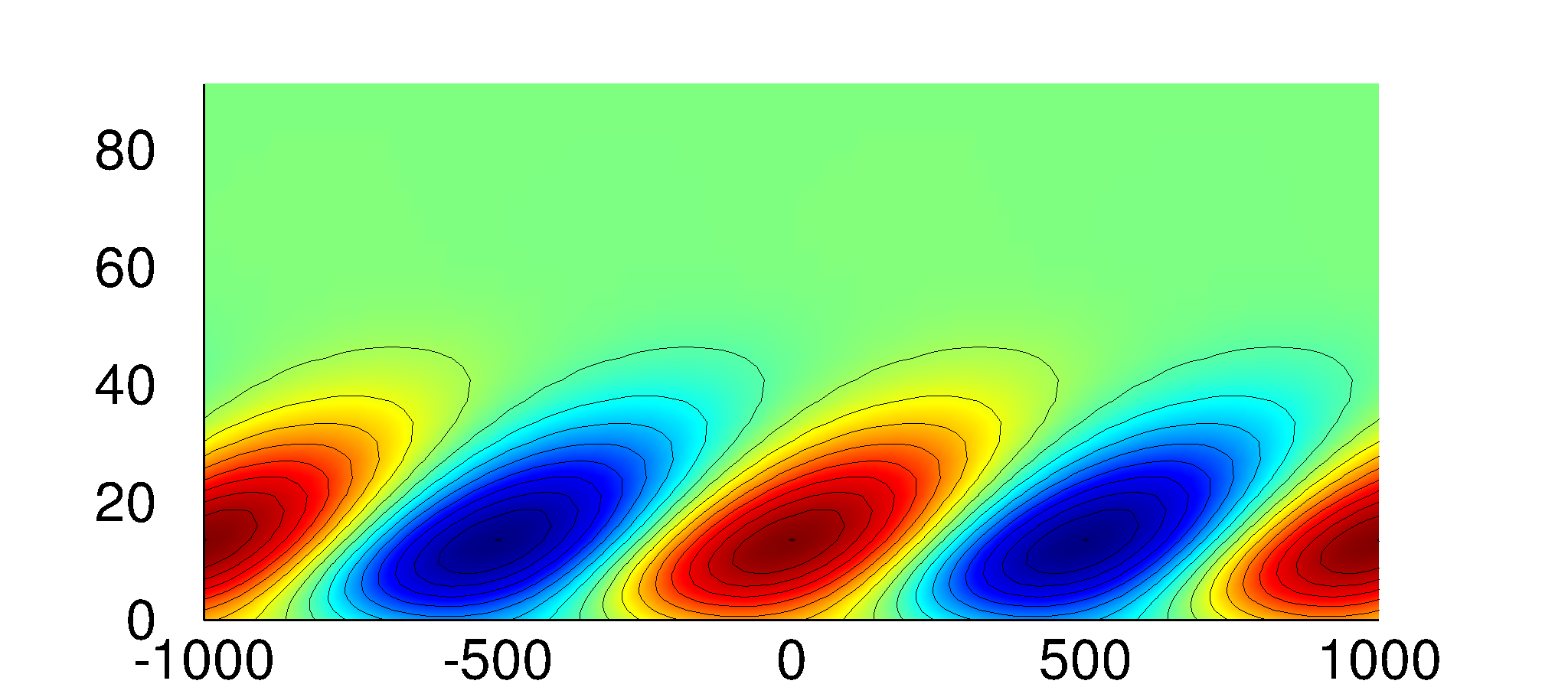}
    \put(0,40){(b)}
    \put(-1,20){{\Large $y^+$}}
    \put(49,-3){{\Large $x^+$}}
    \end{overpic}
    }
    \vspace{0.4cm}    
    \caption{Leading resolvent mode corresponding to the near-wall (NW) cycle for: (a) a rigid wall; (b) a highly compliant wall $G=0.01$. {The stated comparison uses $Re_\tau=180$, $t^+=100$ and $c^+=10$, at the rigid-wall near-wall peak, approximately $(\lambda_x^+,\lambda_z^+)=(1000,125)$.} }
\label{fig:modeShapesIsotropic}
\end{figure}

Figure~\ref{fig:modeShapesIsotropic} compares the leading response
modes at the location (in wavelength space) of the rigid-wall near-wall peak for a rigid wall and a compliant panel ($G=0.01$). The rigid-wall mode
exhibits the familiar near-wall streak structure. For $G=0.01$,
the principal streak-like structure remains identifiable but is
shifted closer to the compliant wall and has a non-zero streamwise
velocity component at the mean interface. 
These differences illustrate how the coupled interface conditions
modify the spatial structure of the leading response mode.

\begin{figure}
    \centering
    \begin{overpic}{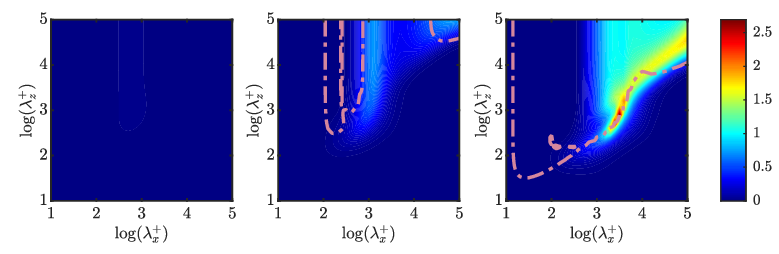}
    \put(2,31){(a)}
    \put(31.3,31){(b)}
    \put(60.5,31){(c)}    
    \end{overpic}
\caption{Logarithmic compliant-to-rigid relative gain $\Gamma_1=\log_{10}(\sigma_{1,\mathrm{comp}}^2/\sigma_{1,\mathrm{rigid}}^2)$ for $Re_\tau=180$, $t^+=100$ and $c^+=10$: (a) $G=1$, (b) $G=0.1$, and (c) $G=0.01$. Positive and negative values denote amplification and attenuation relative to the rigid wall, respectively. The dash-dotted line separates the stable and unstable regimes from linear stability analysis.}
\label{fig:GainRigidCompliant1}
\end{figure}

\begin{figure}
    \centering
    \begin{overpic}{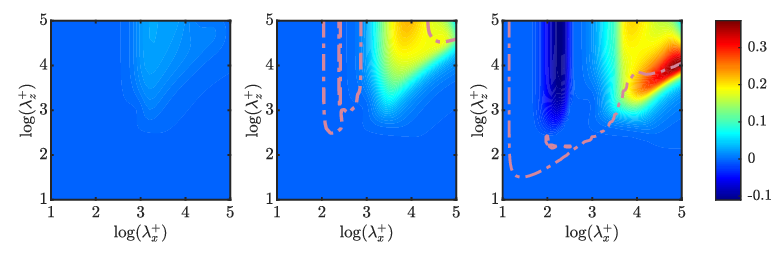}
    \put(2,31){(a)}
    \put(31.3,31){(b)}
    \put(60.5,31){(c)}    
    \end{overpic}
    \caption{Logarithmic compliant-to-rigid relative gain for the second singular value $\Gamma_2=\log_{10}(\sigma_{2,\mathrm{comp}}^2/\sigma_{2,\mathrm{rigid}}^2)$ for $Re_\tau=180$, $t^+=100$ and $c^+=10$: (a) $G=1$, (b) $G=0.1$, and (c) $G=0.01$. Positive and negative values denote amplification and attenuation relative to the rigid wall, respectively. The dash-dotted line separates the stable and unstable regimes from linear stability analysis.}
\label{fig:GainRigidCompliant2}
\end{figure}

While regions of high $\mathcal{E}$ have been associated with
energetic turbulent motions in rigid-wall channel flow
\citep{Moarref_Sharma_Tropp_McKeon_2013}, we also examine how the
compliant wall modifies the absolute resolvent gain.
The logarithmic compliant-to-rigid squared-gain ratios for the first
and second modes, $\Gamma_1$ and $\Gamma_2$, are shown in
figures~\ref{fig:GainRigidCompliant1} and
\ref{fig:GainRigidCompliant2}, respectively.
For the $G=1$ panel, little modification is observed for $\Gamma_1$. 
For the $G=0.1$ and $G=0.01$ panels, the leading gain is either
increased or remains close to the rigid-wall value over most of the
wavelength plane. No substantial attenuation is apparent in the
displayed relative-gain maps.
For the $G=0.1$ case, modest amplification is observed along vertical bands with wavelength greater than $\lambda_x^+=1000$; this region of amplification lies outside the unstable region which is denoted by dash-dotted pink lines. Some amplification is also observed for very high values of streamwise and spanwise length-scales; however, this amplification lies within the unstable region. For $G=0.01$, the region of elevated relative gain lies within the
temporally unstable region and follows the stability boundary. These
frequency-domain gains should therefore be interpreted with the
stability limitations discussed in
\S~\ref{sec:globalStability}.

The second relative gain, $\Gamma_2$, shown in
figure~\ref{fig:GainRigidCompliant2}, varies with scale and exhibits
both amplification and attenuation for the softest panel,
$G=0.01$. Attenuation occurs in a band centered near
$\lambda_x^+\approx200$, while stronger amplification appears at
larger streamwise and spanwise wavelengths. Both regions lie within
the temporally unstable region. The $G=0.1$ panel exhibits
amplification over a similar range, but primarily within the stable
region. These results demonstrate that wall compliance exerts a scale-dependent influence on energetic motions, selectively attenuating or amplifying specific response modes compared to the rigid-wall baseline.

\begin{figure}
    \centering
    \begin{overpic}{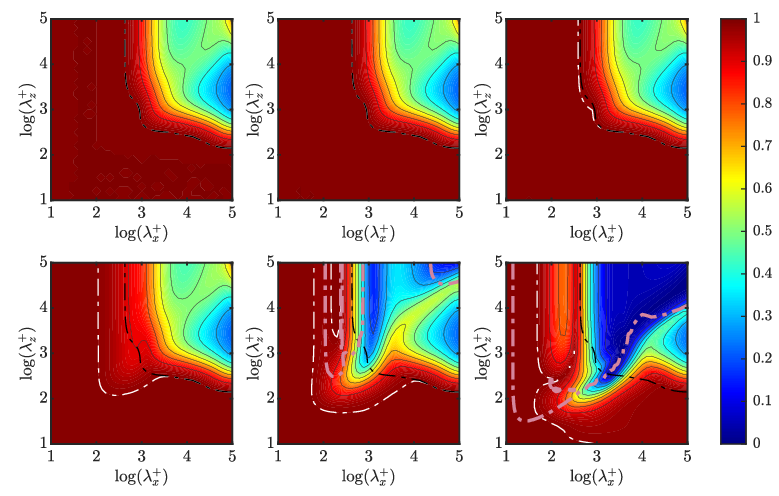}
    \put(2,63){(a)}
    \put(31.5,63){(b)}
    \put(61,63){(c)}    
    \put(2,31.5){(d)}
    \put(31.5,31.5){(e)}
    \put(61,31.5){(f)}    
    \end{overpic}
\caption{Ratio of the squares of the first two singular values $ \Theta= \sigma_2^2 / \sigma_1^2$ contours for $Re_\tau=180$ and material stiffness: (a) rigid; (b) $G=100$; (c) $G=10$; (d) $G=1$; (e) $G=0.1$; (f) $G=0.01$. The dash-dotted pink line indicates the boundary between stable and unstable response regimes from linear stability analysis.}
\label{fig:EnergyRatioGSweep}
\end{figure}

%\subsubsection{Pairing of the leading singular values}
In rigid-wall channel flow, the leading singular values commonly occur in nearly equal pairs, especially for smaller wavelengths~\citep{Moarref_Sharma_Tropp_McKeon_2013}. We quantify this using a ratio 
\begin{equation}
\Theta = \sigma_2^2/\sigma_1^2, 
\end{equation}
which approaches unity when the first two modes appear as a leading pair. This ratio is plotted for a rigid-wall and compliant walls with varying material stiffness values in figure~\ref{fig:EnergyRatioGSweep}. For the comparisons below, we identify the leading gains as
approximately paired when $\Theta\geq0.99$. The change in $\Theta$ is small for stiff panels but becomes pronounced for softer panels with $G \leq 1$, where a substantial range of wavelengths remains paired for the rigid wall but not for the compliant wall. White and black dash-dotted lines indicate $\Theta = 0.99$ for the compliant and rigid walls, respectively. Naturally, only the black dashed-dotted line is shown for the rigid case in  figure~\ref{fig:EnergyRatioGSweep}\,(a). For the $G=0.1$ and $G=0.01$ cases, which admit an unstable eigenvalue over a range of wavelengths, the stable and unstable regions are demarcated with a pink dash-dotted line. 

To examine this change, we divide the wavelength plane into three
regions using the $\Theta=0.99$ contours. The leading gains are
paired for both walls in region A, unpaired for both walls in region
B, and paired only for the rigid wall in region C. 

\begin{figure}
  \centering
  \begin{overpic}{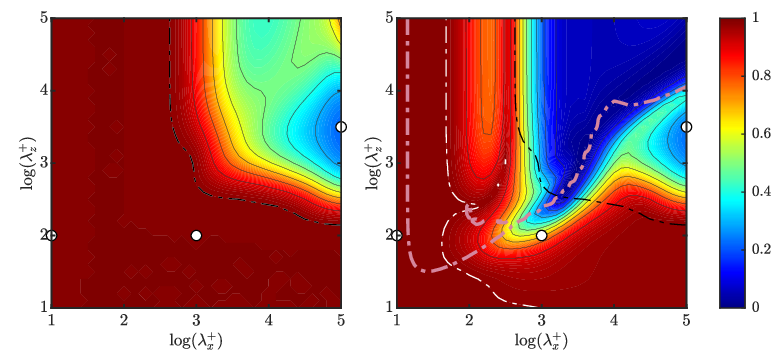}
    \put(2,45){(a)}
    \put(47,45){(b)}
    \put(7,20){\tikz[baseline=(char.base)]{
        \node[shape=circle, draw=black, fill=white, inner sep=1pt] (char) {A};}}
    \put(51.3,20){\tikz[baseline=(char.base)]{
        \node[shape=circle, draw=black, fill=white, inner sep=1pt] (char) {A};}}
    \put(39,25){\tikz[baseline=(char.base)]{
        \node[shape=circle, draw=black, fill=white, inner sep=1pt] (char) {B};}}
    \put(82,25){\tikz[baseline=(char.base)]{
        \node[shape=circle, draw=black, fill=white, inner sep=1pt] (char) {B};}}
    \put(65,11.5){\tikz[baseline=(char.base)]{
        \node[shape=circle, draw=black, fill=white, inner sep=1pt] (char) {C};}}
  \end{overpic}
  \caption{Singular value ratio $\Theta = \sigma_2^2 / \sigma_1^2$ at $Re_\tau = 180$ for (a) rigid and (b) compliant ($G = 0.01$) walls. (a) A black dash-dotted line separates region A (paired leading modes) from region B (unpaired modes). (b) White and black dash-dotted lines denote the $\Theta=0.99$ contours for the compliant and rigid walls, respectively,
dividing the wavelength plane into regions A (paired in both cases), B (unpaired in both cases), and C (paired only for the rigid wall). White circles mark the locations of mode shapes shown in figure~\ref{fig:modeShapesRegions}. The dash-dotted pink line indicates the boundary between stable and unstable regimes from linear stability analysis.}
\label{fig:S1S2-annotated}\end{figure}

\begin{figure}
    \centering
    \begin{overpic}{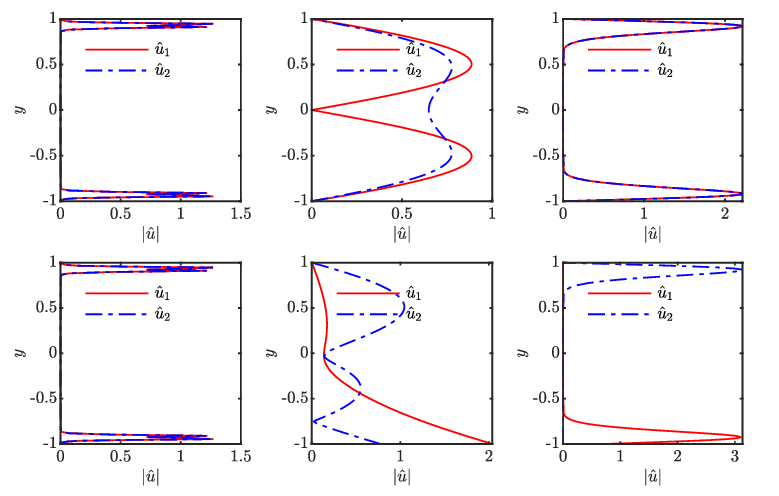}
    \put(3,62){(a)}
    \put(35.5,62){(b)}
    \put(68,62){(c)}    
    \put(3,31){(d)}
    \put(35.5,31){(e)}
    \put(68,31){(f)}    
    \end{overpic}
\caption{Magnitudes of the first two resolvent response modes associated with (a,d) region A, (b,e) region B, and (c,f) region C. Panels (a--c) correspond to the rigid wall and (d--f) to the compliant wall ($G=0.01$). The plotted modes correspond to locations indicated by white circles in figure~\ref{fig:S1S2-annotated}.}
\label{fig:modeShapesRegions}
\end{figure}

Figure~\ref{fig:S1S2-annotated}\,(a,b) shows these regions for
the rigid wall and the $G=0.01$ compliant panel, respectively.
Only regions A and B appear in the rigid-wall plot. Region C for the $G=0.01$ panel includes scales typically associated with the energetic near-wall cycle and therefore identifies a range in which wall compliance modifies the paired nature of the leading modes for a dynamically important mechanism. 
Figure~\ref{fig:modeShapesRegions} compares the magnitude of the
streamwise velocity component, $|\hat u|$, for the first two response
modes in each region. The selected wavelengths are marked by white
circles in figure~\ref{fig:S1S2-annotated} and lie within the
stable region.
In region A, the plotted magnitudes of the first two modes are nearly identical for each wall condition. They are distinct in region B, although both the rigid-wall modes retain the centerline symmetry of the channel, whereas neither of the compliant-wall modes do. Moreover, both compliant-wall modes in region B exhibit a significant non-zero velocity component at the mean interface. In region C, the rigid-wall modes remain paired, whereas the compliant-wall modes occupy opposite halves of the channel. 
The mode adjacent to the compliant wall has a small nonzero velocity at the mean interface. 
Compliance therefore changes not only the magnitude of the leading gains but also their spatial organization and symmetry. 
An important point to be noted here is that \citet{Luhar_Sharma_McKeon_2015} considered the lower half of the channel with symmetry imposed at the centerline. Comparisons with
that formulation are therefore most direct in region A of
figure~\ref{fig:S1S2-annotated}, where the leading gains remain
paired.

While panel stiffness is one of the most important design parameters for a compliant coating, other structural parameters, such as material thickness and viscoelastic damping, and flow parameters, such as Reynolds number, also play important roles in the interaction. Next, we analyze the role of material thickness as a design parameter and its combined effect with the Reynolds number.  


\subsubsection{Panel thickness and Reynolds-number scaling}
\label{subsec:thicknessRe}

\begin{figure}
    \centering
    \begin{overpic}{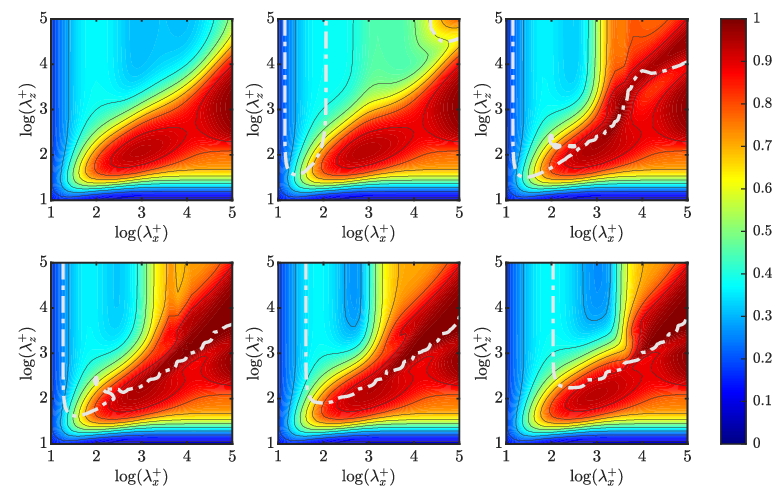}
    \put(2,63){(a)}
    \put(31.5,63){(b)}
    \put(61,63){(c)}    
    \put(2,31.5){(d)}
    \put(31.5,31.5){(e)}
    \put(61,31.5){(f)}    
    \end{overpic}
    \caption{Fractional energy ($\mathcal{E}= (\sigma_1^2 + \sigma_2^2)/\sum^N_{j=1} \sigma_j^2$) contours, for $Re_\tau=180$, $G=0.01$, and material thickness: (a)$t^+=1$, (b)$t^+=10$, (c)$t^+=10^2$, (d)$t^+=10^3$, (e)$t^+=10^4$, and (f)$t^+=10^5$. The dash-dotted gray line separates the stable and unstable regimes from linear stability analysis.}
\label{fig:fractionalEnergyTSweep}
\end{figure}
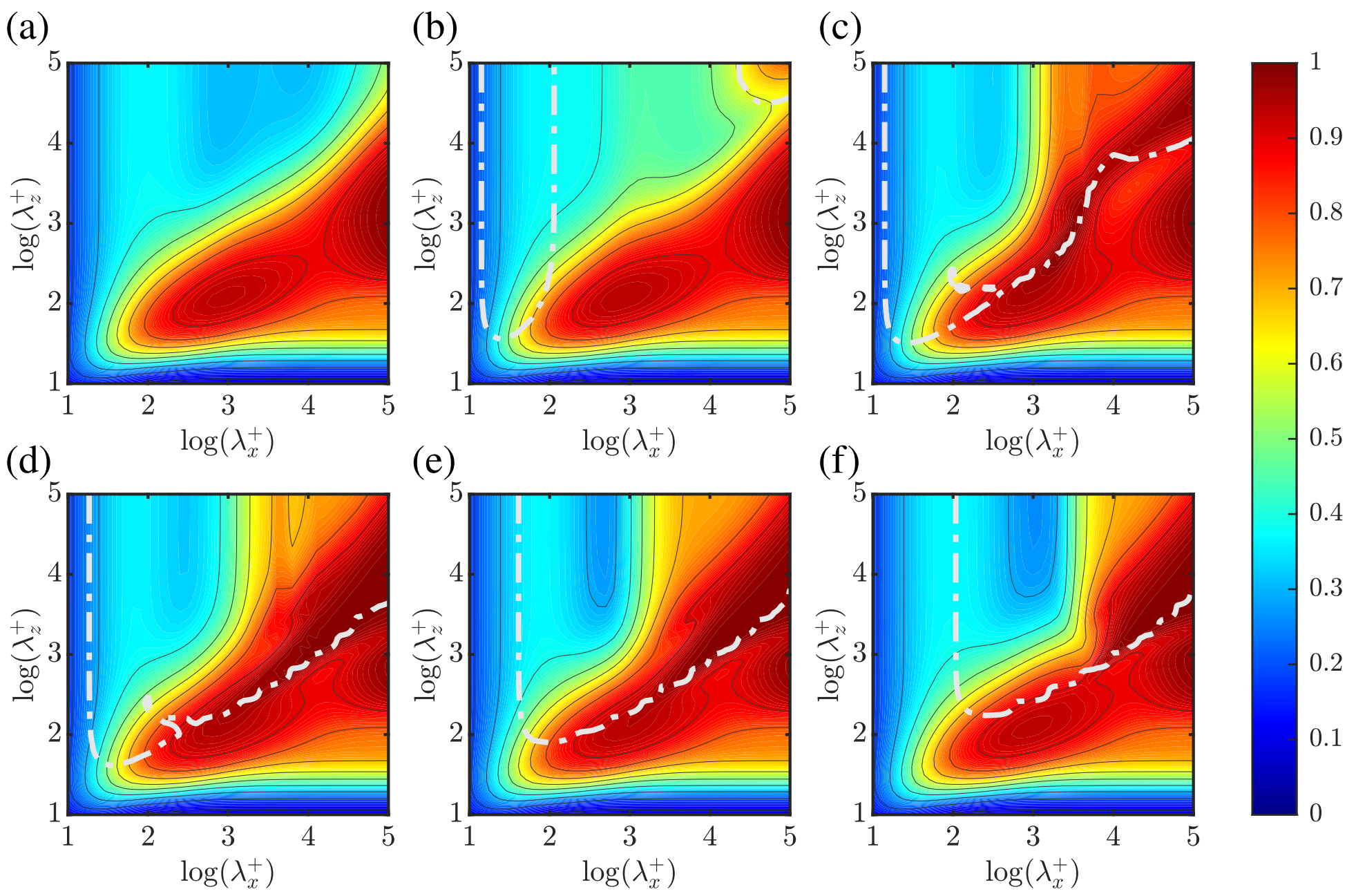

Panel thickness introduces a structural length scale $t$, positioning the thickness-associated response scale $\lambda_x \approx 3t$ relative to the energetic flow scales. Figure~\ref{fig:fractionalEnergyTSweep} shows the fractional energy contours for a soft panel ($G = 0.01$) across various thicknesses. Again, gray dash-dotted lines are used to mark the boundary between the stable and unstable regions, with regions above the line being unstable. For all cases other than the baseline cases shown in figure~\ref{fig:gstabNuTComposite}, the stability demarcating line is computed on a coarser $21 \times 21$ parametric grid, rather than the $41 \times 41$ grid used for the resolvent plots. Since the demarcating line serves only as an approximate guide for interpreting the resolvent results, a coarser parametric grid is adopted to reduce computational cost. 
Even for the soft $G=0.01$ material, the thinnest panel,
$t^+=1$, produces a fractional-energy distribution close to the
rigid-wall result. Increasing thickness modifies this distribution
and introduces additional regions of high $\mathcal E$. At
$t^+=10^5$, the modification becomes less pronounced within the
plotted wavelength range. The largest changes occur for
$t^+=10^2$, $10^3$, and $10^4$.
These trends suggest that the relative position of the structural
and flow scales influences the coupled response. 
Unstable regions are observed for all cases other than $t^+=1$. However, only a limited region in the wavelength space is found to admit an unstable eigenvalue for the $t^+=10$ case. The unstable regions cover a much larger portion of the wavelength space in the other cases with higher panel thickness; however, this effect also appears slightly more subdued for the panel with the largest thickness, $t^+=10^5$.

\begin{figure}
    \centering
    \begin{overpic}{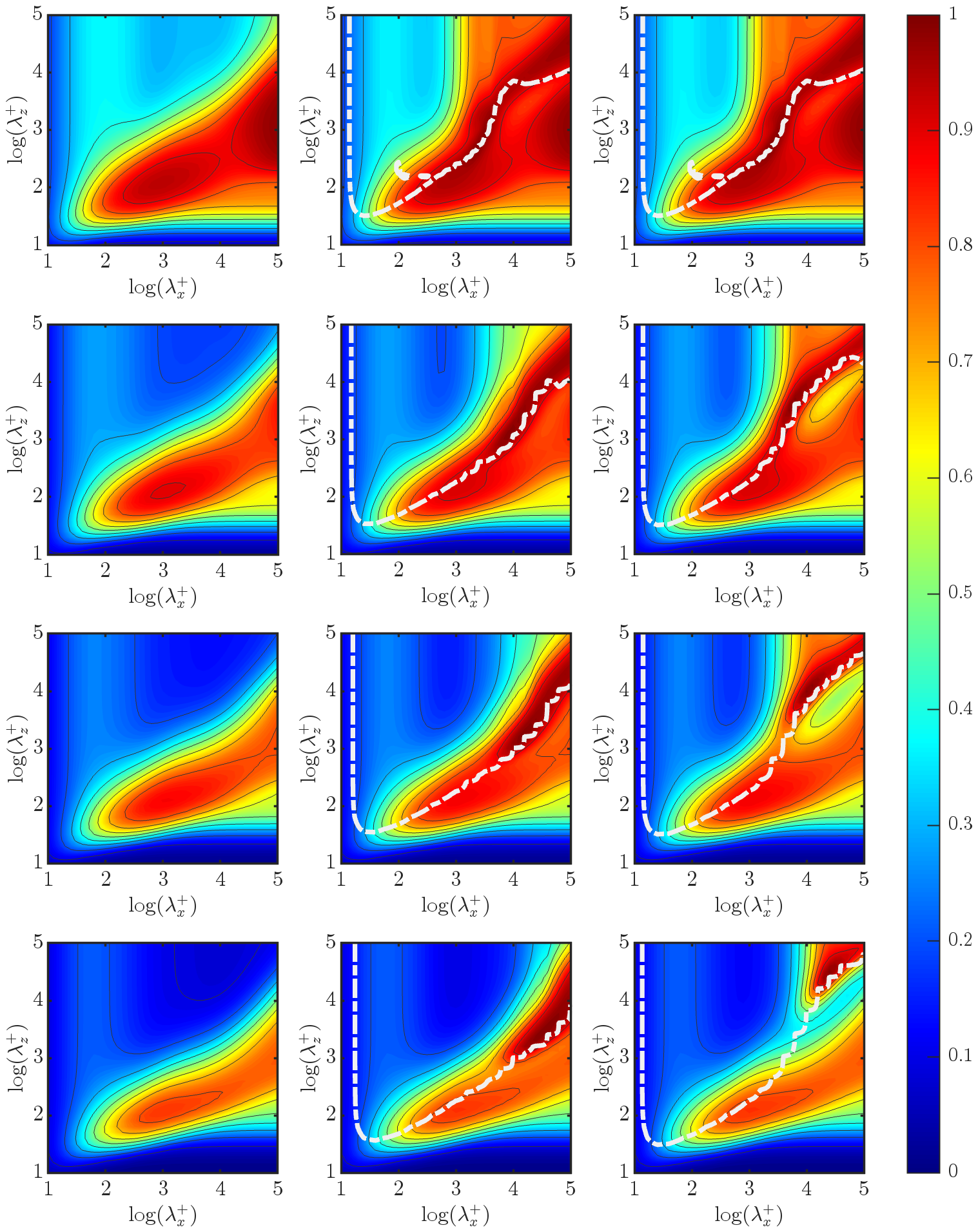}
    \put(-0.5,99.5){(a)}
    \put(24,99.5){(b)}
    \put(48,99.5){(c)}    
    \put(-0.5,75){(d)}
    \put(24,75){(e)}
    \put(48,75){(f)}    
    \put(-0.5,50){(g)}
    \put(24,50){(h)}
    \put(48.5,50){(i)}    
    \put(-0.5,25){(j)}
    \put(24,25){(k)}
    \put(48,25){(l)}    
    \end{overpic}
\caption{Fractional energy ($\mathcal{E}= (\sigma_1^2 + \sigma_2^2)/\sum^N_{j=1} \sigma_j^2$) contours for phase speed $c^+=10$, and Reynolds number: (a-c) $Re_\tau = 180$, (d-f) $Re_\tau = 550$, (g-i) $Re_\tau = 950$, and (j-l) $Re_\tau = 2000$. The material stiffness and thickness are given as (a,d,g,j)  rigid, (b,e,h,k)  $G=0.01$ and $t^*/\delta=0.5555$, and (c,f,i,l) $G=0.01$ and $t^+=100$. The dash-dotted line separates the stable and unstable regimes from linear stability analysis.}
\label{fig:ReStudies}\end{figure}
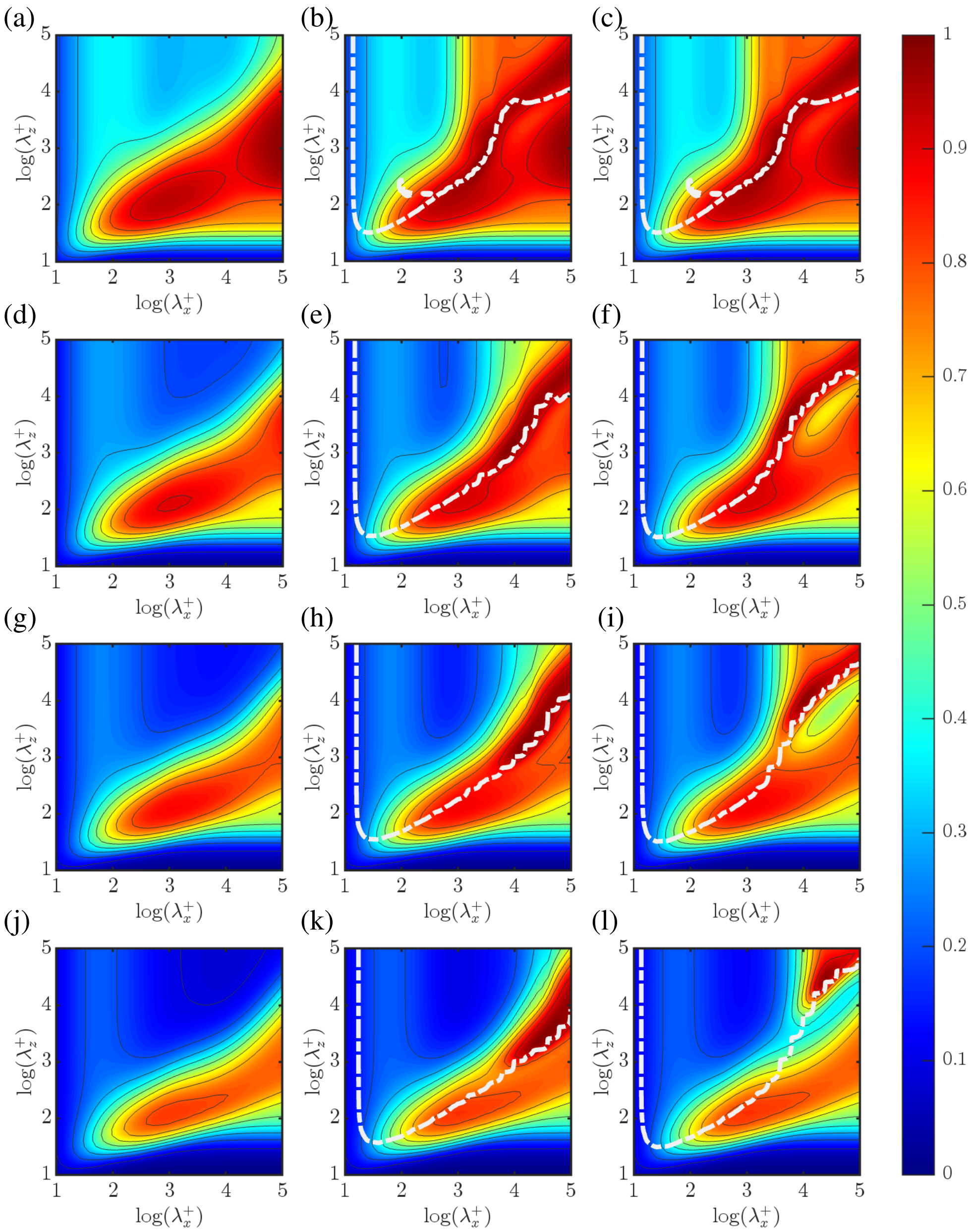

The Reynolds number determines how a fixed structural thickness compares with the inner and outer flow scales. We consider $Re_\tau=180$, $550$, $950$ and $2000$, using the corresponding mean profiles from DNS data~\citep{delAlamo_Jimenez_2003,DEL_ALAMO_JIMENEZ_ZANDONADE_MOSER_2004,hoyas_and_jimenez}. Figure~\ref{fig:ReStudies} compares the rigid wall
with a soft panel ($G=0.01$) under two thickness scalings. The
middle column holds the outer-scaled thickness fixed at
$t=t^*/\delta=0.5555$, while the right column holds the inner-scaled
thickness fixed at $t^+=100$. The two choices coincide at
$Re_\tau=180$, so the same result appears in panels (b) and (c).
Each row is associated with a particular $Re_\tau$. As before, dash-dotted gray lines demarcate the stable and unstable regions. 

For fixed outer-scaled thickness, a region of high $\mathcal E$
appears near the stability boundary. At lower Reynolds numbers,
this region extends toward wavelengths associated with the near-wall
cycle, whereas at higher Reynolds numbers it shifts toward larger
wavelengths. For fixed $t^+=100$, the panel becomes progressively
thinner relative to the channel half-height, reaching
$t^*/\delta=0.05$ at $Re_\tau=2000$.
As $Re_\tau$ increases beyond 180, the two thickness scalings
produce different fractional-energy distributions and stability
boundaries. For fixed $t^+=100$, the compliance-induced features
move away from the near-wall-cycle region toward larger wavelengths
more rapidly. A local minimum in $\mathcal E$ also appears in the
stable region adjacent to the stability boundary at
$Re_\tau=550$ and $950$.
These results illustrate how structural and flow scales jointly
influence the compliance-induced response. They also demonstrate
the value of an efficient framework for exploring the coupled
parameter space.

\subsubsection{Effect of viscoelastic damping}
\label{dampingVariation}
We next examine the effect of material damping on the coupled
fluid--solid response. Damping is represented in Rayleigh form as
\begin{equation}
    \mathbf{C}=\alpha\mathbf{M}+\beta\mathbf{K},
\end{equation}
where $\alpha$ and $\beta$ control the mass- and stiffness-proportional contributions. For a structural mode with undamped natural frequency
$\omega_n$, the modal damping ratio is $\zeta_n=(\alpha/\omega_n+\beta\omega_n)/2$.
Thus, mass-proportional damping contributes more strongly to low-frequency structural modes, whereas stiffness-proportional damping becomes increasingly important for higher-frequency modes.

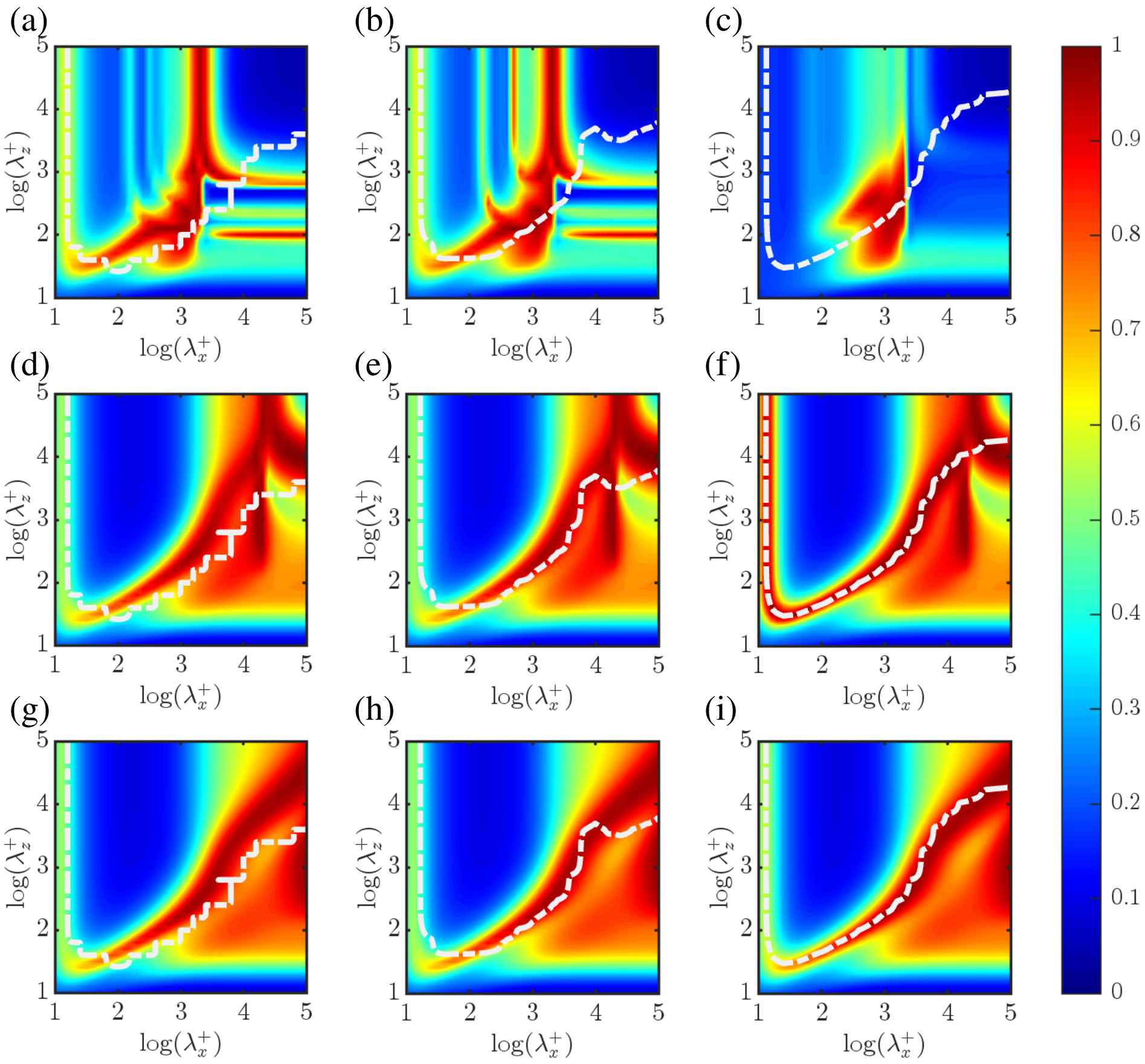
\begin{figure}
    \centering
    \begin{overpic}{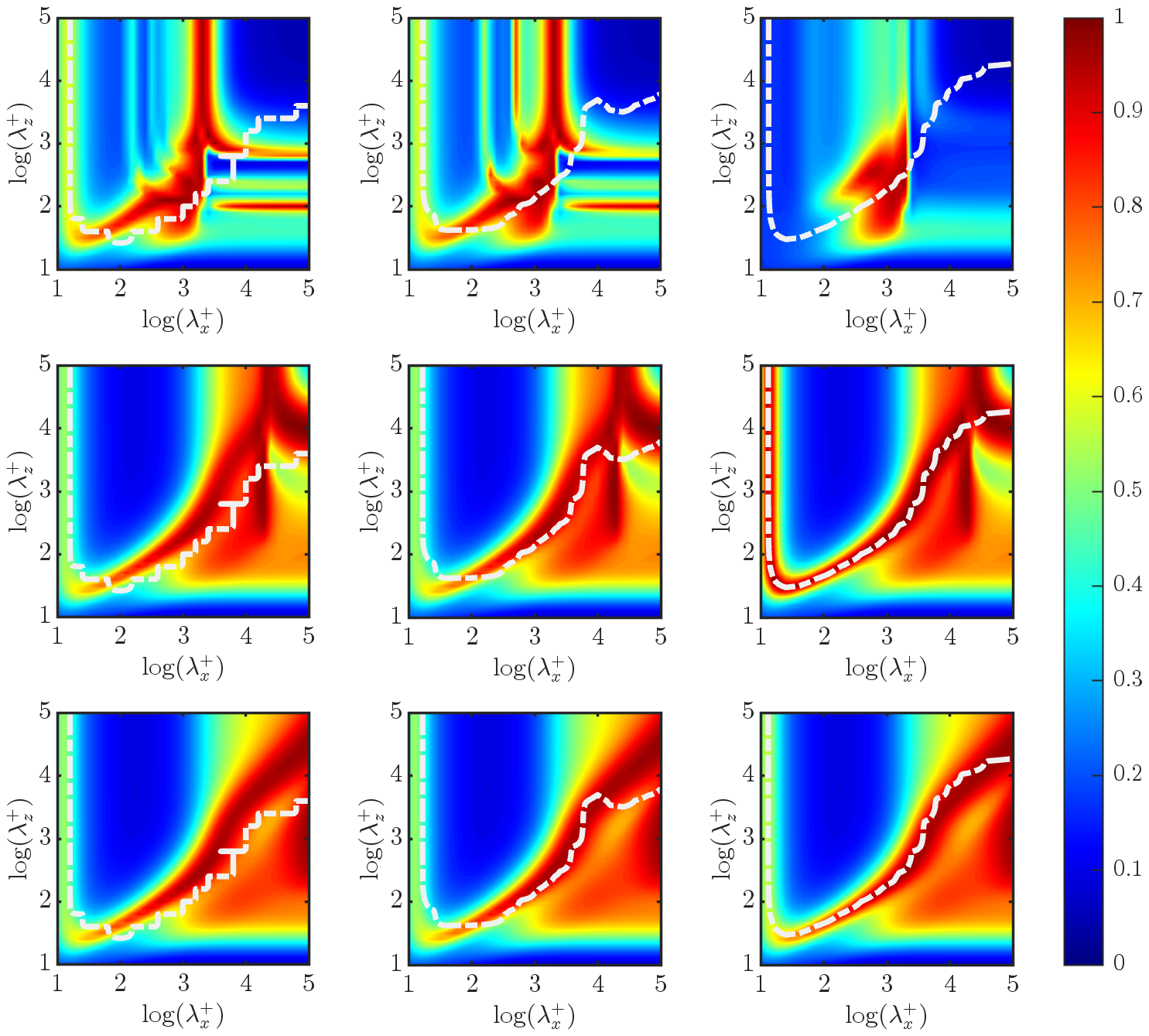}
    \put(1,90){(a)}
    \put(31,90){(b)}
     \put(61.5,90){(c)}
    \put(1,60){(d)}
    \put(31,60){(e)}
     \put(61.5,60){(f)}
    \put(1,30){(g)}
    \put(31,30){(h)}
     \put(61.5,30){(i)}
    \end{overpic}
    \caption{Fractional energy ($\mathcal{E}= (\sigma_1^2 + \sigma_2^2)/\sum^N_{j=1} \sigma_j^2$) contours for $Re_\tau=180$, $G=0.01$, $t^+=100$, for frequency $\omega$: (a-c) $\omega=10$, (d-f) $\omega=1$, and  (g-i) $\omega=0.1$. The mass- and stiffness-proportional Rayleigh coefficients are (a,d,g)  $\alpha=0, \beta=0$, (b,e,h)  $\alpha=0.5, \beta=0$, and (c,f,i) $\alpha=0, \beta=0.5$. The dash-dotted line separates the stable and unstable regimes from linear stability analysis.}
\label{fig:alpbetaomgDamping}
\end{figure}

The relative impacts of mass- and stiffness-proportional damping are assessed in figure~\ref{fig:alpbetaomgDamping}, which displays fractional energy contours across different frequencies for undamped, purely mass-damped, and purely stiffness-damped panels at $G=0.01$; gray dash-dotted lines separate the stable and unstable regions. At $\omega=10$, the undamped and mass-damped cases are largely comparable, whereas the stiffness-damped configuration demonstrates substantial deviations and a significantly smoother energy distribution. At the lower frequencies, $\omega=1$ and $\omega=0.1$, the
Fractional energy distributions are similar across the three damping cases. 
For the low-frequency cases, there is alignment between the stability demarcating line and the high-$\mathcal{E}$ region. 
This alignment also appears in the undamped and mass-damped cases
at $\omega=10$; however, the stiffness-damped case shows less
correspondence with the stability boundary.
For the cases considered, the low-frequency fractional-energy
distribution is weakly sensitive to the damping parameters.
At the highest frequency examined, stiffness-proportional damping
has a stronger effect and produces a smoother distribution.

Overall, the analysis in \S\,\ref{soliddynamics} and \S\,\ref{isotropicresolvent} indicates that even for a single-layer viscoelastic coating, the interaction with turbulent channel flow involves a complex interplay of structural (stiffness, thickness, and viscoelastic damping) and flow (Reynolds number) parameters and their associated scales of motion. The interaction is even more involved for multilayer panels with material properties varying in the wall-normal direction, as discussed next in \S\,\ref{subsec:anisotropicresolvent}.

\subsection{Intrinsic dynamics and coupled response of multi-layer viscoelastic panels}
\label{subsec:anisotropicresolvent}

\begin{figure} 
\centerline{
\begin{tikzpicture}
\vspace{0.5 cm}
\definecolor{yellowochre}{rgb}{0.757,0.561,0.196}
\definecolor{volcanoorange}{rgb}{0.8157,0.2824,0.0627}
\definecolor{darkred}{rgb}{0.5216,0.0941,0.0510}

\node[align=center, text width=1 cm] at (0.1,1.9) {(a)};
\filldraw[yellowochre!80, draw=yellowochre!80!black] (0.25,0.5) rectangle (2.75,0.9);
\filldraw[volcanoorange!80, draw=volcanoorange!80!black] (0.25,0.9) rectangle (2.75,1.3);
\filldraw[darkred!80, draw=darkred!80!black] (0.25,1.3) rectangle (2.75,1.7);
%text 1
\node[align=center, text width=3 cm] at (1.5,0.25) {discrete; stiff on top};
\node[align=center, text=white, text width=3 cm] at (1.5,1.5) {G=100};
\node[align=center, text=white, text width=3 cm] at (1.5,1.1) {G=1};
\node[align=center, text width=3 cm] at (1.5,0.7) {G=0.01};

%box 2
\node[align=center, text width=1 cm] at (3.6,1.9) {(b)};
\filldraw[darkred!80, draw=darkred!80!black] (3.75,0.5) rectangle (6.25,0.9);
\filldraw[volcanoorange!80, draw=volcanoorange!80!black] (3.75,0.9) rectangle (6.25,1.3);
\filldraw[yellowochre!80, draw=yellowochre!80!black] (3.75,1.3) rectangle (6.25,1.7);
%text 1
\node[align=center, text width=3 cm] at (5,0.25) {discrete; soft on top};
\node[align=center, text width=3 cm] at (5,1.5) {G=0.01};
\node[align=center, text=white, text width=3 cm] at (5,1.1) {G=1};
\node[align=center,  text=white, text width=3 cm] at (5,0.7) {G=100};

%box 3
\node[align=center, text width=1 cm] at (7.1,1.9) {(c)};
\shade[top color=darkred!80, middle color=volcanoorange!80 , bottom color=yellowochre!80] (7.25,0.5) rectangle (9.75,1.7);
\draw[darkred!80, draw=darkred!80!black] (7.25,0.5) rectangle (9.75,1.7);
\node[align=center,  text=white, text width=3 cm] at (8.5,1.5) {high G};
\node[align=center, text width=3 cm] at (8.5,0.7) {low G};
\node[align=center, text width=3 cm] at (8.5,0.25) {graded; stiff to soft};

%box 4
\node[align=center, text width=1 cm] at (10.6,1.9) {(d)};
\shade[top color=yellowochre!80, middle color=volcanoorange!80 , bottom color=darkred!80] (10.75,0.5) rectangle (13.25,1.7);
\draw[darkred!80, draw=darkred!80!black] (10.75,0.5) rectangle (13.25,1.7);
\node[align=center, text width=3 cm] at (12,1.5) {low G};
\node[align=center, text=white, text width=3 cm] at (12,0.7) {high G};
\node[align=center, text width=3 cm] at (12,0.25) {graded; soft to stiff};
\end{tikzpicture}}
    \caption{Configurations for vertically inhomogeneous surfaces. The profiles     indicate variation of material properties, specifically material stiffness,    across the thickness (height) of the panel: (a) Configuration I,    (b) Configuration II, (c) Configuration III, and (d) Configuration IV.}
    \label{anisotropicConfig}
\end{figure}
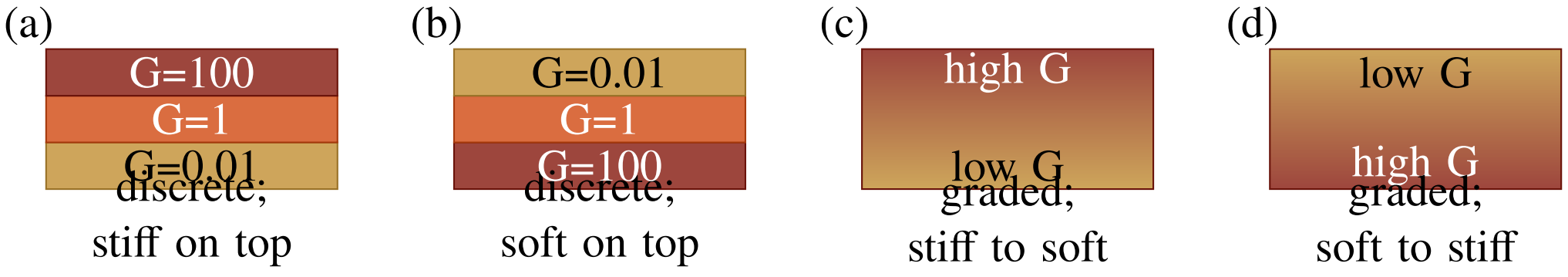

While this method is easily extensible to various kinds of inhomogeneities and anisotropies, such as the presence of periodic substructures (in the wall-parallel, i.e., horizontal directions) or exhibiting smaller-scale variations in streamwise/spanwise directions that can be accounted for via a homogenization approach, this work only focuses on material inhomogeneity in the vertical i.e., wall-normal direction, that arises due to layering. We focus on wall-normal variation in stiffness, which is relevant to both naturally layered bio-materials and engineered multilayer or functionally graded coatings. Four representative configurations are considered. Configurations I and II contain three discrete layers of equal thickness. Configuration I places the stiffest material at the fluid-facing surface, with $(G_\mathrm{top},G_\mathrm{middle},G_\mathrm{bottom})=(100,1,0.01)$; Configuration II reverses this ordering. Configurations III and IV use a continuous stiffness gradient, with the stiffer and softer ends adjacent to the fluid, respectively. In these cases, $G$ changes linearly through the material thickness from $G = 100 \to 1$ or $G = 1 \to 0.01$ for Configuration III and from $G = 1 \to 100$ or $G = 0.01 \to 1$ for Configuration IV. All four configurations are illustrated in figure~\ref{anisotropicConfig}.

Configurations I and II consist of three bonded material layers whose elastodynamic responses are coupled through their interfaces. Their combined dynamics may therefore allow the surface response to be tailored to flow structures at vastly different scales.
This is attractive because isotropic compliant panels optimized to modify VLSMs favorably can amplify motions at other scales~\citep{Luhar_Sharma_McKeon_2015}. {Configurations III and IV instead represent functionally graded materials (FGMs), in this case with stiffness varying continuously through the panel thickness; spatial variation in material stiffness has previously been demonstrated in the case of compliant silicone-based materials~\citep{young_2024}}

\subsubsection{Solid response}\label{sec:anisotropicSolid}
\begin{figure}
    \centering
    \begin{overpic}{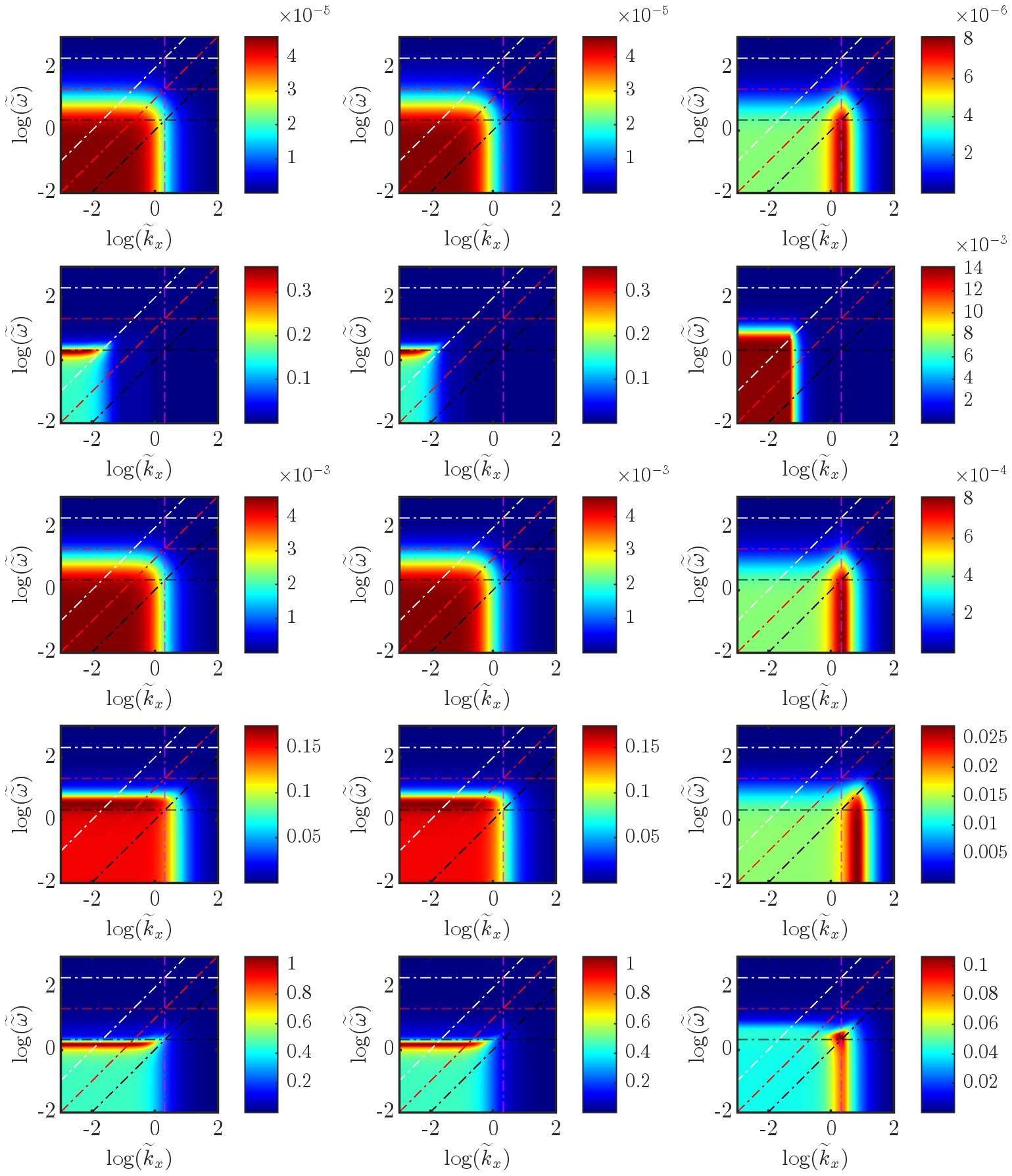}
    \put(1.5,97){(a)}
    \put(30.2,97){(b)}
    \put(59,97){(c)}
    \put(1.5,77.6){(d)}
    \put(30.2,77.6){(e)}
    \put(59,77.6){(f)}    
    \put(1.5,58.3){(g)}
    \put(30.2,58.3){(h)}
    \put(59,58.3){(i)}
    \put(1.5,38.5){(j)}
    \put(30.2,38.5){(k)}
    \put(59,38.5){(l)}
    \put(1.5,19.5){(m)}
    \put(30.2,19.5){(n)}
    \put(59,19.5){(o)}
    \end{overpic}
        \caption{Singular values $\sigma_j$ of the solid transfer matrix $\mathcal{M}^{-1}$ for (a-c) uniform stiffness panel ($G=100$), (d-f) Configuration I ($G_\text{top} = 100, G_\text{middle} = 1, G_\text{bottom} = 0.01$), (g-i) uniform stiffness panel ($G=1$), (j-l) Configuration II ($G_\text{top} = 0.01, G_\text{middle} = 1, G_\text{bottom} = 100$), and (m-o) uniform stiffness panel ($G=0.01$). Diagonal dash-dotted lines indicate the shear speed for different values of material stiffness: $G=100$ (white), $G=1$ (red), $G=0.01$ (black). Horizontal dash-dotted lines indicate the characteristic frequency $\widetilde{\widetilde{\omega}}_{c}$ for different values of material stiffness: $G=100$ (white), $G=1$ (red), $G=0.01$ (black). The vertical magenta line indicates the characteristic wavenumber $k_{x,c} t$. Singular values $\sigma_1$ (left panel), $\sigma_2$ (center panel), and $\sigma_3$ (right panel).}
\label{fig:Eig-VertAniso-5-tmp-matching-parameters-coupled-case}
\end{figure}

The surface-impedance operator $\mathcal M\in\mathbb C^{3\times3}$ maps surface deformation to traction, whereas $\mathcal{M}^{-1}$ maps traction to deformation. A singular value decomposition of $\mathcal{M}^{-1}$ therefore identifies the traction patterns that produce the largest surface response.
For the present solid-centric analysis, the material thickness is used as the reference length scale, while the shear-wave speed of the $G=0.01$ material serves as a common velocity scale across all cases. We define,
\begin{equation}
{
    \widetilde{\widetilde{\omega}}
    =
    \left(
\frac{\omega^*t^*}
         {u^*_{s,0.01}}\right), \qquad 
    \widetilde{k}_x
    =
    k^*_x t^*, \qquad \widetilde{\widetilde{c}}=\frac{c^*}{u^*_{s,0.01}},
    \label{eq:solidScales}
    }
\end{equation}
where $u^*_{s,0.01}$ is the dimensional shear-speed of a material with $G=0.01$. For the $G=0.01$ material the definitions of $\widetilde{\omega}$ ($\widetilde{c}$) and $\widetilde{\widetilde{\omega}}$ ($\widetilde{\widetilde{c}}$) coincide. 
Figure~\ref{fig:Eig-VertAniso-5-tmp-matching-parameters-coupled-case} compares the three singular values of the inverse surface-impedance operator $\mathcal M^{-1}$ for uniform panels with $G=100$, $1$, and $0.01$, and for the three-layer Configurations I, $(G_{\mathrm{top}},G_{\mathrm{middle}},G_{\mathrm{bottom}})=(100,1,0.01)$, and II, $(0.01,1,100)$. When $\mathcal M$ is invertible, its inverse has reciprocal singular values in reverse order: $\sigma_j(\mathcal M^{-1})=1/\sigma_{4-j}(\mathcal M)$ for $j=1,2,3$, when both spectra are sorted in descending order. The singular values are evaluated at $\widetilde{k}_z=0$ and plotted against $\widetilde{k}_x$ and $\widetilde{\widetilde{\omega}}$. 

The vertical dash-dotted magenta line denotes the characteristic wavelength $\widetilde{\lambda}_x=3$, or $\widetilde{k}_{x,c}=2\pi/3$, associated with dominant finite-thickness panel waves~\citep{Benschop_Greidanus_Delfos_Westerweel_Breugem_2019,Esteghamatian_Katz_Zaki_2022,Lu_Xiang_Zaki_Katz_2024}. Characteristic frequencies based on non-dimensional shear speeds for the three constituent materials ($G=100$, $G=1$, and $G=0.01$), $\widetilde{\widetilde{u}}_s$, and characteristic wavenumber $\widetilde{k}_{x,c}$, which satisfy the shear-wave reference dispersion relation $\widetilde{\widetilde{\omega}}_c= \widetilde{k}_{x,c} \widetilde{\widetilde{u}}_s$, are indicated with horizontal dash-dotted lines; the dash-dotted white, red, and black horizontal lines indicate the characteristic frequencies for $G=100$, $1$, and $0.01$, respectively. Similarly, the white, red, and black diagonal dash-dotted lines indicate the non-dimensional shear speeds $\widetilde{\widetilde{u_s}}$ for $G=100$, $1$, and $0.01$, respectively. 

For the first two singular values, all uniform panels exhibit
strong responses at low wavenumbers, primarily below the
characteristic wavenumber $\widetilde{k}_{x,c}$. These responses
are also confined to relatively low frequencies. For the softest panel ($G=0.01$), the frequency bound aligns closely with the characteristic value $\widetilde{\widetilde{\omega}}_{c,0.01}=\widetilde{k}_{x,c} \widetilde{\widetilde{u}}_{s,0.01}$. However, stiffer panels ($G=1$ and $G=100$) deviate from this trend, exhibiting frequency bounds substantially lower than $\widetilde{\widetilde{\omega}}_{c,1}$ and $\widetilde{\widetilde{\omega}}_{c,100}$, respectively. 
This lower frequency limit is consistent with the effect of
stiffness-proportional Rayleigh damping, which increases with
frequency and has a stronger influence for stiffer panels.
In contrast, the layered panels contain materials with different
shear-wave speeds and exhibit different response limits.
Configuration II, with the softest layer at the fluid interface,
shifts the cut-off wavenumber slightly above $\widetilde{k}_{x,c}$
and has a higher frequency bound than the uniform $G=0.01$ panel.
Configuration I, with the stiffest layer at the interface, shifts
the cut-off wavenumber well below $\widetilde{k}_{x,c}$. Its
frequency bound is substantially lower than that of the uniform
$G=100$ panel but remains approximately consistent with the
$G=100$ shear-wave reference at the shifted cut-off wavenumber.

For both the uniform $G=0.01$ panel and Configuration I, the cut-off frequency remains low enough that stiffness-proportional damping does not suppress the resonance. 
For these first two modes, the response approximately follows the
shear-wave reference associated with the fluid-facing layer,
while layering changes the effective panel thickness. Here,
effective thickness denotes an apparent structural length scale
inferred from the characteristic response wavenumber, rather than
the physical panel thickness. A stiff surface layer increases this
effective thickness, whereas a soft surface layer decreases it.

The third singular value behaves differently. Uniform panels show a high-response band near $\widetilde{k}_{x,c}$, which peaks at $(\widetilde{k}_{x,c},\widetilde{\widetilde{\omega}}_{c})$ for $G=0.01$ but is prematurely truncated by damping in stiffer cases. Configuration II shifts this band to slightly larger wavenumbers. Conversely, Configuration I develops a distinct low-wavenumber, low-frequency response with a cut-off below $\widetilde{k}_{x,c}$, dictated by the $G=100$ shear speed. Aside from this specific mode in Configuration I, layered-panel responses remain consistent with the modified effective-thickness interpretation.

\begin{figure}
    \centering
    \begin{overpic}{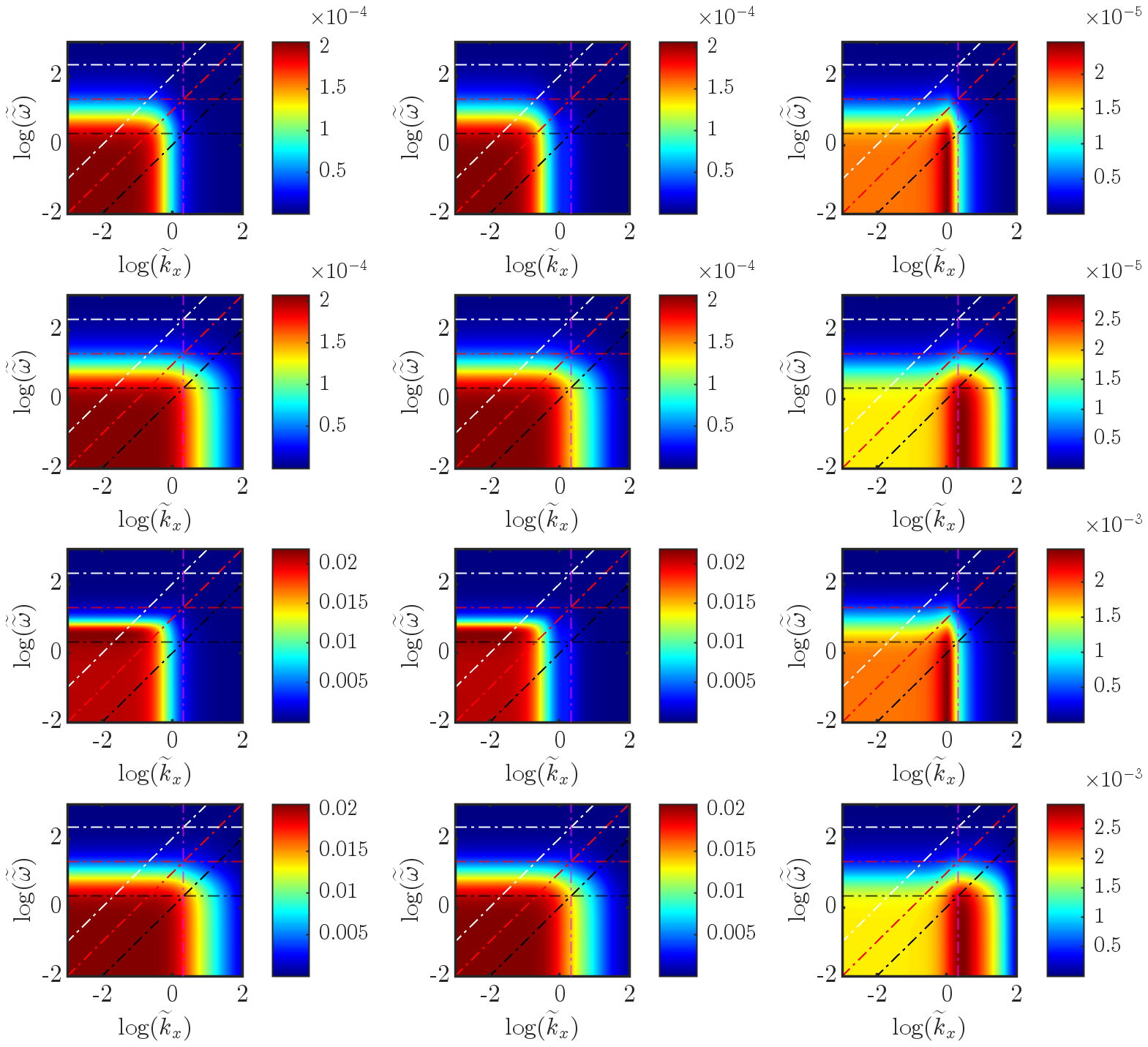}
    \put(1.85,89){(a)}
    \put(36,89){(b)}
    \put(69.75,89){(c)}
    \put(1.85,67){(d)}
    \put(36,67){(e)}
    \put(69.75,67){(f)}
    \put(1.85,45){(g)}
    \put(36,45){(h)}
    \put(69.75,45){(i)}
    \put(1.85,22){(j)}
    \put(36,22){(k)}
    \put(69.75,22){(l)}
    \end{overpic}
    \caption{Singular values $\sigma_j$ of the solid transfer matrix $\mathcal{M}^{-1}$ for (a-c) Configuration III ($G_\text{top} = 100, G_\text{bottom} = 1$), (d-f) Configuration IV ($G_\text{top}= 1, G_\text{bottom} = 100$), (g-i) Configuration III ($G_\text{top} = 1, G_\text{bottom} = 0.01$), and  (j--l) Configuration IV ($G_\text{top}= 0.01, G_\text{bottom} = 1$). Diagonal dash-dotted lines indicate the shear speed for different values of material stiffness: $G=100$ (white), $G=1$ (red), $G=0.01$ (black). Horizontal dash-dotted lines indicate the characteristic frequency $\widetilde{\widetilde{\omega}}_{c}$ for different values of material stiffness: $G=100$ (white), $G=1$ (red), $G=0.01$ (black). The vertical magenta line indicates the characteristic wavenumber $k_{x,c} t$. Singular values $\sigma_1$ (left panel), $\sigma_2$ (center panel), and $\sigma_3$ (right panel). }
\label{fig:Eig-VertAnisoCont-1-linear-matching-parameters-coupled-case}\end{figure}

Next, figure~\ref{fig:Eig-VertAnisoCont-1-linear-matching-parameters-coupled-case} evaluates continuous graded panels: Configuration III (stiffer at the fluid interface) and Configuration IV (softer at the interface). 
As with the discrete layered panels, the first two modes exhibit
cut-off wavenumbers below $\widetilde{k}_{x,c}$ for the
stiffer-on-top gradient and above $\widetilde{k}_{x,c}$ for the
softer-on-top gradient.
The third mode responses also diverge, with Configuration III exhibiting a concentrated high-response region narrowly below $\widetilde{k}_{x,c}$, while Configuration IV creates a broader band above $\widetilde{k}_{x,c}$. Ultimately, both discrete and graded panels provide a route to tailored compliant coatings, allowing the structural response to be dynamically tuned for targeted flow features. 

\begin{figure}
    \centering
    \begin{overpic}{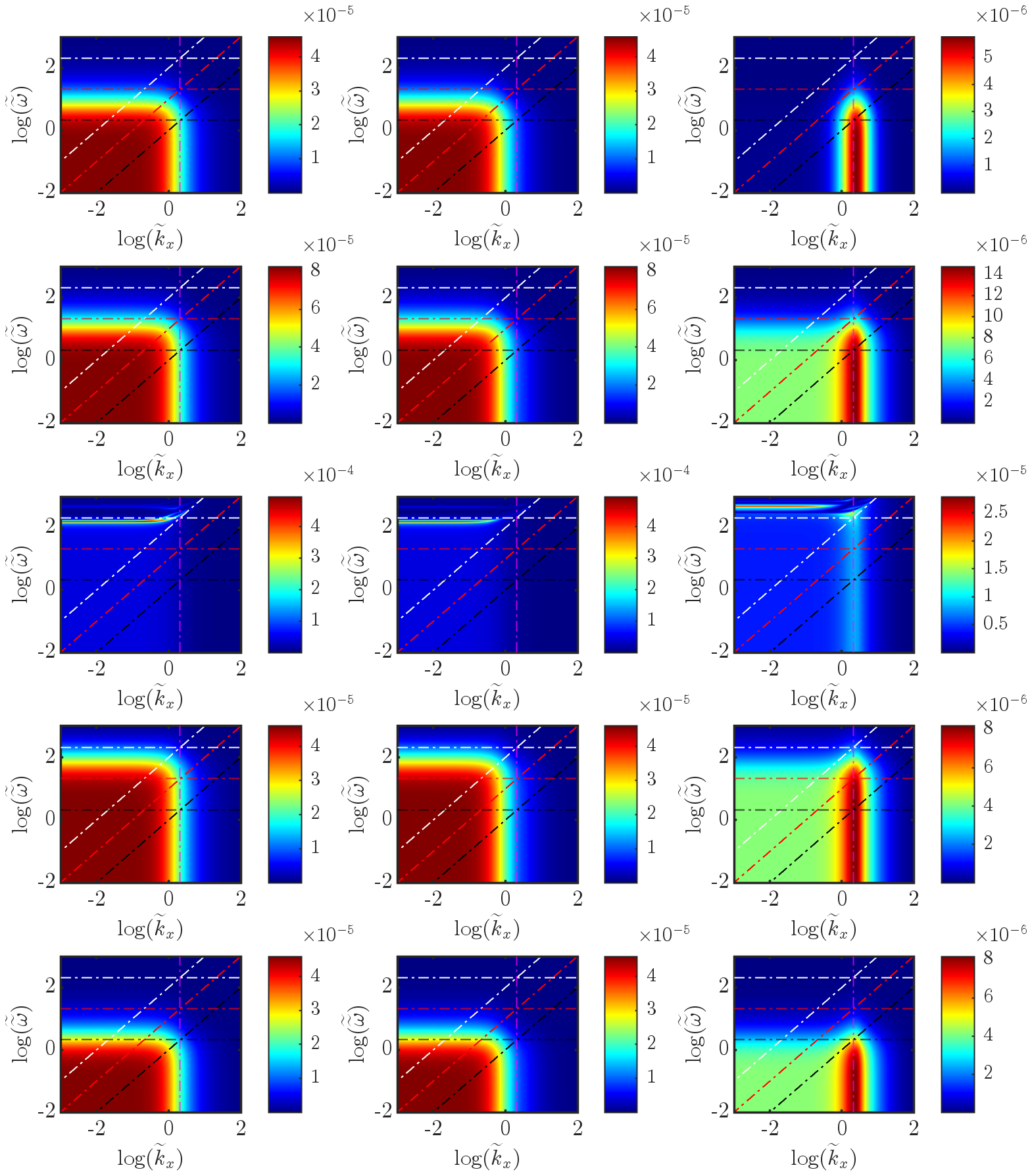}
    \put(1.5,98){(a)}
    \put(31,98){(b)}
    \put(60,98){(c)}
    \put(1.5,79.3){(d)}
    \put(31,79.3){(e)}
    \put(60,79.3){(f)}    
    \put(1.5,59){(g)}
    \put(31,59){(h)}
    \put(60,59){(i)}
    \put(1.5,38.5){(j)}
    \put(31,38.5){(k)}
    \put(60,38.5){(l)}
    \put(1.5,19.5){(m)}
    \put(31,19.5){(n)}
    \put(60,19.5){(o)}
    \end{overpic}
    \caption{Singular values $\sigma_j$ of the solid transfer matrix $\mathcal{M}^{-1}$ for uniform stiffness panel ($G=100$). Relative to the baseline uniform panel described in the text, the five rows examine: (a-c) Poisson's ratio $\nu_p = 0.499$, (d-f) thickness  ($t^*/\delta=1$), (g-i) Damping with a constant loss factor $\tan(\phi_s)= 0.075$, (j-l) Rayleigh Damping ($\alpha = 0.2$, $\beta = 0.01$), and (m-o) Rayleigh Damping ($\alpha = 0.2$, $\beta = 0.3$). Diagonal dash-dotted lines indicate the shear speed for different values of material stiffness: $G=100$ (white), $G=1$ (red), $G=0.01$ (black). Horizontal dash-dotted lines indicate the characteristic frequency $\widetilde{\widetilde{\omega}}_{c}$ for different values of material stiffness: $G=100$ (white), $G=1$ (red), $G=0.01$ (black). Vertical magenta line indicates the characteristic wavenumber $k_{x,c}t$. Singular values $\sigma_1$ (left panel), $\sigma_2$ (center panel), and $\sigma_3$ (right panel).  
    }
\label{fig:Eig-VertAniso-5-tmp-matching-parameters-coupled-case-OneParameterVariation}
\end{figure}

Figure~\ref{fig:Eig-VertAniso-5-tmp-matching-parameters-coupled-case-OneParameterVariation} explores parametric sensitivities relative to a baseline uniform panel ($G=100$, $t^+=100$ i.e. $t = 0.5555$, $\nu_p=0.45$, $\alpha=0.2$, and $\beta=0.1$). We find that near-incompressibility ($\nu_p=0.499$) tightly concentrates the third mode around $k_{x,c}$, while increased thickness ($t=t^*/\delta=1$) modifies the structural frequency scales and the associated stiffness-proportional damping, shifting the
normalized cut-off frequency. Furthermore, the high-frequency extent of the response is strongly influenced by the damping model: replacing Rayleigh damping with a constant loss factor ($\tan(\phi_s)=0.075$) reveals pronounced response peaks near $\widetilde{\widetilde{\omega}}_{c,100}=\widetilde{k}_{x,c} \widetilde{\widetilde{u}}_{s,100}$. 
As shown in
figure~\ref{fig:Eig-VertAniso-5-tmp-matching-parameters-coupled-case-OneParameterVariation}\,(j--l)
and (m--o), decreasing $\beta$ raises the response cut-off frequency
relative to the baseline, whereas increasing $\beta$ lowers it.
Thus, stiffness-proportional damping strongly influences the
predicted response bandwidth for the configurations considered.

Taken together, these results show that the material adjacent to the fluid primarily sets the propagation speed, while the internal stiffness distribution changes the effective thickness and spectral bandwidth, and viscoelastic damping further restricts the accessible frequencies. The following coupled analysis therefore examines whether these intrinsic changes persist when the multi-layer panels interact with the turbulent flow.

\subsubsection{Coupled solid-fluid response}
\begin{figure}
    \centering
    \begin{overpic}{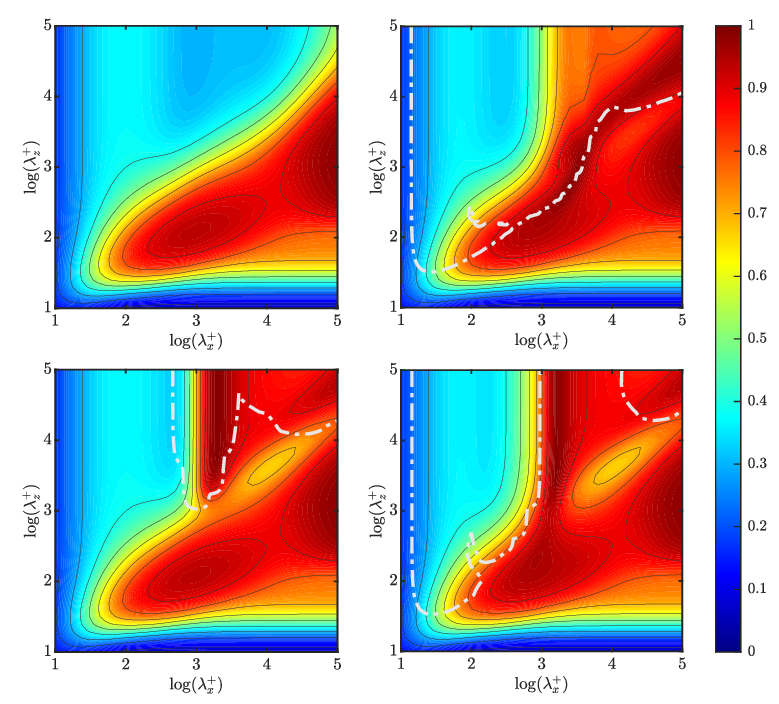}
    \put(3,88.5){(a)}
    \put(47.5,88.5){(b)}
    \put(3,44){(c)}
    \put(47.5,44){(d)}
    \end{overpic}
  \caption{Fractional energy ($\mathcal{E}= (\sigma_1^2 + \sigma_2^2)/\sum^N_{j=1} \sigma_j^2$) contours for $Re_\tau = 180$: (a) $G=100$, uniform, (b) $G=0.01$, uniform, (c) Configuration I ($G=100$, $G=1$, and $G=0.01$)  , and (d) Configuration II ($G=0.01$, $G=1$, and $G=100$). {All cases use $t^+=100$ and $c^+=10$.}} 
\label{fig:vertAnisotropy}
\end{figure}

\begin{figure}
    \centering
    \begin{overpic}{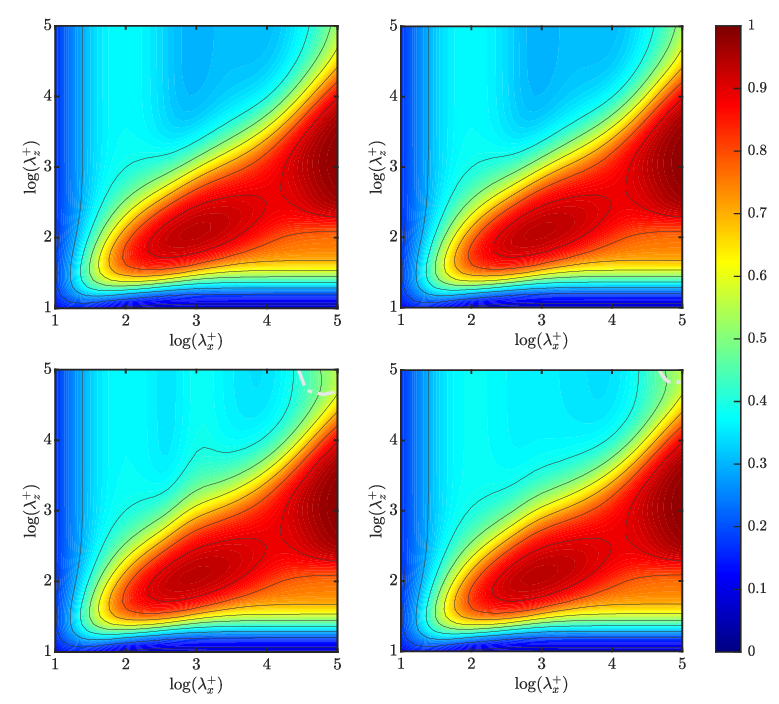}
    \put(3,88.5){(a)}
    \put(47.5,88.5){(b)}
    \put(3,44){(c)}
    \put(47.5,44){(d)}
    \end{overpic}
  \caption{Fractional energy ($\mathcal{E}= (\sigma_1^2 + \sigma_2^2)/\sum^N_{j=1} \sigma_j^2$) contours for $Re_\tau = 180$: (a)  Configuration III ($G_\text{top}=100$ and $G_\text{bottom}=1$), (b) Configuration IV ($G_\text{top}=1$ and $G_\text{bottom}=100$), (c) Configuration III ($G_\text{top}=1$ and $G_\text{bottom}=0.01$)  , and (d) Configuration IV ($G_\text{top}=0.01$ and $G_\text{bottom}=1$). {For all four cases, the total panel thickness is  $t^+=100$.}} 
\label{fig:vertAnisotropyCONT}
\end{figure}

Next, we evaluate the multi-layer panels coupled to the turbulent flow ($Re_\tau=180$, $t^+=100$, $\nu_p=0.45$, $\alpha=0.2$, $\beta=0.1$). The coupled responses for the discrete three-layer configurations (I and II) and the uniform panels ($G=100$ and $G=0.01$) are shown in figure~\ref{fig:vertAnisotropy}, while the corresponding results for the four graded materials are provided in figure~\ref{fig:vertAnisotropyCONT}. 
For all cases, the fractional energy contours {are evaluated} at a constant phase-speed $c^+=10$. 

The uniform $G=100$ panel remains close to the rigid-wall limit, whereas the uniform $G=0.01$ panel produces significant modification of fractional energy distribution as discussed earlier in \S\,\ref{subsec:stiffnessVariation}. 
Configuration I, with the stiff layer at the fluid-facing surface,
modifies the fractional-energy contours at large scales, while
the region associated with the near-wall cycle remains largely
unaffected. A vertical band of high $\mathcal{E}$ is largely
confined to the unstable region. Configuration II, with the soft
layer at the surface, exhibits a larger unstable region at small
streamwise wavelengths. In this case, a vertical band of high
$\mathcal{E}$ lies on the stable side of the stability boundary.
Configuration II also produces a stronger modification of the
near-wall region, with a distorted and displaced peak relative to
the stiff panel. Thus, the two configurations modify the
fractional-energy distribution in different regions of wavelength
space.
Moreover, the unstable region is substantially smaller than that
of the uniform $G=0.01$ panel. These results show that multilayer
panels can produce responses distinct from those of the uniform
panels considered here.

Among the panels considered, the graded panels produce weaker
changes in $\mathcal{E}$ than the discrete arrangements.
Configurations III and IV with stiffness gradients
$G=100\to1$ and $G=1\to100$, respectively, show little change
relative to the rigid-wall baseline. The softer panels with gradients,
$G=1\to0.01$ and $G=0.01\to1$, produce more pronounced
modifications, although these remain weaker than those of the
discrete arrangements considered.
Even Configuration IV, with $G=0.01$ at the fluid-facing surface increasing gradually to $G=1$ at the support, remains relatively close to the rigid-wall near-wall fractional-energy distribution. All the graded panels are largely stable for the wavelengths considered here, except for a small unstable region at high streamwise and spanwise wavelengths which is observed for Configurations III ($G = 1 \to 0.01$) and IV ($G=0.01 \to 1)$, respectively.  
These differences suggest that, for the profiles examined here,
a continuous stiffness gradient distributes the structural
response across the panel thickness differently from a discrete
change in material properties. The discrete configurations
therefore produce more distinct spectral modifications than the
selected graded profiles.

These findings do not necessarily imply that graded materials are generally less effective. Rather, they show that the particular gradients considered here lead to a smoother structural response and a correspondingly weaker modification of the low-rank flow dynamics. 
The result also underscores the importance of the fluid-facing material: a stiff surface layer can mask substantial compliance beneath it, whereas a soft surface layer alone is insufficient to reproduce the response of a uniformly soft panel.

%===================================================================================
\section{Estimating drag reduction}\label{sec:drag-estimation}

In \S\,\ref{sec:results}, we used the coupled resolvent model to determine how compliant-wall design parameters modify the amplification and structure of energetic motions. Here, we use these changes to obtain a leading-order estimate of
the skin-friction drag. Resolvent methods have previously been used to identify coherent structures~\citep{McKEON_SHARMA_2010,Moarref_Sharma_Tropp_McKeon_2013,Schmidt_Towne_Rigas_Colonius_Bres_2018}, estimate flow statistics based on sparse sensor measurements~\citep{Towne_Lozano-Duran_Yang_2020, Amaral_Cavalieri_Martini_Jordan_Towne_2021,Martini_Cavalieri_Jordan_Towne_Lesshafft_2020, Martini_Jung_Cavalieri_Jordan_Towne_2022, towne_2024_jet_estimation}, and inform flow control strategies~\citep{Yeh_Taira_2019,jung_resolvent-based_2024,zhou_2026}. Here, we adapt the LNS-based drag-prediction methodology of \citet{Moarref_Jovanovic_2012} and \citet{Ran_Zare_Jovanovic_2021} to the present compliant-wall resolvent formulation. 
Resolvent analysis is performed over a range of wavenumber--frequency
triplets $(k_x,k_z,\omega)$, and the cross-spectral density (CSD)
is integrated over the sampled domain to estimate flow statistics.
The resulting drag estimate depends on the coupled solid/fluid dynamics, contained in the resolvent operator, and on the assumed forcing covariance, as explained in \S\,\ref{subsec:SOS}. Eddy-viscosity enhanced resolvent analysis has been shown to improve the prediction of turbulent structures and statistics~\citep{Morra_Semeraro_Henningson_Cossu_2019,Morra_Nogueira_Cavalieri_Henningson_2021,Pickering_Rigas_Schmidt_Sipp_Colonius_2021}. While analytical expressions such as Cess' eddy viscosity profile are available for a standard, rigid wall channel flow, analogous eddy viscosity models for flows over compliant or non-standard walls are not generally available. The framework of \citet{Moarref_Jovanovic_2012} is used here to estimate the eddy-viscosity modification associated with a viscoelastic compliant wall.  

\begin{figure} 
\centerline{
\begin{tikzpicture}
%\draw[step=1cm,gray,very thin] (0,0) grid (13,8);
%box 1
\filldraw[blue!10, draw=blue!80!black, rounded corners=10pt] (1.5,4) rectangle (5.5,6);
%text 1
\node[align=center, text width=3.8 cm] at (3.5,5) {Rigid wall mean velocity $U_r(y)$ and rigid wall eddy viscosity $\nu_{T,r}(y)$ as a starting point};
%box 2
\filldraw[blue!10, draw=blue!80!black, rounded corners=10pt] (7.5,4) rectangle (11.5,6);
%text 2
\node[align=center, text width=3.8 cm] at (9.5,5) {Use resolvent analysis with modeled forcing covariance to estimate second-order statistics};
%box 3
\filldraw[blue!10, draw=blue!80!black, rounded corners=10pt] (1.5,1) rectangle (5.5,3);
%text 3
\node[align=center, text width=3.8 cm] at (3.5,2) {Use rigid-to-compliant changes in second-order statistics to estimate $\Delta \nu_{T,c}(y)$};
%box 4
\filldraw[blue!10, draw=blue!80!black, rounded corners=10pt] (7.5,1) rectangle (11.5,3);
%text 4
\node[align=center, text width=3.8 cm] at (9.5,2) {Use $\nu_{T,c}(y)$ to estimate $U_c(y)$ at fixed bulk velocity and the corresponding drag change};
%arrows
\draw[-stealth, thick] (5.5,5) -- (7.5,5);
\draw[-stealth, thick] (9.5,4) -- (9.5,3.5) -- (3.5,3.5) -- (3.5,3);
\draw[-stealth, thick] (5.5,2) -- (7.5,2);
\end{tikzpicture}}
\caption{Resolvent-based framework used to estimate the drag change induced by a compliant wall.}
\label{fig:DRframework}
\end{figure}
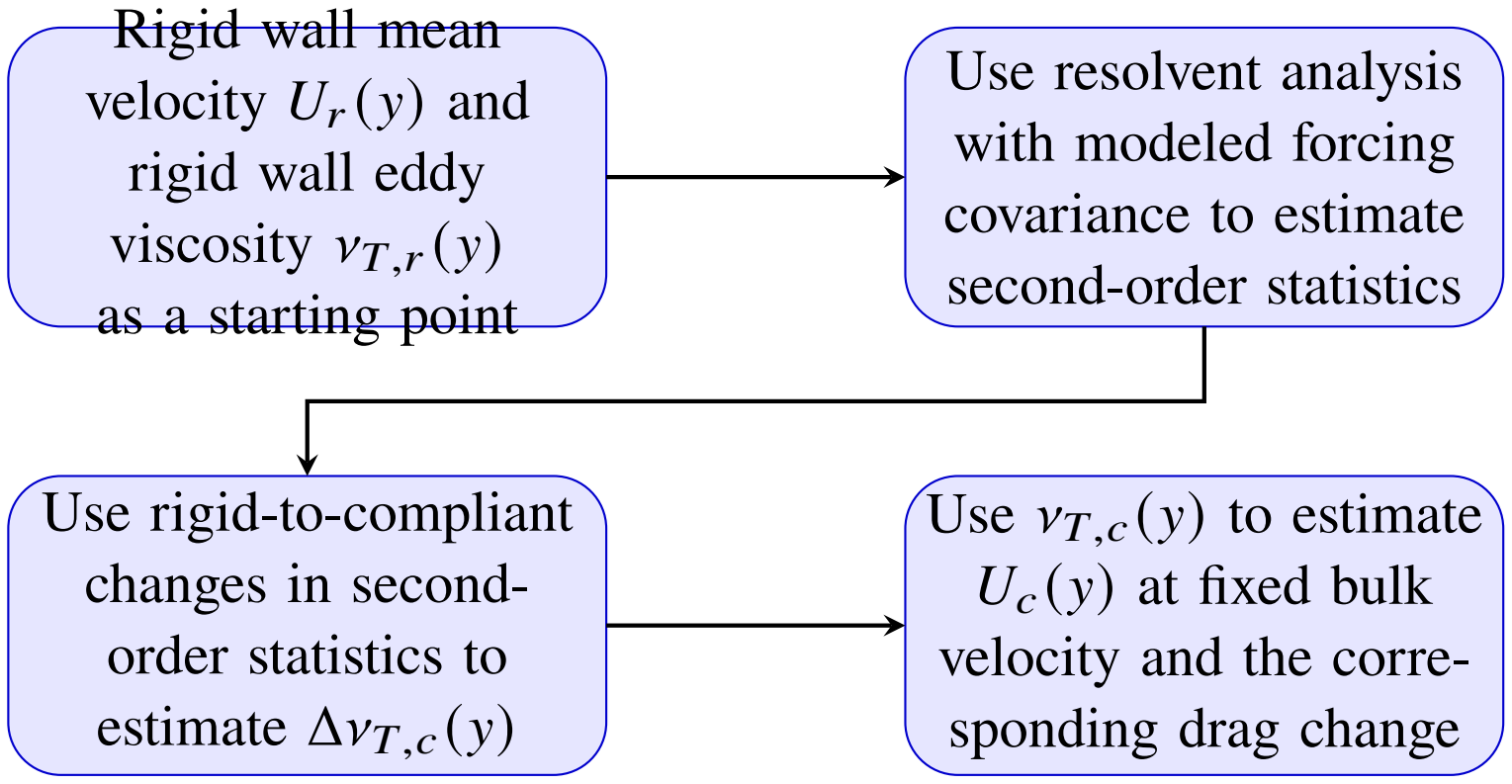

The procedure, summarized in figure~\ref{fig:DRframework}, consists of the following steps:
\begin{enumerate}
    \item initialize with the rigid-wall mean profile, $\overline{U}_r$;
    \item perform resolvent analysis over the relevant spectral domain $(k_x,k_z,\omega)$;
    \item use the resolvent operator and the modeled forcing
        covariance to estimate second-order statistics and the
        compliant-wall correction to the eddy viscosity;
    \item obtain the updated mean profile, $\overline{U}_c$, for a fixed bulk velocity; and
    \item calculate the corresponding change in skin-friction drag due to the change in the mean profile.
\end{enumerate}
Following \citet{Ran_Zare_Jovanovic_2021}, we perform a single
leading-order update without iterating the mean flow and covariance to convergence.

\subsection{Eddy-viscosity correction from resolvent-based second-order statistics}\label{subsec:turbModeling}

The central step is to estimate turbulent kinetic energy ${k}(y)$ and dissipation rate ${\epsilon}(y)$ from resolvent analysis of the mean flow, and then use them to update the eddy viscosity. Following the $k$--$\epsilon$ closure of \citet{Pope_2000} and \citet{jones_1972}, the turbulent eddy viscosity $\nu_T=\nu_T^*/\nu^*$ is written as
\begin{equation}
{\nu_T}=C_\mu {Re_{\tau}}\frac{k^2}{\epsilon},
\end{equation}
where
{\begin{equation}
    k(y)=\frac12\langle u_i'u_i'\rangle,\qquad
\epsilon(y)= 
\frac{1}{Re_\tau}\left[ \left\langle\frac{\partial u_j'}{\partial x_i}
                \frac{\partial u_i'}{\partial x_j}\right\rangle
+\left\langle\frac{\partial u_i'}{\partial x_k}
                 \frac{\partial u_i'}{\partial x_k}\right\rangle \right].
\end{equation}}

{Here, $u_i'$ and $x_i$ are nondimensionalized using $u_\tau^*$ and $\delta$, respectively, while $k=k^*/(u_\tau^*)^2$, $\epsilon=\epsilon^*\delta/(u_\tau^*)^3$; $C_\mu=0.09$.}  
For small mean-flow modifications expected for sufficiently stiff panels, the compliant-wall correction is obtained by retaining terms linear in $\Delta k_c/k_r$ and $\Delta\epsilon_c/\epsilon_r$~\citep{Moarref_Jovanovic_2012},
\begin{equation}
    \Delta\nu_{T,c}(y)=\nu_{T,r}(y)
    \left[2\frac{\Delta k_c(y)}{k_r(y)}
    -\frac{\Delta\epsilon_c(y)}{\epsilon_r(y)}\right],
    \label{dnu_equation}
\end{equation}
where subscripts $r$ and $c$ denote rigid and compliant walls and $\Delta k_c=k_c-k_r$ and $\Delta\epsilon_c=\epsilon_c-\epsilon_r$, respectively.
In equation~\eqref{dnu_equation}, $k_r(y)$ and $\epsilon_r(y)$
are estimated from resolvent-based second-order statistics. The
rigid-wall eddy viscosity, $\nu_{T,r}(y)$, is obtained from the
Cess profile in equation~\eqref{eqn:cess}.
This approximation retains the leading sensitivity of the closure
to compliance and requires the relative changes
$\Delta k_c/k_r$ and $\Delta\epsilon_c/\epsilon_r$ to remain
sufficiently small.

The modified eddy viscosity $\nu_{T,c}$ is then given by:
\begin{equation}
\nu_{T,c}(y)=\nu_{T,r}(y)+\Delta\nu_{T,c}(y)\,.
\label{eq:viscosity_modification}
\end{equation}

\subsection{Modelling second-order statistics and drag estimation}\label{subsec:SOS}
For each wavenumber--frequency combination, equation
\eqref{eqn:velocity-resolvent} gives

\begin{equation}
   \hat{\bm u}_{\bm k}
    =
    \mathbf{C}_u
    \left(-i\omega\mathbf{B}-\mathbf{A}\right)^{-1}
    \mathbf{B}_f 
    \hat{\bm f}_{u,\bm k}
    =
    \mathbf{H}_u (k_x,k_z,\omega)  \hat{\bm f}_{u,\bm k}.
    \label{eqn:velocity-resolvent-DR}
\end{equation}

Henceforth the dependence of $\mathbf H_u (k_x,k_z,\omega)$ on frequency and wavenumbers is omitted for brevity. 
The cross-spectral density (CSD) at a given frequency-wavenumber triplet $(k_x,k_z,\omega)$ is therefore~\citep{Towne_Lozano-Duran_Yang_2020},
\begin{equation}\label{CSD1}
    \mathbf S_{uu}=\mathbf H_u  \mathbf S_{ff}  \mathbf H_u^\dagger,
\end{equation}
where $\mathbf S_{uu}$ and  $\mathbf S_{ff}$ are the cross-spectral densities of the response $\hat{\bm u}_{\bm k}$ and forcing $\hat{\bm f}_{u, \bm k}$. 
Using the full singular value decomposition $\mathbf H_u=\mathbf U\mathbf\Sigma\mathbf V^\dagger$ of the unweighted resolvent operator, the CSD can be expressed as 
\begin{equation}\label{CSD2}
    \mathbf S_{uu}
    =\mathbf U\mathbf\Sigma\mathbf V^\dagger
     \mathbf S_{ff}
     \mathbf V\mathbf\Sigma^\dagger\mathbf U^\dagger.
\end{equation}

Equation~\eqref{CSD2} is an exact algebraic identity for the specified linear operator and forcing CSD when the full SVD is retained; no low-rank truncation has yet been introduced.
Because the forcing covariance is not generally available, we
follow \citet{Moarref_Jovanovic_2012} and model the statistics
using two ingredients. First, we augment the molecular viscosity
in the resolvent operator with the eddy viscosity. 
The inclusion of eddy viscosity implicitly models part of the nonlinear forcing and can also be interpreted through an equivalent effective forcing covariance~\citep{Morra_Nogueira_Cavalieri_Henningson_2021}.
Following~\citet{Moarref_Jovanovic_2012}, the forcing covariance matrix is shaped spatially using the 2D turbulent energy spectrum from DNS~\citep{delAlamo_Jimenez_2003, DEL_ALAMO_JIMENEZ_ZANDONADE_MOSER_2004} hosted at \url{https://torroja.dmt.upm.es/channels/data/spectra/}; the available data at sparse wall-normal locations are interpolated using piecewise cubic Hermite interpolation, as implemented in MATLAB's \texttt{pchip} function. Temporally, the forcing is modeled as white noise, corresponding to a frequency-independent forcing CSD at each wavenumber pair $(k_x, k_z)$.

With the two-dimensional energy spectrum defined as $\widetilde{E}_{\bm k}(y)=\widetilde{E}(k_x,k_z,y)$, we assume the following forcing covariance:
\begin{equation}
    \mathbf S_{ff}(k_x,k_z,\omega)= \gamma
    \begin{bmatrix}
        \widetilde{E}_\mathbf{k}(y) & 0 & 0\\
        0 & \widetilde{E}_\mathbf{k}(y) & 0\\
        0 & 0 & \widetilde{E}_\mathbf{k}(y)
    \end{bmatrix},
\end{equation}
where the same spatial forcing shape is assumed in all three momentum equations without any temporal color.
Here, $\gamma$ is a scaling factor given as,
 
\begin{equation}
    \gamma(k_x,k_z)=
\frac{\displaystyle\int_0^2\widetilde E_{\bm k}(y)\,dy}
     {\displaystyle\int_0^2\widetilde E^{\mathbf H}_{\bm k}(y)\,dy},
\end{equation}
where $\widetilde{E}^\mathbf{H}_\mathbf{k}(y)$ is the 2D energy spectrum computed via resolvent analysis, i.e., the eddy-viscosity-enhanced LNS equations.
The scaling factor $\gamma(k_x,k_z)$ is calibrated using the unscaled, frequency-integrated response of the rigid-wall reference system and is subsequently held fixed for the compliant-wall calculations.
While ideally, spatiotemporal forcing color is required to adequately reconstruct second-order statistics of a turbulent channel flow~\citep{Zare_Jovanovic_Georgiou_2017}, the above approach has been shown to estimate the effect of flow control on eddy viscosity in channel flows in cases involving transverse wall oscillations~\citep{Moarref_Jovanovic_2012} and riblets~\citep{Ran_Zare_Jovanovic_2021,naseri2025turbulencemodelingribletsdomain}.

\begin{figure}
    \centering
    \begin{overpic}{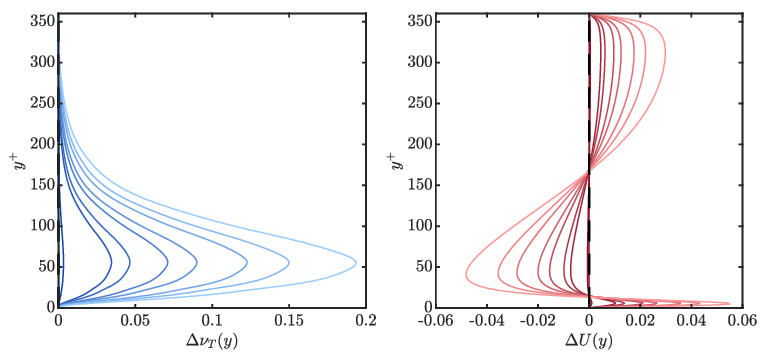}
    \put(0.6,45){(a)}
    \put(49.5,45){(b)}
    \put(27.5,13){$G$}
    \put(18,12){
    \begin{tikzpicture}
        \draw[-stealth] (0,0) -- (-2.6,0.0);
    \end{tikzpicture}
    }
    \put(80,35){$G$}
    \put(75,37){
    \begin{tikzpicture}
        \draw[-stealth]  (1.8,0.6) -- (0,0) ;
    \end{tikzpicture}
    }
    \put(82,9.5){$G$}
    \put(78,8.5){
    \begin{tikzpicture}
        \draw[-stealth] (1,-0.5) -- (-0.2,-0.5);
    \end{tikzpicture}
    }
        \put(64,13){$G$}
    \put(60,12){
    \begin{tikzpicture}
        \draw[-stealth] (-0.2,-0.5) -- (1,-0.5) ;
    \end{tikzpicture}
    }
    \end{overpic}
  \caption{Variation in (a) turbulent viscosity correction $\Delta \nu_T(y)$, and (b) the mean velocity correction $\Delta U(y) = U_c(y) - U_r (y)$, as a function of material stiffness $G$; the arrows indicate the direction of increasing material stiffness $G$. The solid (a) blue and (b) red coloured lines get darker with material stiffness; the black dashed lines indicate the rigid wall results in all cases. The material stiffness values considered are $G=100$, $10$, $1$, $0.75$, $0.5$, $0.4$, $0.3$, $0.25$, and $0.2$.} 
    \label{fig:nuT_U_dU}
\end{figure}
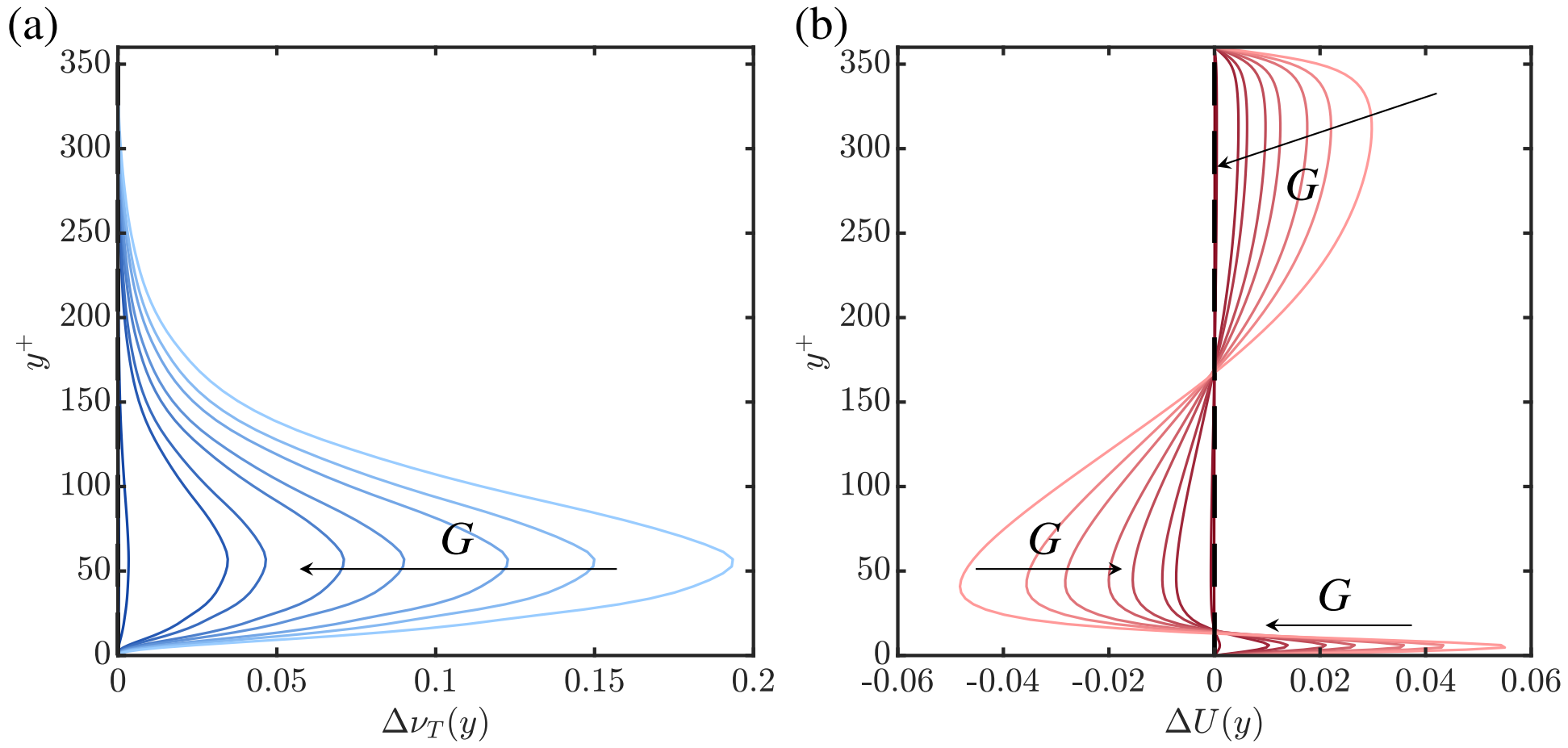

The velocity and velocity-gradient correlations required for $k(y)$ and $\epsilon(y)$ at each wavenumber-frequency triplet can then be extracted from the CSD matrix $\mathbf S_{uu}(k_x,k_z,\omega)$. 
At each spectral point, the diagonal entries of $\mathbf S_{uu}$ provide the componentwise spectral variance densities, while the gradient correlations entering $\epsilon$ are obtained from the full velocity covariance using the corresponding spatial derivatives.
Summing these contributions over the sampled $(k_x,k_z,\omega)$ domain yields the wall-normal profiles $k(y)$ and $\epsilon(y)$ used in the closure in equation \,\ref{dnu_equation}. In this section, the drag reduction calculations are conducted by integrating over the wavenumber--frequency domain $(k_x,k_z,\omega)$ with ranges given by
\begin{equation}
    5 \leq \lambda_x^+ \leq 5000, \qquad 5 \leq \lambda_z^+ \leq 5000, \qquad -25 \leq c^+ \leq 25, 
    \label{integration_range_DR}
\end{equation}
where the wavelength range is sampled by 21 equispaced points in the log space, whereas the phase-speed range is sampled by 81 linearly spaced points with equal intervals. Here, we define the range of the domain based on phase-speed instead of frequency $\omega$, because it is easier to place physical bounds based on the phase-speed $c^+$; the resultant frequency is then computed as 
$\omega=k_x\,c=2\pi Re_\tau c/\lambda_x^+$.
The change in eddy-viscosity $\Delta \nu_{T,c}$ computed using equation~\ref{dnu_equation} produces pronounced differences with a change in $G$ as shown in figure~\ref{fig:nuT_U_dU}\,(a).

The updated compliant-wall mean profile, $\overline{U}_c$, is
obtained using the modified eddy viscosity by solving

\begin{equation}
\frac{d}{dy}\left[(1+\nu_{T,c})\frac{d\overline U_c}{dy}\right]=Re_\tau\overline P_{x,c},
\end{equation}
Following \citet{Ran_Zare_Jovanovic_2021}, the pressure gradient
is adjusted to maintain the rigid-wall bulk velocity, subject to
$\overline U_c(0)=\overline U_c(2)=0$ and
$\frac{1}{2}\int_0^2\overline U_c(y)\,dy=U_{b,r}$ for a channel
of nondimensional height $2$.
We define the normalized pressure-gradient correction as $\Pi_c=(\overline P_{x,c}-\overline P_{x,r})/\overline P_{x,r}$, where $\overline P_{x,r}<0$ is the signed rigid-wall reference gradient. The estimated reduction in bottom-wall skin friction is then
\begin{equation}
DR=1-\frac{\overline U_c'(0)}{\overline U_r'(0)}=\frac{2Re_\tau(\overline P_{x,c}-\overline P_{x,r})-[\overline U_c'(2)-\overline U_r'(2)]}{D_{\mathrm{rigid}}}=-2\Pi_c-\frac{\overline U_c'(2)-\overline U_r'(2)}{D_{\mathrm{rigid}}},
\end{equation}
where $D_{\mathrm{rigid}}=\overline U_r'(0)=-\overline U_r'(2)>0$, and primes denote wall-normal differentiation. 
The change in the mean profile at fixed bulk velocity,
$\Delta \overline U=\overline U_c-\overline U_r$, is shown in
figure~\ref{fig:nuT_U_dU}\,(b).
The pressure-balance expression follows by integrating the mean-momentum equation across the channel, assuming that the modeled eddy viscosity vanishes at both walls. Positive $DR$ indicates a reduction in the modeled bottom-wall skin friction, while the upper-wall term accounts for the change in shear at the top (rigid) wall.

\begin{figure}
    \centering
\includegraphics{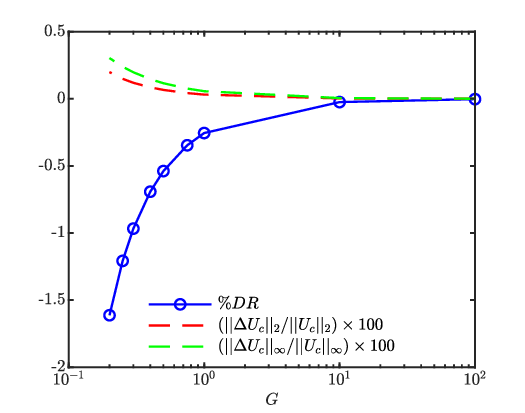}
      \caption{Estimated percentage drag reduction $100DR$ (blue) and normalized mean-profile deviations $100\|\Delta\overline U\|_2/\|\overline U_r\|_2$ (red dashed) and $100\|\Delta\overline U\|_\infty/\|\overline U_r\|_\infty$ (green dashed), as functions of material stiffness $G$. Positive $DR$ denotes a reduction in the estimated compliant-wall skin friction.  Results are presented as percentages.} 
    \label{fig:percentDR}
\end{figure}

Figure~\ref{fig:percentDR} shows the percentage drag reduction, $100DR$, with surface stiffness, $G$; here a positive value indicates drag reduction. The departure of the predicted mean profile from the rigid-wall state is quantified using $\|\Delta\overline U\|_2/\|\overline U_r\|_2$ and $\|\Delta\overline U\|_\infty/\|\overline U_r\|_\infty$. 
As the wall is softened, the estimated skin-friction drag
increases, reaching approximately $1.61\%$ for $G=0.2$. For this
softest retained case, the mean-profile deviation from the
rigid-wall state remains below $0.5\%$.
For highly compliant walls, $G\sim O(0.01)$, the model produces
large eddy-viscosity corrections and mean-flow distortions.
Moreover, unstable eigenvalues occur within the sampled spectral
domain for $G\leq 0.1$, precluding a statistically stationary
response interpretation about the prescribed mean flow.

Within the retained cases, the largest increase occurs at the softest panel. This finding of an increase in drag is consistent with recent findings~\citep{Rosti_Brandt_2017,Esteghamatian_Katz_Zaki_2022,Greidanus_Delfos_Picken_Westerweel_2022,Lu_Xiang_Zaki_Katz_2024}, which indicate that a turbulent boundary layer interacting with a compliant wall may experience an increase in skin-friction drag rather than a reduction. 
Because the formulation is linearized about the rigid-wall mean,
the predictions are most relevant when the mean-profile changes
remain small. The results should therefore be interpreted as
leading-order estimates for the moderate-compliance cases
considered here.
For softer surfaces, an iterative continuation procedure that updates the mean flow as $G$ is reduced could be useful; however, that remains beyond the scope of this study. Within this range of validity, the framework provides a computationally efficient means of screening surface parameters and identifying promising configurations for higher-fidelity simulation or experiment.

\section{Conclusions}
\label{sec:conclusion}
We have developed a two-way coupled resolvent-Galerkin framework for turbulent channel flow over finite-thickness viscoelastic surfaces with through-thickness variation in material properties. 
The fluid is described by the Navier--Stokes equations linearized
about a turbulent mean profile, with the nonlinear terms treated
as forcing to the perturbation dynamics, following
\citet{McKEON_SHARMA_2010}.
The solid is described by the three-dimensional elastodynamic equations and discretized through its thickness. Condensation of the structural system yields a $3\times3$ surface-impedance operator that enters directly into the fluid boundary condition. The formulation couples all three components of surface motion to the specified fluid-traction operator. 
The main methodological contribution is the coupled resolvent--Galerkin treatment of a multilayer solid, in which finite thickness, tangential deformation, material damping, and wall-normal variation of material properties can be included naturally without reducing the coating to a fitted scalar spring-mass-damper model. Moreover, describing the solid as a finite-thickness viscoelastic panel rather than a spring-mass-damper system allows for a more natural comparison with physical experiments and real-world scenarios~\citep{duncan_1987_comparative}. 
The fluid solver was verified against established laminar- and turbulent-channel resolvent results, and the solid solver reproduced the finite-thickness coating response to prescribed harmonic normal and shear tractions, as well as its dependence on material compressibility.
A temporal stability analysis identified growing disturbances
about the prescribed turbulent mean. This analysis distinguishes
the formal resolvent gains of an unstable operator from
covariance-based statistics, whose stationary response
interpretation requires stable finite dynamics.

The solid-only calculations show that a homogeneous panel acts
as a spatiotemporal filter, with its response influenced by the
characteristic wavelength $\lambda_x\approx3t$, the shear-wave
speed, and viscoelastic damping. The formulation also retains cross-coupled tangential and wall-normal traction-displacement responses that are absent from pressure-driven, wall-normal-displacement-only models. {When the panel is coupled to the turbulent flow, the overlap with flow length- and time-scales determines the extent to which these structural scales can be dynamically relevant.} 

In the coupled calculations, stiff panels remain close to the rigid-wall limit. As the panel is softened, the distribution of low-rank resolvent amplification is substantially modified. 
The panels showing substantial changes in the near-wall
fractional-energy distribution also admit unstable modes. For
$G=0.1$, the changes occur largely in the stable region, whereas
for $G=0.01$ they lie closer to the boundary between stable and
unstable regions. 
Wall compliance also modifies the structure and wall-normal
position of the leading near-wall mode and permits a non-zero
streamwise velocity component at the mean interface.
The usual pairing of the leading gains and the centerline
symmetry of the rigid-wall response also break down at scales
relevant to the near-wall cycle.
{The compliant-to-rigid gain ratios show that the leading gain, $\sigma_1$, remains close to, or greater than, its rigid-wall value over much of the wavelength plane, while compliant-to-rigid gain ratios for $\sigma_2$ exhibit both amplification and attenuation, albeit at different scales.} 
Wall compliance can therefore attenuate selected response modes at some scales while amplifying the response at others. A useful design criterion should therefore account for changes
in the gains of both optimal and suboptimal modes across the
relevant spectral range.

Similarly, fractional energy and stability contours for a $G=0.01$ soft panel, as a function of panel thickness, indicate that material stiffness alone does not determine compliance-induced effects, as the fractional energy response approaches the rigid-wall limit at the smallest panel thickness considered. Compliance-induced effects are most pronounced over an intermediate range of thicknesses, suggesting they are maximized when structural and flow scales are comparable. 
The Reynolds-number study also shows that the response depends
on how the panel thickness is scaled. As $Re_\tau$ increases,
fixed $t^+=100$ and fixed $t^*/\delta=0.5555$ produce different
fractional-energy distributions and stability boundaries.
Viscoelastic damping also plays an important role in determining the timescales at which the solid responds. For the Rayleigh model, stiffness-proportional damping has the stronger effect at relevant frequencies, suppressing narrow resonances and lowering the high-frequency threshold below which the material response is active.

Vertical inhomogeneity in the solid panels provides an additional means to shape the overlap between solid and fluid scales. 
For the multilayer and functionally graded configurations
considered here, the response approximately follows the
shear-wave reference associated with the fluid-facing material,
while the stiffness distribution beneath it affects the effective
thickness and spectral bandwidth.
A stiff surface layer shifts the structural cut-off towards lower wavenumbers, consistent with a larger effective thickness, whereas a soft surface layer shifts it towards higher wavenumbers. This interpretation accounts well for the first two structural modes (singular modes of the inverse surface-impedance operator, which maps surface traction to surface deformation) for all panels and for the third structural mode of all but one of the panels considered here; the third structural mode for the stiff-on-top multilayer panel is more complicated and does not follow this trend. 
The discrete multilayer panels substantially modify the
fractional-energy distribution relative to the uniform
baselines. They also exhibit smaller unstable regions than the
uniformly soft $G=0.01$ panel, suggesting a way to modify the
flow response while limiting the unstable range.
The graded panels produce more subdued changes in the coupled
response than the solid-only results might suggest. For the
stiffness profiles examined here, discrete layers produce more
pronounced changes in the fractional-energy distribution than the
graded profiles. Candidate architectures must therefore be
assessed through their coupled response as well as their
solid-only dynamics.

Following the formulation of \citet{Moarref_Jovanovic_2012,Ran_Zare_Jovanovic_2021}, we also connected the mode-by-mode analysis to a leading-order estimate of skin-friction drag. Resolvent-based second-order statistics were used to evaluate changes in turbulent kinetic energy and dissipation, from which an eddy-viscosity correction, a modified mean profile, and the pressure-gradient change at constant bulk flow were obtained. {For the moderate-compliance cases retained in \S~\ref{sec:drag-estimation}, the model predicts a modest skin-friction increase, reaching approximately $1.61\%$ at $G=0.2$.} 
This value is a leading-order estimate that can guide the
screening of candidate configurations. The calculation is
linearized about the rigid-wall mean, uses a first-order
eddy-viscosity correction, and does not iterate the mean flow
and forcing statistics to convergence.

The main design implication is that a compliant coating cannot be characterized adequately by a static modulus alone. Its finite thickness introduces a structural length scale; its stiffness and density set interfacial wave speeds; its damping controls the accessible frequency band; and, for layered materials, the fluid-facing layer and the subsurface architecture play different roles. These properties must be selected together and matched to the wavelengths and phase speeds of the turbulent motions being targeted. This perspective also explains why stronger compliance is not necessarily beneficial: once structural resonances or additional amplified flow scales are introduced, a nominally more responsive surface may move farther from the desired regime.

Future work could include making the drag calculation self-consistent by iterating the mean velocity, eddy viscosity, and forcing covariance with compliant-wall statistics. Moreover, this formulation can be extended to wall-parallel periodic substructures, embedded features, spatially developing, curved, or axisymmetric boundary layers, and favorable or adverse pressure gradients. Complementary high-fidelity simulations and water-tunnel experiments can also be used to assess finite deformation, nonlinear coupling, form drag, and manufacturing constraints for promising designs. The present framework can help narrow the design space and
identify candidates for such simulations and experiments.

\backsection[Acknowledgments]{Portions of this manuscript benefited from the use of artificial intelligence tools to primarily improve readability and grammar. 
}
\backsection[Funding] {This work was supported in part by the Defense Advanced Research Projects Agency (DARPA) under grant number HR00112490348 and by the National Science Foundation (NSF) under grant number CBET-1943810.}

\backsection[Declaration of interests]{The authors report no conflict of interest.}

\backsection[Author ORCIDs]{  R. Bhagwat, https://orcid.org/0009-0009-2462-7442; K. Shoele, https://orcid.org/0000-0003-2810-0065
}

\appendix
\section{Fluid domain operators}\label{appA}
The standard molecular-viscosity linearized Navier--Stokes (LNS) operator $\mathbf A$ in equation~\eqref{forced_lde} is 
\begin{equation}
    \mathbf{A}=  \begin{pmatrix}
\frac{1}{Re_\tau} \nabla^2 - i k_x \bar{u}&-\overline{\frac{\partial u}{\partial y}} &0 &-i k_x \\
0 &\frac{1}{Re_\tau} \nabla^2  - i k_x \bar{u}&0 &-D_y \\
0 &0  &\frac{1}{Re_\tau} \nabla^2 - i k_x \bar{u}&-i k_z \\
-i k_x & -D_y  &-ik_z &0 
\end{pmatrix}, 
\end{equation}
The descriptor matrix multiplying the time derivative in equation~\eqref{forced_lde} is,
\begin{equation}
    \mathbf{B}=  \begin{pmatrix}
I &0 &0 &0 \\
0 &I &0 &0 \\
0 &0 &I &0 \\
0 &0 &0 &0 
\end{pmatrix}. 
\end{equation}
Here, $\nabla^2=D_y^2-k_x^2-k_z^2$ in the Fourier representation.
The identity blocks act on the wall-normal degrees of freedom of
the three velocity components. The last row of $\mathbf A$
imposes incompressibility, and the corresponding row of
$\mathbf B$ is zero.

For the drag calculations in \S\,\ref{sec:drag-estimation}, the
eddy-viscosity-enhanced LNS operator uses the effective-to-molecular
viscosity ratio $\nu_e/\nu=\nu_e^*/\nu^*=1+\nu_T$, where
$\nu_T=\nu_T^*/\nu^*$.
\begin{equation}
    \mathbf{A}_{\nu_e}=  \begin{pmatrix}
\frac{\nu_e/\nu}{Re_\tau} \nabla^2 + \Phi - i k_x \bar{u}&-\overline{\frac{\partial u}{\partial y}} + \frac{i k_x}{\nu Re_\tau} \overline{\frac{\partial \nu_e}{\partial y}}  &0 &-i k_x \\
0 &\frac{\nu_e/\nu}{Re_\tau} \nabla^2  + 2 \Phi - i k_x \bar{u}&0 &-D_y \\
0 & \frac{i k_z}{\nu Re_\tau} \overline{\frac{\partial \nu_e}{\partial y}}   &\frac{\nu_e/\nu}{Re_\tau} \nabla^2  + \Phi - i k_x \bar{u}&-i k_z \\
-i k_x & -D_y  &-ik_z &0 
\end{pmatrix}, 
\end{equation}
where,
\begin{equation}
    \Phi = \frac{1}{\nu Re_\tau} \overline{\frac{\partial \nu_e}{\partial y}} D_y 
\end{equation} 
For a turbulent channel with $0\leq y\leq2$, the rigid-wall
reference eddy viscosity is given by the Cess
profile~\citep{Reynolds_Tiederman_1967}, 
\begin{equation}
    \nu_T (y) = \frac{1}{2} \left[1 + \left( \frac{c_2 Re_\tau}{3} (2y - y^2)(3-4y+2y^2)(1 - e^{(|y-1|-1)Re_\tau / c_1})  \right)^2 \right]^{1/2} - \frac{1}{2}\,.
    \label{eqn:cess}
\end{equation}
We use the fitted values $c_1=46.2$ and $c_2=0.61$ for
$Re_\tau=180$, following \citet{Moarref_Jovanovic_2012}.
\section{Solid domain operators}\label{appA2}
The solid domain operators are assembled following the procedure outlined in \citet{barbosa_2012}. For the isotropic material used in each layer, the constitutive matrix $\mathbf D$ in equation~\eqref{waveeqn} is
\begin{equation}
    \mathbf{D}=  \begin{pmatrix}
\lambda + 2\breve{G} &\lambda &\lambda &0  & 0 &0\\
\lambda & \lambda + 2\breve{G} &\lambda &0  & 0 &0\\
\lambda &\lambda &\lambda + 2\breve{G} &0  & 0 &0\\
0 &0 &0 &\breve{G}  & 0 &0\\
0 &0 &0 &0  & \breve{G} &0\\
0 &0 &0 &0  & 0 &\breve{G}\\
\end{pmatrix}, 
\end{equation}
where $\lambda=\lambda^*/(\rho_f^*{u_\tau^*}^2)$ is the non-dimensional Lam\'e parameter,
\begin{equation}
    \lambda = \frac{2\breve{G}\nu_p}{1-2\nu_p},
\end{equation}
Here, $\nu_p$ is Poisson's ratio and
$\breve G=G^*/(\rho_f^*{u_\tau^*}^2)$ is the nondimensional shear
modulus. The stiffness parameter $G=\breve G/\breve G_{\mathrm{ref}}$
is normalized by $\breve G_{\mathrm{ref}}=121$, the reference
value from \citet{Esteghamatian_Katz_Zaki_2022}.

In this context, $\lambda$ denotes the Lam\'e parameter and is distinct from the temporal eigenvalue, $\lambda_{(e)}$, used in the stability formulation.

The spatial derivative operators $\mathbf L$, $\mathbf L_1$,
$\mathbf L_2$, and $\mathbf L_3$ in
equation~\eqref{L1L2L3Eqn} are

\begin{equation}
    \mathbf{L}^T=  \begin{pmatrix}
\partial_x &0 &0 &0  &\partial_z &\partial_y\\
0 &\partial_y &0 &\partial_z  &0 &\partial_x\\
0 &0 &\partial_z &\partial_y &\partial_x &0\\
\end{pmatrix},  
\end{equation}

\begin{equation}
    \mathbf{L_1}^T=  \begin{pmatrix}
1 &0 &0 &0  & 0 &0\\
0 &0 &0 &0  & 0 &1\\
0 &0 &0 &0  & 1 &0\\
\end{pmatrix}, 
\end{equation}

\begin{equation}
    \mathbf{L_2}^T=  \begin{pmatrix}
0 &0 &0 &0  & 0 &1\\
0 &1 &0 &0  & 0 &0\\
0 &0 &0 &1  & 0 &0\\
\end{pmatrix}, 
\end{equation}
\noindent and
\begin{equation}
    \mathbf{L_3}^T=  \begin{pmatrix}
0 &0 &0 &0  & 1 &0\\
0 &0 &0 &1  & 0 &0\\
0 &0 &1 &0  & 0 &0\\
\end{pmatrix}. 
\end{equation}

The stiffness, mass, and damping matrices form the dynamic-stiffness matrix $\mathbb M=\mathbf K-i\omega\mathbf C-\omega^2\mathbf M$. 
For a linear finite element $i$ with constant density
$\rho_i$ over $y_i\leq y\leq y_{i+1}$, the element mass matrix is
\begin{equation}
   m_i = \int_{h_i}  \rho    N_i^T N_i\, dy = \frac{\rho_i h_i}{6} \begin{bmatrix}
2\mathbf{I} & \mathbf{I}\\
\mathbf{I} & 2\mathbf{I}
\end{bmatrix},
\end{equation}
\noindent where $\mathbf I$ is the $3\times3$ identity matrix,
$\rho_i=\rho_{s,i}^*/\rho_f^*$ and $h_i=h_i^*/\delta$
are the non-dimensional density and thickness of element $i$,
and $\mathbf N_i$ is its block shape-function matrix
for the three displacement components. 
The two block rows and columns of the element mass matrix
correspond to the three displacement components at nodes $i$
and $i+1$. Assembly over the $N_s$ elements gives the global
mass matrix $\mathbf M$ of size $3(N_s+1)\times3(N_s+1)$,
which acts on the nodal displacement amplitudes
$\hat{\boldsymbol X}\in\mathbb C^{3(N_s+1)}$.

Similarly, the local spectral stiffness matrix in its general form is given as, 
\begin{equation}
\mathbf K_i=\int_{h_i}\mathbf B_i^\dagger\mathbf D\mathbf B_i\,dy,
\qquad
\mathbf B_i=ik_x\mathbf L_1\mathbf N_i+
\mathbf L_2\mathbf N_{i,y}+ik_z\mathbf L_3\mathbf N_i .
\end{equation}
{where $\dagger$ denotes conjugate transpose. For the expressions below,
the local nodes are ordered in increasing $y$, with
$\eta=y-y_i\in[0,h_i]$ and $y_{i+1}=y_i+h_i$. Thus,
$\mathbf N_i=[(1-\eta/h_i)\mathbf I,\,(\eta/h_i)\mathbf I]$
and $\mathbf N_{i,y}=h_i^{-1}[-\mathbf I,\,\mathbf I]$.}
The element stiffness is the sum of the directional contributions
$\mathbf K_i=\sum_{j=1}^3\sum_{k=1}^3\mathbf K_i^{(j,k)}$,
with $k_1=k_x$ and $k_3=k_z$. The closed forms below assume
that $\mathbf D$ is constant within each element.

If $j \neq 2$ and $k \neq 2$, 
{\begin{equation}
    \mathbf K_i^{(j,k)}=k_jk_k\int_{h_i}\mathbf N_i^T\mathbf D_{j,k}\mathbf N_i\,dy
=\frac{k_jk_kh_i}{6}
\begin{bmatrix}2\mathbf D_{j,k}&\mathbf D_{j,k}\\
\mathbf D_{j,k}&2\mathbf D_{j,k}\end{bmatrix},
\end{equation}}
if $j = 2$ and $k \neq 2$, 
{\begin{equation}
    \mathbf K_i^{(2,k)}=ik_k\int_{h_i}\mathbf N_{i,y}^T\mathbf D_{2,k}\mathbf N_i\,dy
=\frac{ik_k}{2}\begin{bmatrix}-\mathbf D_{2,k}&-\mathbf D_{2,k}\\
\mathbf D_{2,k}&\mathbf D_{2,k}\end{bmatrix},
\end{equation}}
if $j \neq 2$ and $k = 2$, 

\begin{equation}
    \mathbf K_i^{(j,2)}=-ik_j\int_{h_i}\mathbf N_i^T\mathbf D_{j,2}\mathbf N_{i,y}\,dy
=\frac{ik_j}{2}\begin{bmatrix}\mathbf D_{j,2}&-\mathbf D_{j,2}\\
\mathbf D_{j,2}&-\mathbf D_{j,2}\end{bmatrix},
\end{equation}

and if $j = 2$ and $k = 2$, 
{\begin{equation}
    \mathbf K_i^{(2,2)}=\int_{h_i}\mathbf N_{i,y}^T\mathbf D_{2,2}\mathbf N_{i,y}\,dy
=\frac{1}{h_i}\begin{bmatrix}\mathbf D_{2,2}&-\mathbf D_{2,2}\\
-\mathbf D_{2,2}&\mathbf D_{2,2}\end{bmatrix}.
\end{equation}}

Here $\mathbf N_{i,y}$ denotes the derivative of the shape-function matrix in the wall-normal direction. For all $j$ and $k$, $\mathbf{D}_{j,k}$ is defined as 
\begin{equation}
    \mathbf{D}_{j,k} = \mblj^T\,\mbd\,\mblk\,.
\end{equation}

The element stiffness matrices are assembled into the global
matrix $\mathbf K$ of size $3(N_s+1)\times3(N_s+1)$.
The global damping matrix $\mathbf C$ is obtained from the
Rayleigh model in equation~\eqref{RayleighDampingEqn}.

%=====================================================================
\section{Stability analysis and verification of the eigenvalue calculations}
\label{app:gstabProcedure}

\subsection{Fluid--solid descriptor system}

After imposing the substrate constraint, the structural equations can be written in the spectral domain as
\begin{equation}
 (\mathbf K_s+\lambda_{(e)}\mathbf C_s+\lambda_{(e)}^2\mathbf M_s)\bm X_s
 =\mathbf L_T\bm q_f,\qquad
 \mathbf C_s=\alpha\mathbf M_s+\beta\mathbf K_s.
 \label{eq:gstabSolidPolynomial}
\end{equation}
Introducing the solid velocity
$\bm V_s=\lambda_{(e)}\bm X_s$ gives the first-order
eigenvalue problem in equation~\eqref{eq:gstabPencil}, with
the uncondensed matrix pair
\begin{equation}
 \mathbf A_g=
 \begin{bmatrix}
 \mathbf A_{f,\Gamma}&\mathbf B_{\Gamma X}&\mathbf B_{\Gamma V}\\
 \mathbf0&\mathbf0&\mathbf I\\
 \mathbf L_T&-\mathbf K_s&-\mathbf C_s
 \end{bmatrix},\qquad
 \mathbf E_g=
 \begin{bmatrix}
 \mathbf E_{f,\Gamma}&\mathbf0&\mathbf0\\
 \mathbf0&\mathbf I&\mathbf0\\
 \mathbf0&\mathbf0&\mathbf M_s
 \end{bmatrix}.
 \label{eq:gstabPencilBlocks}
\end{equation}
The fluid mass operator $\mathbf E_{f,\Gamma}$ has {zero rows} for the
algebraic constraints.
With $n_f=4N_y$ fluid unknowns and $n_s=3N_s$ unconstrained solid displacements after clamping one end, the unreduced pencil has size $(n_f+2n_s)\times(n_f+2n_s)$. The blocks $\mathbf B_{\Gamma X}$ and $\mathbf B_{\Gamma V}$ are $n_f\times n_s$, and $\mathbf L_T$ is $n_s\times n_f$. 
The blocks $\mathbf B_{\Gamma X}$ and $\mathbf B_{\Gamma V}$
are nonzero only in the interface kinematic rows. The block
$\mathbf L_T$ applies the three fluid-traction components to
the corresponding solid surface degrees of freedom.
The fluid blocks $\mathbf A_{f,\Gamma}$ and
$\mathbf E_{f,\Gamma}$ both have size $n_f\times n_f$.
The interface kinematic constraints enter through the
corresponding rows of $\mathbf A_g$, with zero rows in
$\mathbf E_g$.

The interface rows impose
\begin{equation}
 \hat u(0)=\hat V_{\mathrm{surf},x}
           -\overline U'(0)\hat X_{\mathrm{surf},y},\qquad
 \hat v(0)=\hat V_{\mathrm{surf},y},\qquad
 \hat w(0)=\hat V_{\mathrm{surf},z}.
 \label{eq:gstabKinematic}
\end{equation}

The fluid-traction vector $\widehat{\bm T}_f$ acting on the solid
with outward normal $\bm n_s=\bm e_y$ is

{
\begin{equation}
 \widehat{\bm T}_f=
 \left.\begin{bmatrix}
 \frac{\nu_e/\nu}{Re_\tau}(\mathrm D_y\hat u+i k_x\hat v)\\
 -\hat p+2\frac{\nu_e/\nu}{Re_\tau}\mathrm D_y\hat v\\
 \frac{\nu_e/\nu}{Re_\tau}(\mathrm D_y\hat w+i k_z\hat v)
 \end{bmatrix}\right|_{y=0},
 \label{eq:gstabTraction}
\end{equation}
}
In these analyses, the wall-unit wavelengths are related to the outer-scaled wavenumbers by
$k_x=2\pi Re_\tau/\lambda_x^+$ and $k_z=2\pi Re_\tau/\lambda_z^+$;
the dimensionless time is $\mathcal T=\mathcal T^*u_\tau^*/\delta$.

\subsection{Extraction of the finite spectrum and projection onto compatible states}

The QZ decomposition of the generalized eigenvalue pencil
$(\mathbf A_g,\mathbf E_g)$ returns homogeneous eigenvalue pairs
$(a_j,b_j)$. For $b_j\neq 0$, the corresponding finite eigenvalue is
\begin{equation}
    {\lambda_{(e)}}_j=\frac{a_j}{b_j}.
\end{equation}
A pair with $b_j=0$ and $a_j\neq 0$ represents an eigenvalue at
infinity, which is associated with an algebraic constraint of the
descriptor system. By contrast, the pair $(a_j,b_j)=(0,0)$ does not
define an eigenvalue and is therefore classified as indeterminate.

The indeterminate case is checked separately before the finite and
infinite eigenvalues are classified. A non-indeterminate pair is
regarded as infinite when
\begin{equation}
    \frac{|b_j|}
    {\max\left(|a_j|,|b_j|\right)}
    \leq \tau_n,
    \label{eq:gstabPairThreshold}
\end{equation}
where the numerical tolerance is set to $\tau_n=2\times10^{-14}$. 
All remaining pairs are retained as finite eigenvalues.
The criterion in equation~\eqref{eq:gstabPairThreshold} is
invariant under a common nonzero rescaling of $(a_j,b_j)$.
Pairs with both entries small relative to the pencil scaling
require a separate check for possible singularity or poor scaling.
An independent projection is used to check the finite spectrum
while retaining the pressure contribution to the traction
acting on the solid. To construct this projection, we first
separate the pressure degrees of freedom from the remaining
state variables, denoted by $\bm z$:
\begin{equation}
    \mathbf E_d\dot{\bm z}
    =
    \mathbf A_d\bm z+\mathbf B_p\bm p,
    \qquad
    \mathbf C_\Gamma\bm z=\bm 0,
    \label{eq:gstabDescriptorReduced}
\end{equation}
where $\bm p$ contains the pressure degrees of freedom and
$\mathbf C_\Gamma\bm z=\bm 0$ represents incompressibility and
the algebraic boundary and interface constraints.
The matrix $\mathbf B_p$ includes the pressure contributions
to both the fluid equations and the structural traction.

Here, orthonormal null-space bases are first constructed using singular-value
decompositions. Specifically, the columns of $\mathbf Z$ span the null
space of $\mathbf C_\Gamma$, whereas the columns of $\mathbf W$ span
the left null space of $\mathbf B_p$:
\begin{equation}
    \mathbf C_\Gamma\mathbf Z=\mathbf 0,
    \qquad
    \mathbf W^\dagger\mathbf B_p=\mathbf 0.
    \label{eq:gstabNullSpaces}
\end{equation}
The substitution $\bm z=\mathbf Z\bm y$ enforces the compatibility
constraint exactly. Premultiplication by $\mathbf W^\dagger$ then eliminates
the pressure term and gives the reduced generalized eigenvalue problem
\begin{equation}
    \left(\mathbf W^\dagger\mathbf A_d\mathbf Z\right)\bm y
    =
    \lambda_{(e)}
    \left(\mathbf W^\dagger\mathbf E_d\mathbf Z\right)\bm y,
    \qquad
    \bm z=\mathbf Z\bm y.
    \label{eq:gstabProjection}
\end{equation}
Let $n_z$ be the dimension of $\bm z$ and $n_e$ the number of
rows in $\mathbf E_d\dot{\bm z}
=\mathbf A_d\bm z+\mathbf B_p\bm p$. With
$r_\Gamma=\operatorname{rank}\mathbf C_\Gamma$ and
$r_p=\operatorname{rank}\mathbf B_p$, the bases have dimensions
$\mathbf Z\in\mathbb C^{n_z\times(n_z-r_\Gamma)}$ and
$\mathbf W\in\mathbb C^{n_e\times(n_e-r_p)}$.
The projected pencil is square when
$n_e-r_p=n_z-r_\Gamma$.

The finite spectra obtained from the original and projected formulations
are first compared by requiring the same number of finite eigenvalues.
The distance between an original pair $(a_i,b_i)$ and a pair from the projected system $(\widehat a_j,\widehat b_j)$ is measured using the \emph{chordal distance},
\begin{equation}
    d_{ij}
    =
    \frac{
        \left|a_i\widehat b_j-\widehat a_j b_i\right|
    }{
        \sqrt{|a_i|^2+|b_i|^2}
        \sqrt{|\widehat a_j|^2+|\widehat b_j|^2}
    }.
    \label{eq:gstabChordal}
\end{equation}
The chordal distance is invariant under nonzero rescaling of
either pair and is defined for finite and infinite eigenvalues
represented by non-indeterminate pairs. To obtain a one-to-one
match, we pair each original finite eigenvalue with the nearest
unmatched projected finite eigenvalue in chordal distance.
For repeated or tightly clustered eigenvalues, agreement is
additionally assessed by comparing the corresponding deflating
subspaces in common state coordinates.

After the eigenvalue problem is solved in projected coordinates,
the full right and left eigenvectors are reconstructed,
including the pressure components and adjoint constraint
components, before the residuals are evaluated using the
original matrices.
For the reconstructed right eigenvector $\bm x_j$ and its
projected eigenvalue pair $(\widehat a_j,\widehat b_j)$, the
normalized residual is
\begin{equation}
    r_j^R
    =
    \frac{
        \left\|
            \widehat b_j\mathbf A_g\bm x_j
            -
            \widehat a_j\mathbf E_g\bm x_j
        \right\|_2
    }{
        \left(
            |\widehat b_j|\|\mathbf A_g\|_2
            +
            |\widehat a_j|\|\mathbf E_g\|_2
        \right)
        \|\bm x_j\|_2
    }.
    \label{eq:gstabResidual}
\end{equation}
The corresponding left residual is evaluated from the adjoint
homogeneous eigenvalue equation,
\begin{equation}
    r_j^L
    =
    \frac{
        \left\|
            \overline{b_j}\mathbf A_g^\dagger\bm\ell_j
            -
            \overline{a_j}\mathbf E_g^\dagger\bm\ell_j
        \right\|_2
    }{
        \left(
            |b_j|\|\mathbf A_g\|_2
            +
            |a_j|\|\mathbf E_g\|_2
        \right)
        \|\bm\ell_j\|_2
    },
    \label{eq:gstabLeftResidual}
\end{equation}
where $\bm\ell_j$ is the left eigenvector.
The acceptance criteria for the algebraic residuals and agreement between the computed finite spectra are
\begin{equation}
    \max_j\left\{r_j^R,r_j^L\right\}<10^{-13},
    \qquad
    \max_i d_{i,\pi(i)}<10^{-7},
    \label{eq:gstabAcceptance}
\end{equation}
where $\pi(i)$ is the index of the projected eigenvalue matched to the $i$th eigenvalue of the original spectrum. 
For the spectra reported here, the maximum normalized residual
over all left and right eigenvectors is below $10^{-13}$, and
the maximum matched chordal distance is below $1.5\times10^{-11}$.
Both satisfy the tolerances in equation~\eqref{eq:gstabAcceptance}.
These checks assess algebraic accuracy and agreement between the
two formulations at the chosen spatial discretization. 

\bibliographystyle{unsrtnat}
\bibliography{refs}
\end{document}